\documentclass{aastex6}

\newcommand\ia{\'{\i}}

\begin{document}

\title{Orbits for eighteen visual binaries and two double-line
  spectroscopic binaries observed with HRCAM on the CTIO SOAR 4m
  telescope, using a new Bayesian orbit code based on Markov Chain
  Monte Carlo\footnote{Based on observations obtained at the Southern
    Astrophysical Research (SOAR) telescope, which is a joint project
    of the Minist\'{e}rio da Ci\^{e}ncia, Tecnologia, e
    Inova\c{c}\~{a}o (MCTI) da Rep\'{u}blica Federativa do Brasil, the
    U.S. National Optical Astronomy Observatory (NOAO), the University
    of North Carolina at Chapel Hill (UNC), and Michigan State
    University (MSU).}}

\author{Rene A. Mendez}
\affil{Department of Astronomy\\
Facultad de Ciencias F\'{\i}sicas y Matem\'aticas, Universidad de Chile\\
Casilla 36-D, Santiago, Chile}

\author{Ruben M. Claveria, Marcos E. Orchard, and Jorge F. Silva}
\affil{Department of Electrical Engineering\\
Information and Decision Systems Group (IDS)\\
Facultad de Ciencias F\'{\i}sicas y Matem\'aticas, Universidad de Chile \\
Beauchef 850, Santiago, Chile}

\begin{abstract}
We present orbital elements and mass sums for eighteen visual binary
stars of spectral types B to K (five of which are new orbits) with
periods ranging from 20 to more than 500~yr. For two double-line
spectroscopic binaries with no previous orbits, the individual
component masses, using combined astrometric and radial velocity data,
have a formal uncertainty of $\sim0.1 M_\odot$. Adopting published
photometry, and trigonometric parallaxes, plus our own measurements,
we place these objects on an H-R diagram, and discuss their
evolutionary status. These objects are part of a survey to
characterize the binary population of stars in the Southern
Hemisphere, using the SOAR 4m telescope+HRCAM at CTIO. Orbital
elements are computed using a newly developed Markov Chain Monte Carlo
algorithm that delivers maximum likelihood estimates of the
parameters, as well as posterior probability density functions that
allow us to evaluate the uncertainty of our derived parameters in a
robust way. For spectroscopic binaries, using our approach, it is
possible to derive a self-consistent parallax for the system from the
combined astrometric plus radial velocity data (``orbital parallax''),
which compares well with the trigonometric parallaxes. We also present
a mathematical formalism that allows a dimensionality reduction of the
feature space from seven to three search parameters (or from ten to
seven dimensions - including parallax - in the case of spectroscopic
binaries with astrometric data), which makes it possible to explore a
smaller number of parameters in each case, improving the computational
efficiency of our Markov Chain Monte Carlo code.
\end{abstract}


\section{Introduction} \label{sec:intro}

The laws of physics, applied to stars in hydrostatic equilibrium,
indicate, through the well-known Vogt-Russell theorem
(\cite{ka72,kiet12}), that the most fundamental parameter determining
the evolutionary path and the internal structure of stars of a given
chemical composition is their initial mass content (for a general
review see \cite{maet01}, and the textbooks by \cite{ib13} for details
of the physical models). Luckily, nature has been generous, providing
us with the means of determining this otherwise elusive property of
such distant objects through the observation of binary stars, and the
application of Kepler's laws of motion, which is the only direct
method to determine the mass of a stellar
system\footnote{Gravitational microlensing might eventually become
  another potentially very precise method \citep{ghet04,go14}, albeit
  so far has been restricted to few cases, see
  e.g. \cite{goet04}.}. This is particularly so, considering that
roughly half of solar-type stars in the solar neighborhood are in
binary systems, with a separation distribution that peaks at $\sim$60
AU, and which follows a log-normal distribution
(\cite{duma91,raet10,diet12,gaet14,maet14,yuet15,fuet17}).

One of the most fundamental relationships depicting the dependency on
mass of the star's properties is the mass-luminosity relation (MLR),
first discovered empirically in the early 20th century, and later on
``explained'' by the theory of stellar structure (\cite{ed24}). The
MLR has a statistical dispersion which cannot be explained exclusively
by observational errors in luminosity (or mass) - it seems to be an
intrinsic dispersion caused by differences in age and chemical
composition from star to star (e.g., for an effort to determine an
empirical low-metallicity empirical MLR in the solar vicinity see
\cite{hoet15}). As was emphasized long ago (\cite{vale88}),
improvements in the MLR requires not only good masses via the study of
binary stars, but also high precision trigonometric parallaxes.

\begin{figure}
\plotone{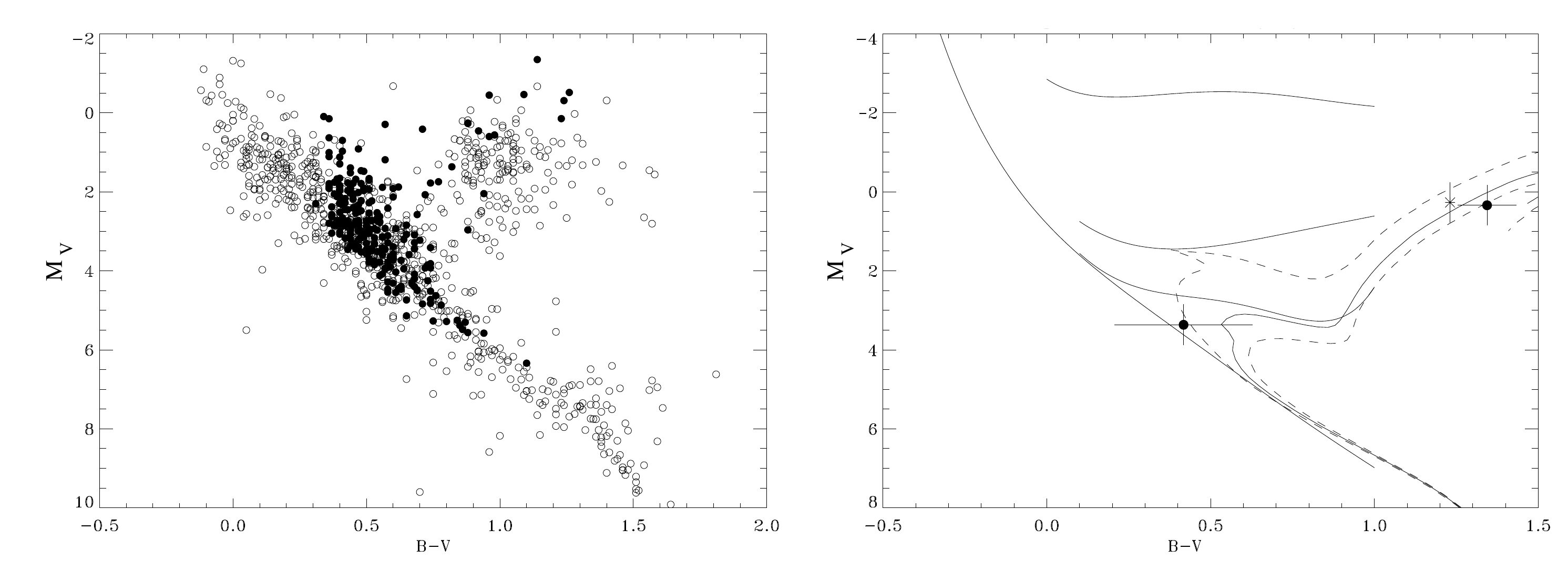}
\caption{Left panel: H-R diagram for the double stars discovered by
  Hipparcos within 250 pc of the Sun, the key sample being followed up
  in our speckle survey.  Stars with high-precision metallicities in
  the literature are shown as filled circles. This sample will
  dramatically improve our knowledge of the MLR and stellar
  astrophysics. By comparison, there are only 260 stars with $\le$1\%
  uncertainty in the semi-major axis, drawn from the latest update of
  the USNO Orbit Catalog, and less than half of those have
  metallicities. These are systems that can yield high-precision
  masses at present. Right panel: H-R diagram for the northern system
  HIP 15737 (HDS~423). The WIYN speckle photometry reveals the system
  as consisting of a giant and near--main-sequence star (the asterisk
  marks the position of the combined light of the system).  Using
  isochrone fitting, the age derived is $4.6 \pm 0.1$~Gyr (both
  figures courtesy of Elliot Horch).}\label{fig:sample}
\end{figure}

The prospect of exquisite-precision trigonometric parallaxes that will
be enabled by the {\it Gaia} satellite (\cite{ga16}) dramatically
changes the landscape of observational stellar astrophysics: If one
considers the Hipparcos double stars that lie within 250~pc of the
Solar system (\cite{liet97}), a parallax determined by {\it Gaia}
would (conservatively) yield an uncertainty well under 1\% for all
these objects. In this volume, there are 1,112 Hipparcos double star
discoveries (see Figure~\ref{fig:sample}, left panel) and 325
spectroscopic binaries from the Geneva-Copenhagen spectroscopic survey
(\cite{noet04}) south of DEC=+30$^\circ$. These two samples are an
important source of new binaries from which it will be possible to
derive masses, component luminosities, and effective temperatures in
the coming years. They have been systematically observed in the
Northern Hemisphere at the WIYN Telescope by Horch, van Altena, and
their collaborators (see, e.g., \cite{hoet11,hoet17}).  On the other
hand, Tokovinin has shown the capabilities of the instrument HRCAM at
the SOAR 4m telescope in northern Chile\footnote{For details of the
  instrument see
  http://www.ctio.noao.edu/$\sim$atokovin/speckle/index.html} for
binary-star research, producing significant results (see, e.g.,
\cite{to12, toet15}).

In 2014 we started a systematic campaign to observe the two primary
samples mentioned in the previous paragraph with HRCAM at SOAR, in
order to confirm their binary nature in the case of the Hipparcos
``suspected binaries'', and to add observational data on the confirmed
and spectroscopic binaries, to eventually compute their orbits, and
masses. So far we have been granted twelve observing nights in three
consecutive semesters at SOAR, and have observed more than seven
hundred objects from our sample. At the current rate of nights per
semester, the whole list of targets would have been observed at least
once in about three to four semesters. Historical astrometric data,
when available, will be combined to compute or improve on existing
orbits, but for many Hipparcos binaries we would have only the
Hipparcos point and the SOAR observation, therefore sustained
observations will be required to compute orbits. In addition to having
confirmed and resolved many systems, from the current observations we
have discovered twenty inner or outer subsystems in previously known
binaries, as well as one quadruple system (\cite{toet15,
  toet16}). This work will complement and significantly extend the
WIYN Northern sky speckle program, allowing us to compile an all-sky,
volume-limited speckle survey of these two primary samples. Surveying
objects out to 250~pc from the Sun, and in the Northern and Southern
hemisphere, will allow us to sample a larger volume in terms of
Galactocentric distances and distances from the Galactic plane than
what is possible with the Northern sample alone, permitting us to
encompass a broader range in metallicity and Galactic populations
(i.e., the thin \& thick disk components).

The WIYN speckle program is producing excellent results in terms of
binary statistics, component colors, orbits, and masses: It has
achieved the basic goal of characterizing the Hipparcos sample within
250~pc. However, extending that work to the Southern Hemisphere will
significantly improve the science in terms of the metallicity range
that can be studied (given the Galactic distribution of the
sources). When complete, our survey will open the door to many
sensitive tests of stellar evolution theory (see
Figure~\ref{fig:sample}, right panel), and a large number of new
points on the mass-to-luminosity relationship (MLR). With this we will
truly be able to investigate effects such as metallicity and age on
the MLR for the first time.  In cases where one component has evolved
off of the main sequence, age determinations will also be possible
(see, e.g., \cite{daet09}), a hint of this is shown in
Figure~\ref{fig:hrdiag} on Section~\ref{sec:comments} of our paper.

The Hipparcos binaries are also important in terms of binary
statistics.  \citet{hova11} have identified two distinct groups of
binaries (belonging to the thin and thick disk Galactic populations)
from Hipparcos and the Geneva spectroscopic catalog on the basis of
kinematics, location relative to the plane of the Galaxy, and
metallicity. Our Speckle sample extends out to 250~pc, a distance that
is needed for a clear differentiation of the two groups as shown in
the WIYN work, which reaches the peak of the \citet{duma91} period
distribution at that distance. Our observations of these targets are
needed to characterize their orbital motion in as many cases as
possible (and a 20-year time baseline provided by Hipparcos is an
excellent start in that regard), and - just as importantly - to obtain
accurate magnitudes and colors of the components.  Once accurate
information is known about these stars, a comparative study between
the two samples can be completed. This would include, {\it e.g.\ },
the period distribution as a function of metallicity and the
mass-ratio distribution as inferred from photometry and the projected
separation distribution, and (eventually) a correlation between period
and eccentricity or other orbital parameters. These statistics are
expected to be related to dynamical interactions at the time of star
formation and in the post-formation environment, and could be quite
different for thin and thick disk samples. If so, this could provide
important information about the formation of these systems in the
Galaxy.

We note that even though Gaia will deliver excellent parallaxes, it
will not resolve equal-brightness systems with separations smaller
than $\sim$0.1~arcsec (final processed data $\sim$2022). A single
sweep of the on-board star mapper will detect stars 0.3/0.7~arcsec
(along/cross scan) apart (\cite{albo16}). For the barely resolved
systems, the image ``blob'' at $\sim0.1$~arcsec separation will
introduce larger astrometric residuals which could only be alleviated
by having a proper orbit for the binary system (a similar issue
occurred with the Hipparcos satellite). Therefore, apart from
getting at the masses (main driver of our speckle survey), the
complementarity with {\it Gaia} is another strong reason to survey
nearby binaries now, thus making an all-sky survey of this sort very
timely.

In this paper we report orbits for eighteen visual binaries, and two
spectroscopic binaries observed in the context of our survey. The
orbits have been computed using a newly developed Bayesian code using
Markov Chain Monte Carlo techniques (MCMC hereafter). In
Section~\ref{sec:sample} we introduce the basic properties and
characteristics of our sample, while in Section~\ref{sec:orbits} we
introduce our methodology for computing the orbits and present the
results for the visual and spectroscopic binaries. In
Section~\ref{sec:comments} we provide an annotated list of comments
for each object on our sample. Finally, in Section~\ref{sec:conc} we
present our conclusions.

\section{Sample selection and basic properties} \label{sec:sample}

Our sample was derived from the published speckle data by Tokovinin
and collaborators (\cite{toet10, toet14,toet15,toet16}), which
includes objects from the samples indicated in the introduction, plus
objects from Tokovinin's own sample of nearby F- and G-type
dwarf-stars within 67 pc of the Sun (\cite{to14}). The selection
considered those orbital pairs for which their observed-computed
ephemeris ([O-C]) in either angular separation ([O-C]$_\rho$) or
position angle ([O-C]$_\theta$) evaluated at the epoch of our SOAR
Speckle data was too large in comparison with the internal precision
of our data and thus indicated that their orbits should be revised or
improved with the addition of our new data points. We added binaries
where first-time orbits could be computed using the SOAR
observations. Some objects of this initial list were later removed due
to either a lack of historical data or due to the impossibility of
improving on their orbits.

Historical astrometric measurements and computed orbital parameters
(if available) for all these binaries, compiled as part of the
Washington Double Star Catalogue effort (WDS hereafter,
\cite{haet01}\footnote{Updated regularly, and available at
  http://www.usno.navy.mil/USNO/astrometry/optical-IR-prod/wds/orb6}),
were kindly provided upon request by Dr. William Hartkopf from the US
Naval Observatory. 

As for the uncertainty (or equivalent weight) of each observation
(necessary for the orbit calculations, see Section~\ref{sec:orbits}),
we adopted the value indicated in the WDS entries when available, or
estimated the errors depending on the observation method
(interferometric {\it vs.} digital or photographic or micrometer
measurements).

\floattable
\begin{deluxetable}{cccccccccccc}
\tablecaption{Compiled photometry \label{tab:photom}}
\tabletypesize{\scriptsize}
\tablecolumns{12}
\tablewidth{0pt}
\tablehead{
\colhead{WDS name} & \colhead{Discoverer designation} & \colhead{HIP number} & 
\colhead{$V_{Sim}$\tablenotemark{a}} &
\colhead{$V_{Hip}$\tablenotemark{b}} &
\colhead{Source$_{V_{Hip}}$\tablenotemark{c}} &
\colhead{$(V-I)_{Hip}$\tablenotemark{b}} &
\colhead{Source$_{(V-I)_{Hip}}$\tablenotemark{d}} &
\colhead{$V_P$\tablenotemark{e}} &
\colhead{$V_S$\tablenotemark{e}} &
\colhead{$V_{Sys}$} & 
\colhead{$\Delta I$\tablenotemark{f}}
}
\startdata
16115$+$0943 & FIN 354   & 79337  & 6.519 & 6.52 & H & $0.30\pm 0.01$  & L & 7.19  &  7.52 & 6.59 & 0.7 \\ 
17305$-$1446 & HU 177    & 85679  & $8.76 \pm 0.01$ & 8.72  & H & $0.51\pm 0.02$ & L & 8.5   &  9.6 & 8.16 & 1.0 \\
17313$+$1901 & COU 499   & 85740  & $8.96\pm 0.01$ & 8.98 & G &  $0.42\pm 0.01$ & H &  8.7   &  8.7  & 7.95 & 0.5 \\
17533$-$3444 & BU 1123   & 87567  & 6.17  & 6.14 & G & $0.00\pm 0.01$ & L &  6.86  &  6.92 & 6.14 & 0.3 \\
18003$+$2154 & A 1374AB  & 1566-1708-1\tablenotemark{g}  & $8.53 \pm 0.01$ & $8.611 \pm 0.012$\tablenotemark{h} & --- & --- & --- & 8.9 & 10.9  & 8.74 & 1.2 \\
18099$+$0307 & YSC 132Aa,Ab & 89000  & 5.69  & 5.67 & H & $0.56 \pm 0.01$ & L & 6.1   &  7.1 & 5.74 & 0.7 \\
18108$-$3529 & B 1352    & 89076  & $9.02 \pm 0.02$  & 8.98 & H & $0.74 \pm 0.03 $ & L & 9.87  &  9.88 & 9.12 & 0.3 \\
18191$-$3509 & OL 18     & 89766  & $8.55 \pm 0.02$ & 8.51 & H & $ 0.99\pm 0.02 $ & L & 9.17  &  9.73 & 8.66 & 0.6 \\
18359$+$1659 & STT 358AB & 91159  & 6.21  & 6.21 & H & $ 0.63 \pm 0.07 $& F & 6.94  &  7.08 & 6.26 & 0.1 \\
18537$-$0533 & A 93      & 92726  & 8.78  & 8.78 & G &$0.77\pm 0.01$ & H   & 9.16  & 10.15 & 8.79 & 1.2 \\
18558$+$0327 & A2162     & 92909  & $ 7.610 \pm 0.009 $ & 7.07 & G & $ 0.17 \pm 0.01 $ & H &  7.73  &  8.00 & 7.10 & 0.4 \\
19027$-$0043 & STF 2434BC & 93519  & 8.81  & 8.80 & G & $ 0.77 \pm 0.00 $ & R & 8.44  &  8.93 & 7.91 & 1.2 \\
19350$+$2328 & A 162     & 96317  & $ 7.94 \pm 0.01 $  & 7.93 & H & $ 0.23 \pm 0.02 $ & L &  8.73  &  8.77 & 8.00 & 0.1 \\  
20073$-$5127 & RST 1059  & 99114  & $ 8.11 \pm 0.01 $  & 8.13 & H & $0.34 \pm 0.02 $  & L &  8.89  &  9.03 & 8.21 & 0.5 \\
20514$-$0538 & STF 2729AB & 102945 & 6.07  & 5.99 & G & $0.53 \pm 0.00 $ & H & 6.40  &  7.43 & 6.04 & 1.1 \\
20597$-$5211 & I 669 AB  & 103620 & $ 8.32 \pm 0.01 $  & 8.33 & G & $ 0.87 \pm 0.00 $ & H & 9.01  &  9.51 & 8.48 & 0.4 \\
21504$-$5818 & HDS 3109  & 107806 & $ 7.89 \pm 0.01 $  & 7.89 & H & $ 0.78 \pm 0.01 $ & L & 8.56  &  9.07 & 8.03  & 0.3 \\
22156$-$4121 & CHR 187   & 109908 & 4.810 & 4.79 & G & $0.83 \pm 0.02 $  & A  & 5.20  &  6.68 & 4.95 & 2.2 \\ 
22313$-$0633 & CHR 111   & 111170 & 6.615 & 6.15 & G & $0.64 \pm 0.02 $ & A & 6.3   &  8.6  & 6.18 & 1.9 \\
23171$-$1349 & BU 182AB  & 114962 & 8.14  & 8.16 & G & $ 0.62 \pm 0.01 $ & A & 8.77  &  9.08 & 8.16 & 0.6 \\
\enddata
\tablenotetext{a}{From SIMBAD.}
\tablenotetext{b}{From Hipparcos catalogue.}
\tablenotetext{c}{G = ground-based, H=HIP, T=Tycho.}
\tablenotetext{d}{'A' for an observation of $V-I$ in Cousins' system;
  'F', 'H' and 'K' when $V-I$ derived from measurements in other
  bands/photoelectric systems; 'L' when $V-I$ derived from
  Hipparcos and Star Mapper photometry; 'R' when colors are unknown or
  uncertain.}
\tablenotetext{e}{From WDS.}
\tablenotetext{f}{From our own measurements in the $I$-band. When one more than one, it is the average, excluding uncertain (:) values.}
\tablenotetext{g}{Tycho number}
\tablenotetext{h}{This is the $V_T$ mag from the Tycho catalogue. No colors provided.}
\end{deluxetable}

In Table~\ref{tab:photom} we present the list of objects in our final
sample, along with their published photometry from the literature. The
first three columns give the WDS name, discoverer designation, and
sequential number in the Hipparcos catalogue (HIP), then we have the
apparent $V$ magnitude for the system listed in the SIMBAD database
($V_{Sim}$) which is, in itself, a compilation from many sources. The
fifth and sixth columns give the $V$ magnitude on the Hipparcos
catalogue ($V_{Hip}$) and its source respectively, while the seventh
and eighth columns list the values for the color ($(V-I)_{Hip}$) and
source respectively, also from the Hipparcos catalogue. The ninth and
tenth columns give the $V$ magnitudes for the primary ($V_P$) and
secondary ($V_S$) respectively as listed in the WDS catalogue, the
integrated apparent magnitude for the system $V_{Sys}$ is in the
eleventh column (see next paragraph). Finally, in the twelveth column
we report our measured magnitude difference $\Delta I \equiv I_S -I_P$
between secondary and primary.

The quality of the available photometry is variable, as can be readily
seen by comparing the fourth and fifth columns of the table. One could
also check the consistency between the combined magnitude of the
system ($V_{Sim}$ or $V_{Hip}$ in the table) with the equivalent total
magnitude derived from the photometry reported for each component from
the WDS (denoted $V_{Sys}$), since we should have that $V_{Sys} = -2.5
\times \log \left(10^{-0.4 \cdot V_P} + 10^{-0.4 \cdot V_S} \right)$,
which is shown in the penultimate column of
Table~\ref{tab:photom}. Checking the self-consistency of the
photometry is important when addressing the compatibility of the
dynamical and trigonometric parallaxes, or when comparing the
astrometric mass sum to the dynamical masses (see
Section~\ref{sec:massum}). The photometry (both, magnitudes and
colors) is also used later on to place the individual components in an
H-R diagram (Section~\ref{sec:comments}). We note that, although
listed as part of the Hipparcos catalogue, $(V-I)_{Hip}$ was actually
not measured during the Hipparcos mission (see footnote {\it d} on
Table~\ref{tab:photom}), these values result from empirical
transformations whose validity has been questioned in some cases
(\cite{plet03}). However, judging from the quoted color uncertainties,
it seems that its precision is better than our own measured $\Delta I$
(see next paragraph) and, therefore, in the absence of other sources
of colors, we can use them as an indicative value of the system's
color.

\begin{figure}
\plotone{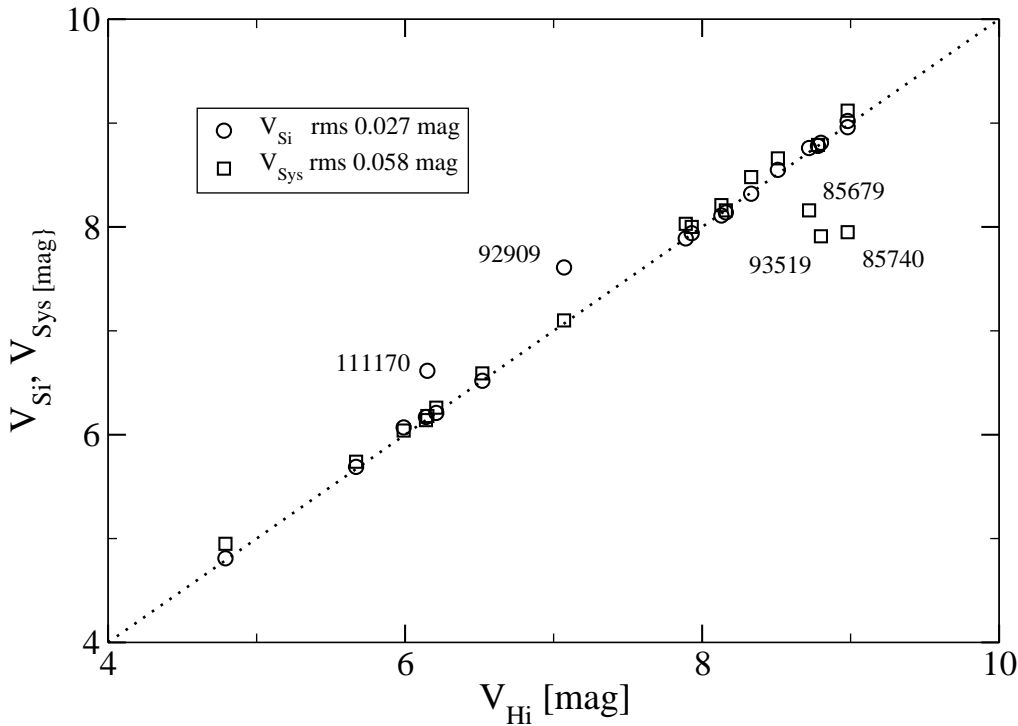}
\caption{Comparison of the photometric values presented in
  Table~\ref{tab:photom}. We indicate the outlier objects (Hipparcos
  number) that exhibit a very large difference in the reported
  magnitudes. A linear fit between $V_{Sim}$ and $V_{Sys}$ {\it vs.}
  $V_{Hip}$ (excluding the outliers) gives the rms values indicated in
  the legend, this is an indication of the uncertainty of the
  photometry. The dotted line is not a fit, it only shows the expected
  one-to-one relationship. \label{fig:photom}}
\end{figure}

In Figure~\ref{fig:photom} we show a comparison of the values
presented in Table~\ref{tab:photom}. A few objects are clearly off
from the expected 45 degrees sequence, we have no explanation for
these differences.  A fit of $V_{Sim}$ {\it vs.} $V_{Hip}$, excluding
the two deviant points indicated in the plot (HIP 92909 and 11170) has
an rms of 0.027~mag, while a fit of $V_{Sys}$ {\it vs.} $V_{Hip}$,
excluding the three deviant points (HIP 85679, 85740, and 93519), has
an rms of 0.058~mag, which we will take as an estimate of the
uncertainty of the photometry in Section~\ref{sec:comments} (see also
Figure~\ref{fig:hrdiag}). These values are approximately consistent
with the color uncertainties reported in
Table~\ref{tab:photom}. Regarding the uncertainty of our $\Delta I$
values, this is more difficult to ascertain, since it depends on a
number of factors such as the angular separation between the
components, the quality of the night when the measurement was
performed, the brightness of the primary, etc. From repeated
measurements on different nights for several of our objects we have
estimated an average uncertainty of $\sigma_{\Delta I} \sim 0.18$~mag,
which we take as the typical error for $\Delta I$ for our sample of
binaries. This value, being much larger than the estimated uncertainty
of the $V$-band magnitudes, limits a finer analysis and interpretation
of the location of the components of these systems in the H-R diagram
(see Figure~\ref{fig:hrdiag}), as explained in
Section~\ref{sec:comments}.

Two of the objects in Table~\ref{tab:photom} are double-lined
spectroscopic binaries (SB2), namely HIP 89000 (YSC 132Aa,Ab), and HIP
111170 (CHR 111). For these, we retrieved their radial velocity
measurements from the 9th Catalogue of Spectroscopic Binary Orbits
(\cite{poet04})\footnote{Updated regularly, and available at
  http://sb9.astro.ulb.ac.be/}. In Section~\ref{sec:specbin} we
perform a joint solution of the astrometric and radial velocity data
of these two systems (see Table~\ref{tab:orbel2}).

Regarding the astrometric calibration, precision, and accuracy of our
astrometric data, the reader is referred to the publications from
which these data were taken, and where these issues are extensively
discussed. In a nutshell, HRCAM on SOAR routinely delivers precisions
of 1-3~mas in angular separation for objects brighter than $V \sim
12$. 
%
%
In our survey, the magnitude difference of the
resolved systems ranges from near-equal brightness components to
$\Delta m \sim 6$~mag, while the range in separation goes from
$\sim$35 mas (telescope diffraction limit in the $V$-band) to 2.0~arcsec
(we have a few resolutions below the diffraction limit, down to
12~mas). For many systems we have also been able to uniquely resolve
the quadrant ambiguity inherent to Speckle imaging from either a
resolved long-exposure image, or from the shift-and-add (or “lucky”)
images. In these latter cases, these quadrants are ``enforced'' in the
orbital solution, which allows, in turn, quadrant resolutions of older
data as well, by imposing self-consistency of the computed orbit (a
process that, one must admit, is somewhat ``subjective'').

\newpage

\section{Orbits} \label{sec:orbits}

For our orbital calculations we have used a newly developed code to
compute orbits, based on MCMC, whose implementation is described in
the next subsections. The main motivation behind this approach is to
exploit features that are inherent to these methods, namely (i) to
provide confidence limits to the derived orbital elements, and (ii) to
generate posterior probability density functions (PDF hereafter) for
each orbital element, as well as for the derived masses. PDFs also
allow us to explore the possible existence of non-unique solutions
given the current data. An additional motivation has been the
possibility to incorporate missing or partial data (e.g., only
position angle $\theta$ but not separation $\rho$, or vice-versa),
which might be particularly important if those data fall in a critical
part of the orbit (e.g., see \cite{clet16}). MCMC methods (see, e.g.,
\cite{he12,mebr14}) are currently widely used in exoplanet research
(akin to a binary system), mostly for the interpretation of radial
velocity data (\cite{tuko09,otet16}), but also for astrometric orbits
(see., e.g., \cite{saet13}).

In the following subsections we present an outline of our specific
MCMC implementation, and the main results as applied to the visual and
spectroscopic binaries in the sample reported here.

\subsection{Orbital adjustment through Eggen's effect}

Long ago, \cite{eg67}, suggested that the $a^3/P^2$ quotient can often
be well-determined even if the individual values of the semi-major
axis $a$ and the period $P$ are not accurately known. Nowadays this is
referred to as \emph{Eggen's effect} (\cite{lu14}). Thus, if the
estimate of $P$ suffers a dramatic change after new observations are
incorporated into the analysis, the corresponding estimate of
parameter $a$ may undergo a shift on its value such that it
compensates that variation, yielding a similar value for the $a^3/P^2$
ratio. Numerical results presented in \cite{lu14} strongly support
this conjecture, as they suggest that, if the orbital coverage exceeds
$40\%$ of the full orbit, a reasonable estimate of $a^3/P^2$ can, in
general, be obtained. Of course, there are additional factors to take
into account, such as the quality of the observations, the specific
orbital section being covered (points near the periastron are
significantly more informative than those that are far from it), and
even the particular orbital configuration being observed --some orbits
may be intrinsically more challenging to examine than others; think of
cases with inclination ($i$) very close to $90^{\circ}$, for example.

As long as the observations provide a minimal orbital coverage of the
object under study, the so-called Eggen's effect opens up the
possibility of estimating the mass by identifying the set of feasible
orbital configurations, even if they involve a wide range of values
for $a$ and $P$. The basic idea is that, rather than calculating the
mass based on a single estimate, one can characterize mass (and its
uncertainty) on the base of this set of feasible values.

Since our aim is not only to find a single feasible estimate for the
orbital elements of the objects under study, but rather to
characterize the uncertainty of its orbital parameters, in this paper
we adopt a Bayesian approach for this problem. From a Bayesian
standpoint, the set of feasible values mentioned in the previous
paragraph takes the form of a posterior PDF. In this work, we
construct the posterior PDF from a set of samples, which are drawn by
means of the technique known as MCMC.

\subsection{Model description} \label{subsectionKepler}

Assuming that phenomena such as mass transfer, relativistic effects,
or even the presence of non visible additional bodies, do not affect
the observed objects to a significant degree, we used a Keplerian
model to describe the orbits of the analyzed binary stars. This model
requires seven parameters (represented in what follows by the vector
$\vec \vartheta$)
to fully characterize the trajectory of a visual binary star (Equation
\ref{eqnVIS}), that is, to compute its ephemeris for any given epoch:

\begin{eqnarray}
    \vec \vartheta = \{P, T, e, a, \omega, \Omega, i \} \label{eqnVIS}
\end{eqnarray}

As to systems for which both astrometric and radial velocity
measurements are available, one can perform a joint analysis by
extending the parameter vector $\vec \vartheta$ presented in
Equation~(\ref{eqnVIS}) in the manner shown in
Equation~(\ref{eqnSPEC}):

\begin{eqnarray}
    \vec \vartheta = \{P, T, e, a, \omega, \Omega, i, V_{CoM}, \varpi, q\}, \label{eqnSPEC}
\end{eqnarray}
where $\{P, T, e, a, \omega, \Omega, i \}$ are the well-known Campbell
elements and $V_{CoM}$, $\varpi$ and $q$ denote the velocity of the
center of mass, the parallax of the system, and the mass ratio
$m_S/m_P$ between the secondary of mass $m_S$ and the primary of mass
$m_p$ respectively. The representation shown in
Equation~(\ref{eqnSPEC}), and used previously, e.g., in \cite{buet15},
has some distinct characteristics. First, it includes parallax
$\varpi$ as one the parameters to be estimated rather than a value
known in advance\footnote{This is usually known as ``orbital
  parallax'' to differentiate it from trigonometric parallax or other
  distance estimates.}, thus putting into practice the somewhat
unexplored possibility of utilizing combined data (i.e., astrometry
and radial velocity) to estimate hypothesis-free parallaxes
(\cite{po00}, \cite{ma15}). Secondly, but as a consequence of
including $\varpi$, it exploits all the restrictions imposed by the
formulae of orbit position (Equations~(\ref{TI_constants}),
(\ref{proj_TI})) and radial velocity (Equations~(\ref{eqRadVelx1}),
(\ref{eqRadVelx2})), possibly leading to more precise inferences about
the parameters. By contrast, in some codes, such as e.g., in ORBIT,
the radial velocity amplitudes for the primary ($K_P$) and the
secondary ($K_S$) (see Equations~(\ref{eqRadVelx1}) and
(\ref{eqRadVelx2})) are taken as free parameters. Those methods have
the advantage of not requiring a parallax value to perform the
estimation, but omit the dependency of $K_P$ and $K_S$ on $a$, $i$ and
$\omega$. Appendix \ref{kepler_appendix} presents a summary of our
actual implementation of the adopted Keplerian model, as well as the
deduction of a novel mathematical formalism for a dimensionality
reduction of the number of components of $\vec \vartheta$ for both
visual and spectroscopic binaries, which we have applied in this work,
further details on this will be given in a forthcoming paper. Our
formulation makes it possible to explore a smaller number of
parameters in each case: Three (instead of seven) for visual
binaries\footnote{This dimensionality reduction for visual binaries
  has been applied before by \cite{haet89} in their ``grid search''
  method.}, and seven (instead of ten) for spectroscopic binaries,
while the remaining orbital parameters are {\it unequivocally}
determined by a least-squares fit to the observed data.

\subsection{Description of the Markov Chain Monte Carlo} \label{subsectionMCMC}

MCMC designates a wide class of sampling techniques that rely on
constructing a Markov Chain\footnote{A Markov Chain is understood as a
  sequence of values (in our case orbital parameters) with a defined
  initial state (initial orbital guess), and whose subsequent values
  depend only on the previous state and a transition probability. In
  MCMC the transition probability is defined by the ``proposal
  distribution'' $q(\cdot)$ and acceptance probability $\mathcal{A}$,
  see Appendix~\ref{appendix_DEMC}.}  that explores the domain of the
target PDF in such a way that it spends most of the time in areas of
high probability (for an introduction to MCMC, the reader is referred
to the tutorial by \cite{anet03}). MCMC provides a means to
efficiently draw samples from distributions with complex analytic
formulae and/or multidimensional domains.
This subsection gives some basic details about the implementation of
the MCMC used in this work. Orbital parameters of interest will be
simply called ``parameters'', whereas those related to the Markov
Chain's proposal distribution will be referred to as
``algorithm-related parameters''.

Our list consists of eighteen visual and two spectroscopic binary
systems, with periods ranging from a few months (in the case of the
spectroscopic binaries) to possibly thousands of years in the case of
some visual binaries. Although a reparametrization of $T$ as
$T^{\prime} = (T-T_0)/P$ (an approach used in this work, and
previously in \cite{lu14}), is useful to restrict the search range of
the time of periastron passage to $[0,1)$, both the initial
  distribution and the search range of the period $P$ remains a
  difficult guess. The dimensionality reduction mentioned in
  Subsection \ref{subsectionKepler} (and fully developed in Appendix
  \ref{sec:dimen}) alleviates to some extent the need of choosing an
  initial guess --for it suppresses parameters $a$, $\omega$,
  $\Omega$, $i$ from the analysis. However, the variability of the
  feasible ranges of $P$ among the studied objects imposes a diversity
  of scales among the posterior distributions. Moreover, the shape and
  orientation of each posterior PDF in the multi-dimensional feature
  space\footnote{Feature space refers to the space of (orbital)
    exploration parameters.} is unique (see a few examples in
  Figure~\ref{fig:examples}, right panel). These factors make it
  difficult to choose a single set of algorithm-related parameters. At
  the same time, the list is long enough to make individual-case
  analysis undesirable.

In an effort to study the visual binaries
under a single unified framework, rather than choosing
algorithm-related parameters for each star separately --a task usually
involving a lot of trial and error iterations--, we adopt the
Differential Evolution MCMC approach presented by \cite{br06}, called
DE-MC hereinafter. DE-MC stems as the combination of a genetic
algorithm called Differential Evolution (\cite{stpr97}), described as
``a simple and efficient heuristic for global optimization over
continuous spaces'' by its authors, and MCMC. The algorithm is based
on the idea of running several Markov Chains in parallel but, instead
of letting them run independently as in the classical MCMC convergence
tests, it lets the chains learn from each other. This aims at dealing
with the problem of choosing an appropriate scale and orientation for
the proposal distribution.

The mutual learning between chains is accomplished by using a proposal
distribution based on the DE ``jumping step'' considered in
\cite{stpr97}. In that scheme, the proposal sample $\vartheta_{proposal}$ of
each chain is obtained by adding to the previous sample
($\vartheta_{previous}$), the difference between the current values of two
other randomly chosen chains (say $R1$, $R2$):

\begin{equation}
\vartheta_{proposal} = \vartheta_{previous} + \gamma \cdot ( \vartheta_{R1} - \vartheta_{R2}) + w,
\end{equation}
where $\vartheta$ represents a point in the feature space, the
coefficient $\gamma$ is a term that modulates the difference vector
(its optimal value depends on the dimension of the feature space,
$d$), and $w$ is an additional perturbation drawn from a distribution
with unbounded support (e.g., a normal distribution) and small
variance with respect to that of the target distribution (for further
details, see \cite{br06}). The $w$ term is aimed at guaranteeing the
irreducibility condition of MCMC and, in practice, this additional
noise is useful to explore the feature space at the level of a small
vicinity. The term $\gamma \cdot ( \vartheta_{R1} - \vartheta_{R2})$,
on the other hand, contributes to make larger leaps, without falling
in zones of low likelihood. The algorithm is proven to meet the
reversibility, aperiodicity, and irreducibility conditions which are
required for MCMC in \cite{br06}. The method is summarized in
pseudo-code in Appendix~\ref{appendix_DEMC}, Figure~\ref{alg_DEMC}.

Since we are interested in characterizing a posterior distribution, a
fitness function $f$ is defined as the posterior PDF, which has the
canonical form of $prior \times likelihood $ (see, e.g.,
\cite{geet13}). Terms from the prior PDF can be dropped, as uniform
priors were used for the three relevant exploration parameters after
dimensionality reduction, namely: $T^{\prime}$ (range $(0, 1)$), $\log
P$ (range ($\log 10$~yr, $\log 5000$~yr)) and $e$ (range $(0,
0.99)$). Thus, the likelihood function can be directly used to compute
the Metropolis-Hastings ratio\footnote{This ratio is defined as
  $\mathcal{A}(x^{(i)},x^{\prime}) = \min \left \{1,
  \frac{p(x^{\prime}) \cdot q(x^{(i)})}{p(x^{(i)}) \cdot
    q(x^{\prime})} \right \}$, where $p(\cdot)$ is the target
  distribution and $q(\cdot)$ is the proposal distribution. The term
  $x^{\prime}$ is the the proposal sample and $x^{(i)}$ is the
  previous sample. The proposal sample $x^{\prime}$ is accepted with
  probability $\mathcal{A}$. See Appendix~\ref{appendix_DEMC}.}
Assuming independent individual Gaussian errors for each observation,
the likelihood function for the $i$-th iteration with orbital
parameters $\vartheta_i$ is defined as\footnote{In \cite{gr05} one can
  find a clear explanation for the adoption of this posterior, applied
  to the case of exoplanet research.}:

\begin{equation} \label{likelihood_function_astro}
f(\vartheta_i) \propto \displaystyle \exp \biggl(-\frac{1}{2}
\Bigl(\sum_{k=1}^{N_{x}} \frac{1}{\sigma_x^2(k)} [X(k) -
  X^{model}(k,i)]^2 + \sum_{k=1}^{N_y} \frac{1}{\sigma_y^2(k)} [Y(k) -
  Y^{model}(k,i)]^2\Bigr)\biggr),
\end{equation}
where $(X(k),Y(k))$ is the k-th observation of the apparent
orbit\footnote{This is the position of the secondary as seen from the
  primary in the plane of the sky (called apparent orbit in
  Appendix~\ref{kepler_appendix}), related to the separation angle
  $\rho$ and position angle $\theta$ by $X= \rho \cos \theta$ and $Y =
  \rho \sin \theta$.} with uncertainties $(\sigma_x(k),\sigma_y(k))$,
$(X^{model}(k,i), Y^{model}(k,i))$ are the computed ephemerides (which
depend on $\vartheta_i$, see Appendix~\ref{kepler_appendix}), and where
we have $N_x$ observations in $X$ and $N_y$ observations in $Y$
(usually $N_x=N_y$). Equation~(\ref{likelihood_function_astro})
results from the assumption that the residual of each data point
follows an independent Gaussian distribution. Note that, as in any
orbital calculation procedure, the weights assigned to each
observational point play a critical role in the solution.

The algorithm-related parameters in this case are: the number of
chains $N_{chains}$; the coefficient $\gamma$; the parameters of the
distribution of $w$. Following the guidelines presented in
\cite{br06}, we fixed $N_{chains} = 10$ (the recommendation is to
choose $N_{chains} > 2 \cdot d$), $\gamma = 2.38/\sqrt{2\cdot
  d}$. Values for $w$ are drawn from $\mathcal{N}(0,\Sigma)$, where
$\Sigma$ is a diagonal matrix with: $\sigma^2_{T^{\prime}} = 0.01$,
$\sigma^2_{\log P} = 0.01\cdot(\log P_{upper} - \log
P_{lower})$\footnote{Because the period $P$ works as a scale
  parameter, we explore it through the $\log P$ space, which is
  equivalent to using a Jeffreys prior, \cite{fo05}.} and $\sigma^2_e
= 0.01$. To obtain the final posterior distribution, we ran DE-MC with
$N_{steps} = 10^6$ for each object, discarding the first $10^5$
samples of each chain (the so-called ``burn-in period''). As shown in
\cite{br06}, each chain distributes as the target distribution, so
that the set of all chains can be treated as a single collection of
samples, once the burn-in period is dropped. Therefore, with our
adopted parameters, this results in a single-chain of $9\times 10^6$
samples.

This is a minimalist description of the algorithm since further
details on the inner working of our MCMC implementation, full tests,
and other applications of the method will be given in a separate
publication. We can however mention that our code has been extensively
tested against simulated and real data, and its results have also been
compared to other codes, including the IDL-driven interactive ORBIT
code developed by \citet{to92}\footnote{The code and user manual can
  be downloaded from
  http://www.ctio.noao.edu/$\sim$atokovin/orbit/index.html} which
employs a $\chi^2$ minimization approach through a Levenberg-Marquardt
parameter exploration method to determine the Campbell orbital
elements by a fit to the data (see Appendix~\ref{Appendix2}), and a
code that uses minimization through a downhill simplex algorithm,
developed by \citet{maho04}, and used extensively by Horch and
collaborators (see., e.g., \citet{hoet15}).

\subsection{Orbital elements for our Visual Binaries} \label{sec:visbin}

The results of our MCMC code, described in the previous subsection,
when applied to the eighteen visual binaries of the sample presented
in Section~\ref{sec:sample} are shown in Table~\ref{tab:orbel1}. For
each object, two sets of numbers for the orbital elements are
provided: The upper row represents the configuration with the smallest
mean square sum of the O-C overall residuals\footnote{Note that since
  we use uniform priors, the posterior distribution is completely
  defined by the likelihood function. Thus, in this case maximum
  likelihood (ML) and maximum a posteriori (MAP) estimators have the
  same value. Moreover, as a consequence of how the likelihood
  function is defined in this work (essentially, the exponential of
  $-1/2$ times the mean square error), minimizing the sum of the O-C
  residuals is equivalent to maximizing the likelihood--the values
  reported in Table \ref{tab:orbel1} are ML estimates.}
, while the lower row shows the median derived from the posterior PDF
of the MCMC simulations, as well as the upper (third) quartile ($Q75$)
and lower (first) quartile ($Q25$) of the distribution in the form of
a superscript and subscript respectively\footnote{It is customary to
  represent the uncertainties in terms of $\sigma$. However, this
  quantity is well defined only for orbits where the PDFs are ``well
  behaved'', and it becomes meaningless for uncertain orbits where the
  PDFs exhibit very long tails (see below for more details). For a
  Gaussian function one can convert from one to the other using the
  fact that $\sigma = (Q75-Q25)/1.349$.}. The orbital coverage, and
the reliability of the fitted orbital parameters, ranges from what one
could consider as almost ``final'' orbits (e.g., HIP 85679, HIP 92909,
HIP 102945, HIP103620, HIP 109908) to those with very poor coverage
and uncertain orbits (e.g., HIP 85740, TYC 1566-1708-1, HIP 89766). In
the penultimate column we give an indication of the ``Grade'' of the
orbit as defined in the WDS, while the last column has the reference
for the latest orbit published for each object (or ``New'' if none was
available). The reason why ML/MAP is preferred over the expected value
is that, for most of the cases studied in this work, the PDFs are
rather disperse and asymmetrical, thus yielding average values that
are not in good agreement with the observations. The only exceptions
are those orbits with good orbital coverage: HIP 85679, HIP 102945,
and HIP 103620 (to mention a few examples), for which the expected
value approximately coincides with the MAP/ML estimate.

\begin{figure}[ht] 
\begin{minipage}[b]{0.5\linewidth}
\centering
\includegraphics[height=.235\textheight]{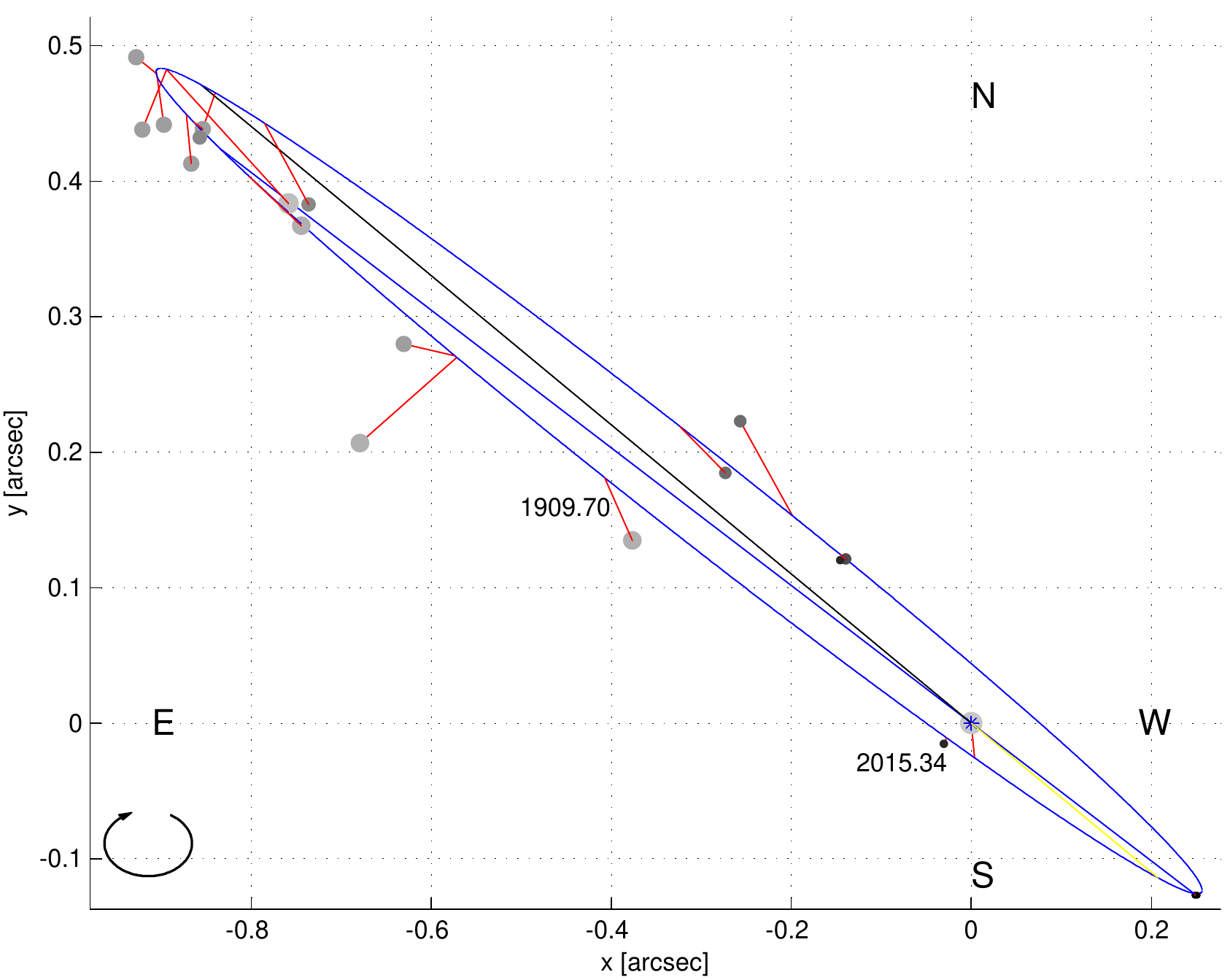}\\
\includegraphics[height=.235\textheight]{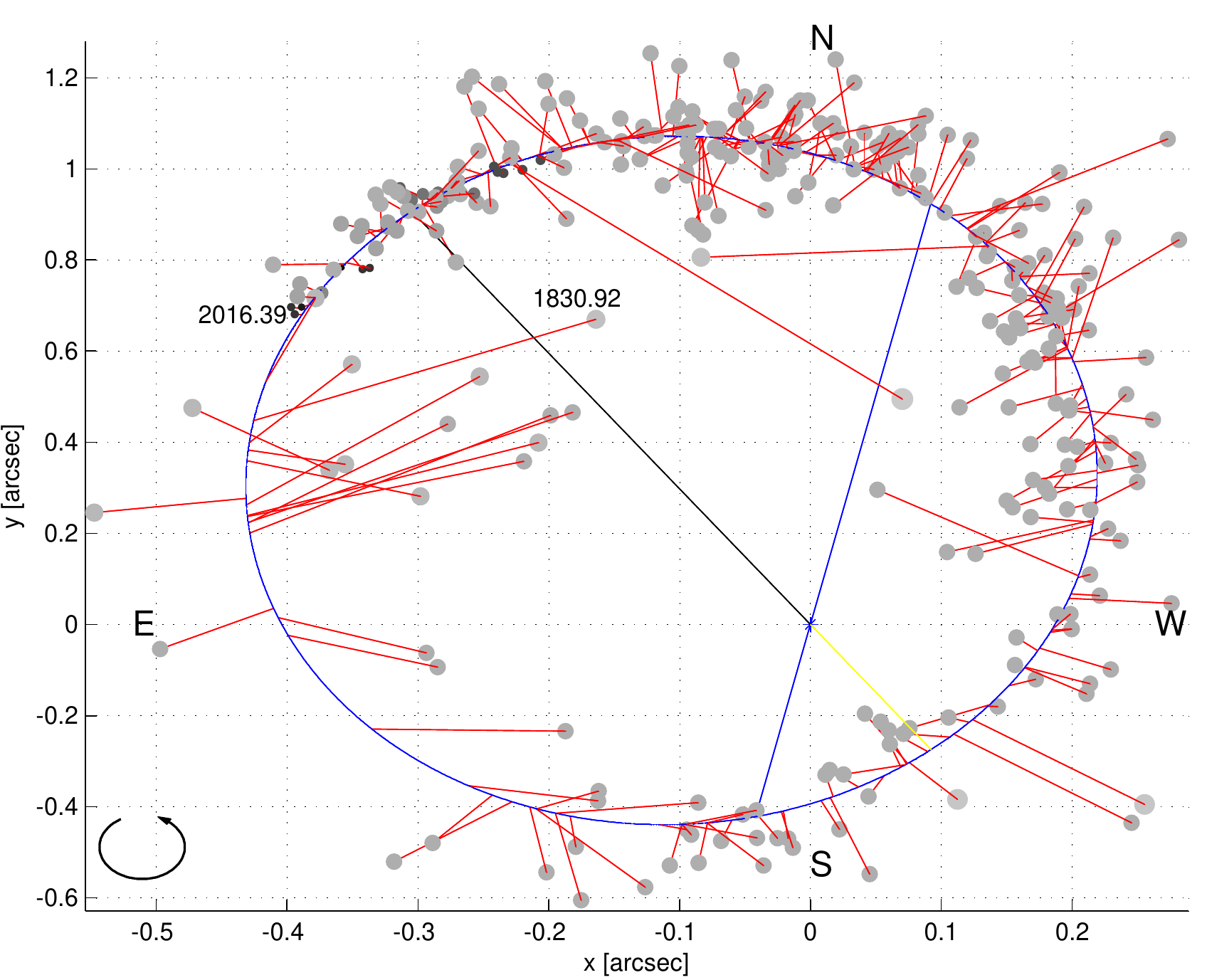}\\
\includegraphics[height=.235\textheight]{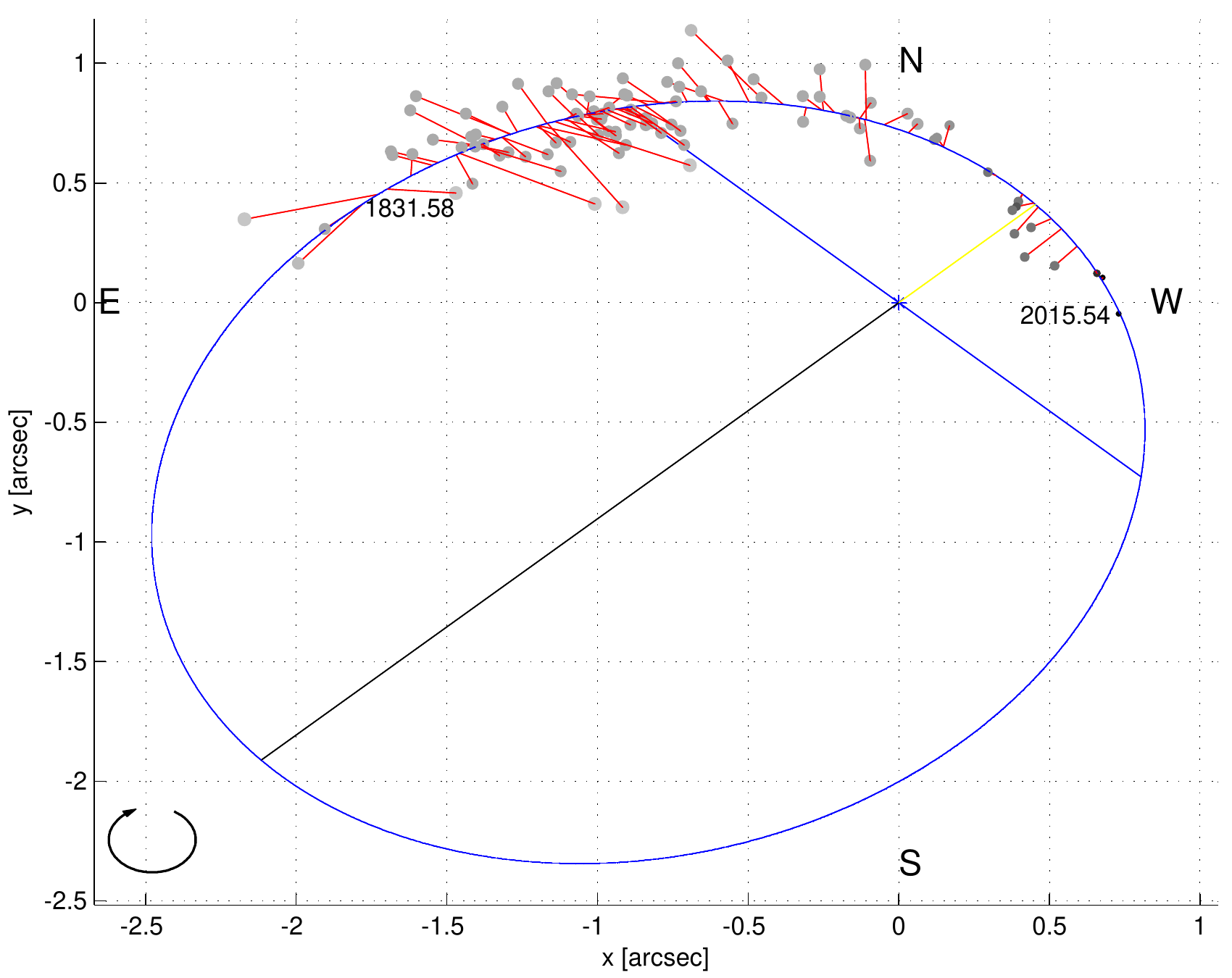}\\
\includegraphics[height=.235\textheight]{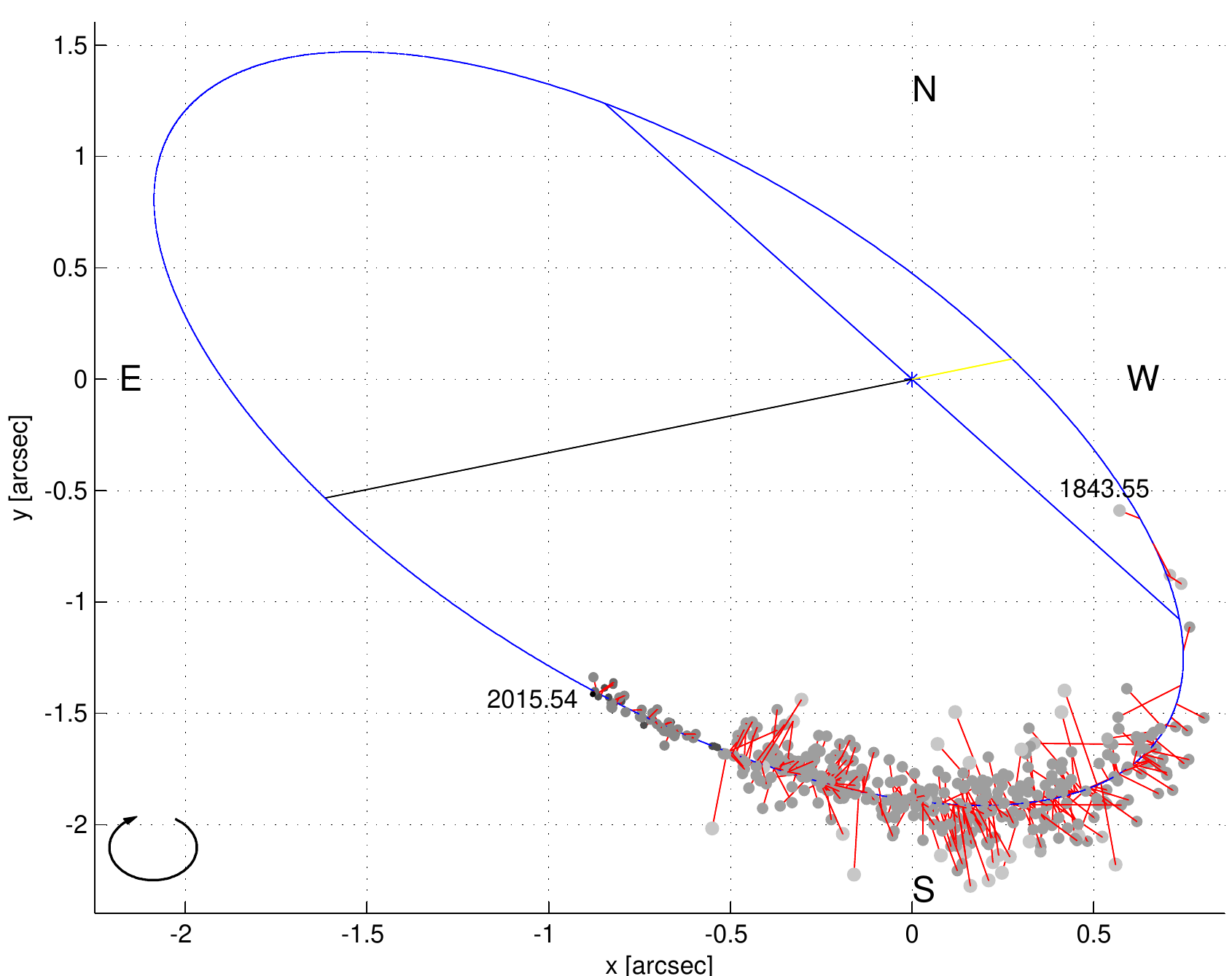}

\end{minipage}
\begin{minipage}[b]{0.5\linewidth}
\centering
\includegraphics[height=.235\textheight]{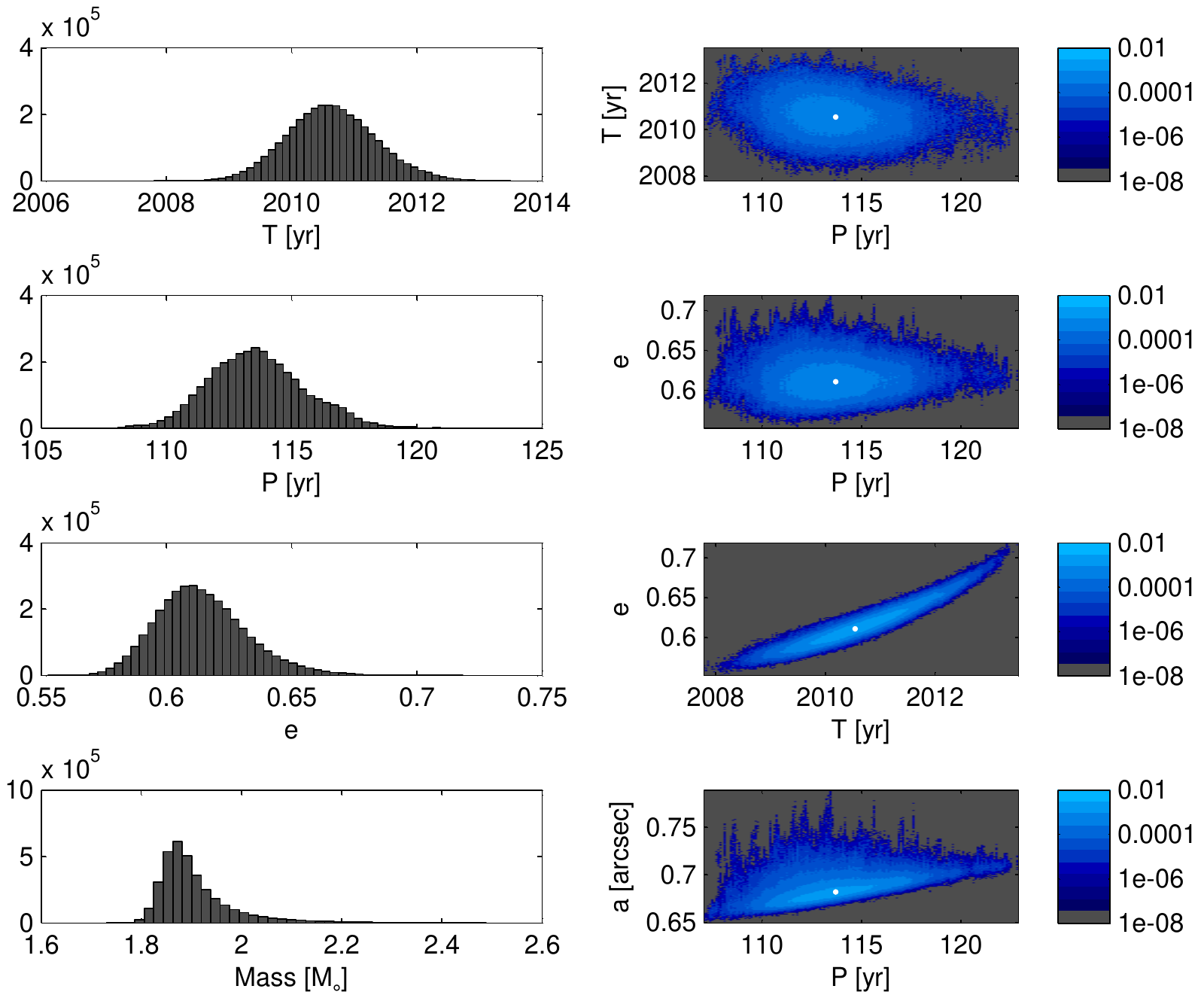}\\
\includegraphics[height=.235\textheight]{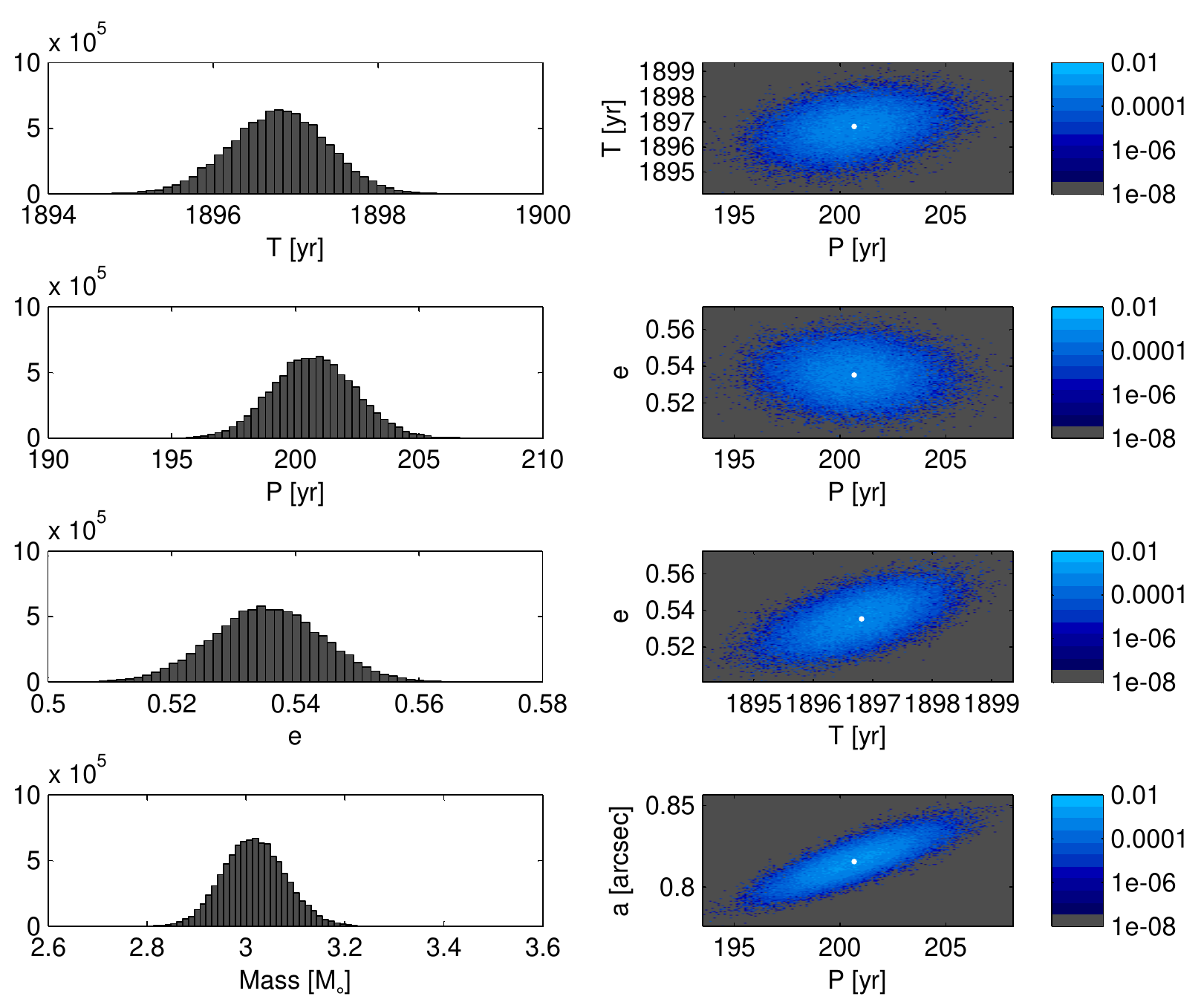}\\
\includegraphics[height=.235\textheight]{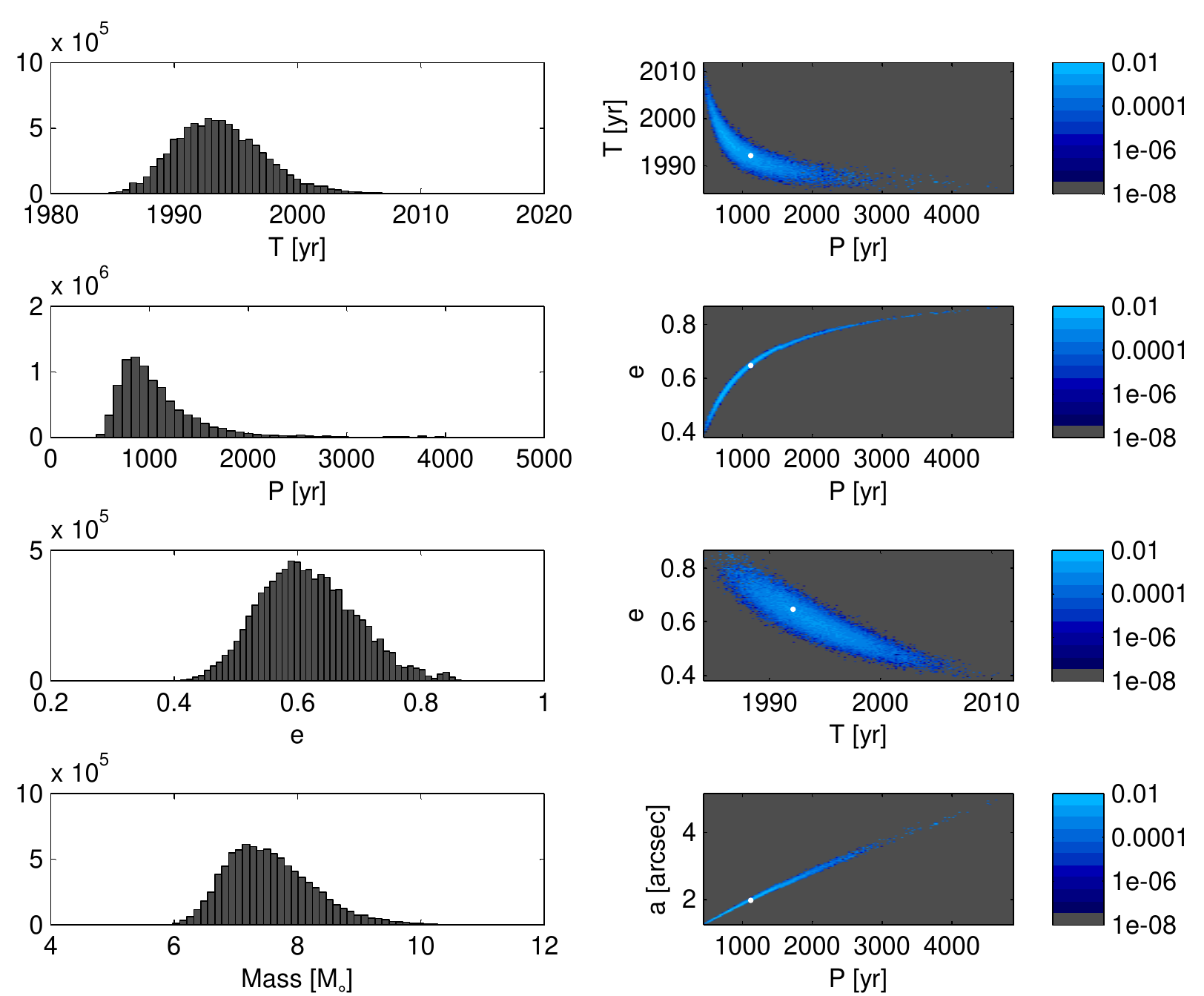}\\
\includegraphics[height=.235\textheight]{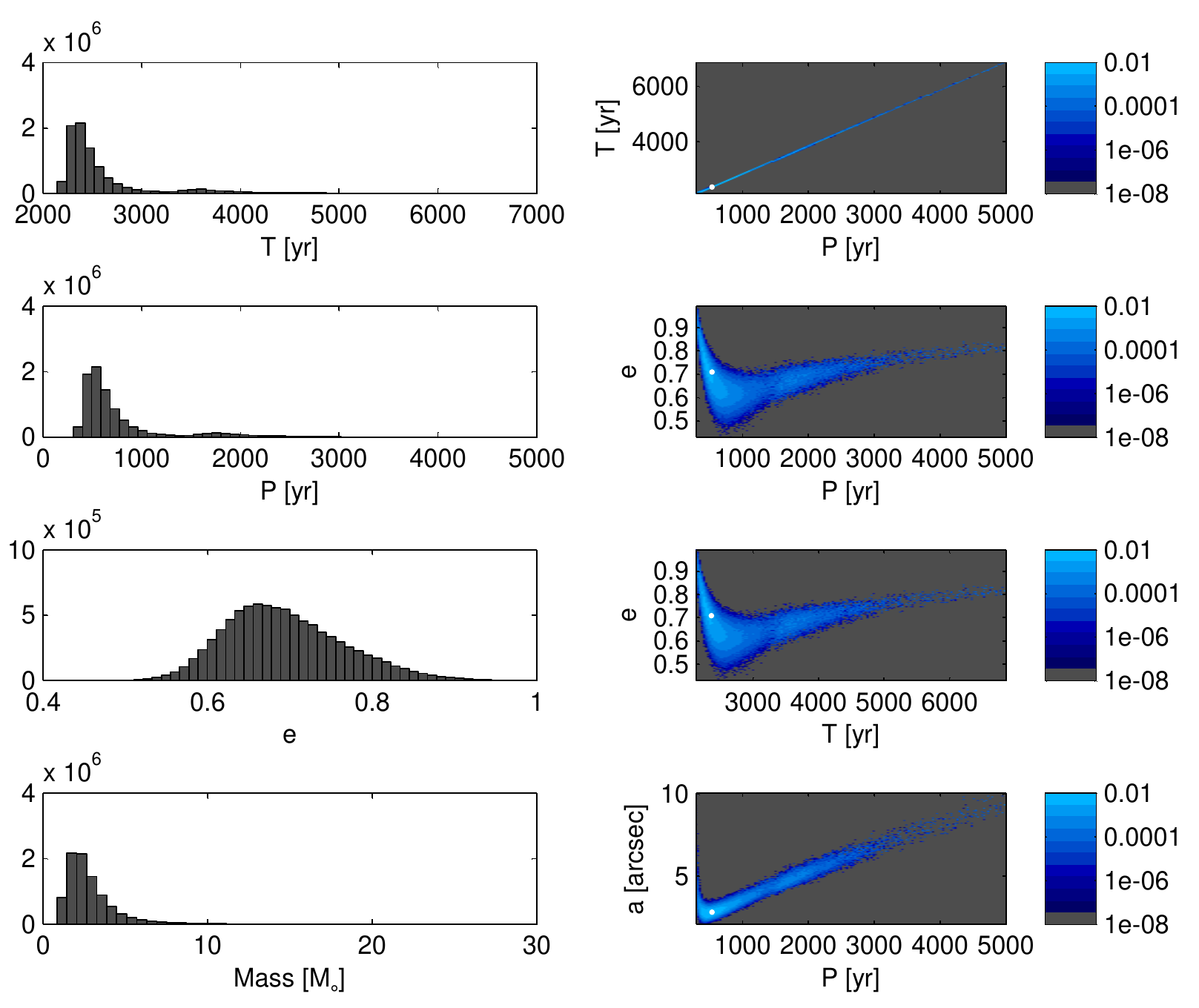}
\end{minipage} 
\caption{Examples of orbital solutions (left panel) and PDFs (right
  panel), from top to bottom: HIP 103620, HIP 102945, HIP 93519, and
  HIP 91159. HIP 103620 and HIP 102945 have well defined orbits and
  tight (usually Gaussian-like) PDFs. HIP 93519, and HIP 91159
  represent two typical examples of objects with incomplete orbital
  coverage, and tangled PDFs. See Section~\ref{sec:comments} for
  further comments on individual objects. On the orbits we indicate
  the first and last epoch of observation. Larger and lighter-grey
  points have less weight in the solution than smaller darker
  points. \label{fig:examples}}
\end{figure}

In Figure~\ref{fig:examples} we show some examples of orbital
solutions and PDFs for objects in our sample. By looking at
Table~\ref{tab:orbel1} and Figure ~\ref{fig:examples} we can say that,
in general, well determined orbits show a ML value that approximately
coincides with the 2nd quartile of the PDF, the inter-quartile range
is relatively well constrained, and the PDFs show a Gaussian-like
distribution, meanwhile poor orbits show PDFs with very long tails (and
therefore large inter-quartile ranges) on which the ML value usually
exceeds by much the 2nd quartile, and the PDFs are tangled.

\floattable
\rotate
\begin{deluxetable}{cccccccccc}
\tablecaption{Orbital elements of our visual binaries \label{tab:orbel1}}
\tabletypesize{\footnotesize}
\tablecolumns{10}
\tablewidth{0pt}
\tablehead{
\colhead{HIP} & \colhead{P} & \colhead{T$_0$} &
\colhead{e} & \colhead{a} & \colhead{$\omega$} & \colhead{$\Omega$} &
\colhead{i} & \colhead{Gr} & \colhead{Orbit}\\
 & \colhead{(yr)} & \colhead{(yr)} &
 & \colhead{($\arcsec$)} & \colhead{($^{\circ}$)} & \colhead{($^{\circ}$)} & \colhead{($^{\circ}$)} & \colhead{Current $\rightarrow$ New} & \colhead{reference\tablenotemark{a}}
}
\startdata
79337 & 60.60 & 2000 & 0.049 & 0.1280 & 93 & 83.36 & 91.13 & 3$\rightarrow$3 & Doc2013d \\
& ${59.31}_{-0.68}^{+0.91}$ & ${1994}_{-27}^{+6}$ & ${0.016}_{-0.009}^{+0.024}$ & ${0.1285}_{-0.0003}^{+0.0002}$ & ${121}_{-31}^{+115}$ & ${83.35}_{-0.01}^{+0.01}$ & ${91.11}_{-0.01}^{+0.01}$ &  \\
85679 & 202	& 1986.53 & 	0.506& 	0.286& 	238.6 & 166.8 & 	151.7 & 5$\rightarrow$3 & USN2002 \\
& ${201}_{-12}^{+14}$	& ${1986.46}_{-0.45}^{+0.44}$& 	${0.503}_{-0.028}^{+0.027}$	& ${0.286}_{-0.008}^{+0.008}$& 	${237.5}_{-6.5}^{+5.6}$& 	${165.8}_{-5.2}^{+4.4}$& 	${150.9}_{-2.8}^{+3.0}$ \\
85740 & 2302 & 1988 & 0.87 & 1.317 & 7 & 59 & 109.3 &  5$\rightarrow$4 & Cou1999b \\
& ${136}_{-31}^{+47}$   & ${2025}_{-42}^{+24}$ & ${0.26}_{-0.14}^{+0.17}$ & ${0.191}_{-0.018}^{+0.050}$  & $9_{-34}^{+47}$    & $64_{-163}^{+4}$   & ${116.8}_{-4.1}^{+3.9}$ \\
87567 & 895 & 1961.74 & 0.750 & 0.641 & 300.3 & 36.7 & 36.3 & 5$\rightarrow$4 & Doc1991b \\
& ${753}_{-154}^{+251}$  & ${1961.52}_{-0.77}^{+0.73}$ & ${0.718}_{-0.049}^{+0.051}$ & ${0.571}_{-0.080}^{+0.123}$  & ${296.0}_{-7.8}^{+6.9}$ & ${40.4}_{-6.0}^{+6.6}$    & ${36.1}_{-1.8}^{+1.8}$ \\
1566-1708-1\tablenotemark{b} & 3431 & 1976.6 & 0.84 & 3.31 & 219 & 153.3 & 125.8 & 5$\rightarrow$5 & USN2002 \\
& ${497}_{-150}^{+294}$ & ${1978.1}_{-2.7}^{+3.4}$ & ${0.41}_{-0.18}^{+0.17}$ & ${0.98}_{-0.17}^{+0.31}$ & ${233}_{-7}^{+10}$ & ${166.0}_{-5.1}^{+5.2}$ & ${124.1}_{-2.0}^{+1.9}$\\
89076 & 427 & 1977 & 0.43 & 0.432 & 7 & 134 & 41.9 & 5$\rightarrow$4 & USN2002 \\
& ${229}_{-36}^{+64}$   & ${1990}_{-31}^{+63}$ & ${0.21}_{-0.11}^{+0.15}$ & ${0.311}_{-0.032}^{+0.058}$  & $5_{-33}^{+51}$    & $98_{-187}^{+16}$  & ${43.4}_{-8.3}^{+8.4}$  \\
89766 & 1119 & 2000 & 0.609 & 3.05 & 114 & 119.38 & 84.81 & X$\rightarrow$5 & NEW \\
& ${309}_{-78}^{+114}$ & ${2049}_{-15}^{+9}$ & ${0.383}_{-0.091}^{+0.176}$ & ${1.96}_{-0.13}^{+0.27}$ & ${209}_{-50}^{+33}$ & ${119.44}_{-0.15}^{+0.16}$ & ${86.73}_{-0.63}^{+0.79}$ \\
91159 & 532 & 2364 & 0.709 & 2.80 & 96 & 34.3 & 110.1  & 4$\rightarrow$4 & Hei1995 \\
& ${595}_{-99}^{+189}$   & ${2380}_{-90}^{+129}$ & ${0.686}_{-0.044}^{+0.053}$ & ${2.94}_{-0.17}^{+0.29}$  & ${100}_{-6}^{+12}$   & ${32.2}_{-4.0}^{+3.7}$    & ${110.5}_{-1.2}^{+1.1}$ \\
92726 & 764 & 1914.7 & 0.746 & 1.20 & 342 & 30 & 35.3 & 5$\rightarrow$4 & Hei1998 \\
& ${549}_{-135}^{+249}$   & ${1911.6}_{-3.7}^{+3.4}$ & ${0.702}_{-0.042}^{+0.051}$ & ${1.03}_{-0.10}^{+0.21}$  & ${338}_{-8}^{+11}$ & ${17}_{-8}^{+11}$     & ${42.9}_{-6.9}^{+6.5}$ \\
92909 & 154.7 & 2008.2 & 0.221 & 0.295 & 63 & 80.1 & 129.3 & 3$\rightarrow$3 & Doc1988c  \\
& ${154.2}_{-8.7}^{+10.7}$   & ${2008.3}_{-2.8}^{+3.0}$ & ${0.222}_{-0.021}^{+0.026}$ & ${0.295}_{-0.014}^{+0.015}$  & ${63}_{-10}^{+11}$    & ${80.2}_{-1.1}^{+1.1}$    & ${129.3}_{-1.8}^{+1.8}$ \\
93519 & 1123 & 1992.1 & 0.647 & 1.97 & 95 & 48 & 151.3 & 5$\rightarrow$5 & Alz1998a \\
& ${975}_{-168}^{+263}$  & ${1993.5}_{-2.3}^{+2.5}$ & ${0.612}_{-0.050}^{+0.056}$ & ${1.82}_{-0.18}^{+0.27}$  & ${97}_{-14}^{+10}$   & ${49}_{-10}^{+6}$    & ${149.0}_{-2.6}^{+2.5}$\\
96317 & 2651 & 1993 & 0.83 & 1.384 & 195 & 76.9 & 70.7 & 3$\rightarrow$3 & Ole1994 \\
& ${279}_{-56}^{+126}$ & ${2015}_{-14}^{+14}$ & ${0.33}_{-0.10}^{+0.12}$ & ${0.294}_{-0.044}^{+0.089}$ & ${234}_{-25}^{+37}$ & ${77.9}_{-4.3}^{+4.6}$ & ${62.8}_{-3.2}^{+3.1}$\\
99114 & 52 & 2017.3 & 0.265 & 0.166 & 159 & 66 & 22.4 & X$\rightarrow$4 & NEW \\
& ${155}_{-19}^{+28}$ & ${2016.8}_{-7.2}^{+5.8}$ & ${0.294}_{-0.043}^{+0.053}$ & ${0.175}_{-0.008}^{+0.013}$ & ${154}_{-78}^{+40}$ & ${77}_{-19}^{+35}$ & ${32.6}_{-6.6}^{+6.4}$ \\
102945 & 200.7 & 1896.8 & 0.535 & 0.816 & 45.8 & 174.32 & 64.06 & 2$\rightarrow$2 & RAO2015  \\
& ${200.7}_{-1.1}^{+1.1}$ & ${1896.8}_{-0.40}^{+0.39}$ & ${0.535}_{-0.005}^{+0.006}$ & ${0.816}_{-0.006}^{+0.006}$ & ${45.9}_{-1.2}^{+1.2}$ & ${174.31}_{-0.39}^{+0.38}$ & ${64.06}_{-0.27}^{+0.26}$  \\
103620 & 113.7 & 2010.54 & 0.611 & 0.682 & 207.2 & 63.07 & 93.69  & X$\rightarrow$3 & NEW \\
& ${113.6}_{-1.2}^{+1.3}$ & ${2010.62}_{-0.46}^{+0.48}$ & ${0.613}_{-0.011}^{+0.012}$ & ${0.684}_{-0.006}^{+0.007}$ & ${207.6}_{-2.6}^{+2.8}$ & ${63.02}_{-0.31}^{+0.31}$ & ${93.66}_{-0.19}^{+0.16}$\\
107806 & 32.8 & 2014.1 & 0.158 & 0.224 & 98 & 127.22 & 87.84 & X$\rightarrow$4 & NEW \\ 
& ${32.9}_{-2.2}^{+2.9}$ & ${2013.9}_{-1.6}^{+1.2}$ & ${0.164}_{-0.021}^{+0.021}$ & ${0.226}_{-0.011}^{+0.015}$ & ${97}_{-22}^{+20}$ & ${127.23}_{-0.09}^{+0.10}$ & ${87.88}_{-0.12}^{+0.14}$ \\
109908 & 19.09 & 1996.39 & 0.562 & 0.169 & 92.5 & 105.0 & 65.70 & 3$\rightarrow$3 & Tok2015c  \\
& ${19.04}_{-0.18}^{+0.16}$ & ${1996.35}_{-0.18}^{+0.17}$ & ${0.561}_{-0.020}^{+0.025}$ & ${0.168}_{-0.002}^{+0.003}$ & ${92.61}_{-0.54}^{+0.67}$ & ${104.2}_{-1.8}^{+1.6}$ & ${65.69}_{-0.53}^{+0.44}$  \\
114962 & 381 & 1927.8 & 0.470 & 0.942 & 92.1 & 44.25 & 86.89 & 4$\rightarrow$3 & Hei1991 \\
& ${388}_{-24}^{+34}$   & ${1928.6}_{-1.5}^{+2.1}$ & ${0.464}_{-0.041}^{+0.040}$ & ${0.957}_{-0.041}^{+0.053}$  & ${93.7}_{-4.3}^{+6.2}$    & ${44.34}_{-0.20}^{+0.21}$    & ${86.93}_{-0.06}^{+0.07}$ \\
\enddata
\tablenotetext{a}{ References taken from the Sixth Catalog of Orbits of
  Visual Binary Stars, available at
  http://ad.usno.navy.mil/wds/orb6/wdsref.html}
\tablenotetext{b}{ Tycho number}
\end{deluxetable}

\newpage
\subsection{Mass sums and dynamical parallaxes} \label{sec:massum}

The orbital parameters and their uncertainties presented in the
previous subsection allow us to compute mass sums for those objects
that have a published trigonometric parallax. In Table~\ref{tab:para}
we indicate, for each of our visual binaries, their spectral type from
the literature (second column), trigonometric parallax (third column,
on the second line of that column we show the parallax uncertainty)
and the astrometric mass sum (in solar masses, last column of the
table) obtained from the published parallax and the period and
semi-major axis from Table~\ref{tab:orbel1}. For each object, in the
upper line we indicate the mass sum from the maximum likelihood (ML)
solution, while in the second line we report the second quartile
(median), with the first quartile as a subscript and the third
quartile as a superscript. The quartiles reported for the mass sum
comes from the MCMC orbital results alone, i.e., they do not include
the error in the trigonometric parallax.

As a test of the reliability of our orbits, we have also computed
dynamical parallaxes (fourth column), primary (fifth column),
secondary (sixth column), and total (seventh column) dynamical masses
on Table~\ref{tab:para}. To compute all these values we have adopted
the photometry values for the primary ($V_P$) and secondary ($V_S$)
from Table~\ref{tab:photom}, the values of $P$ and $a$ from
Table~\ref{tab:orbel1}, and the MLR for main sequence stars from
\cite{hemc93}, who provide an easy-to-evaluate mass {\it vs.}  $M_V$
polynomial relationship for objects below
1$M_{\odot}$\footnote{Several of our objects on Table~\ref{tab:para}
  have masses above 1$M_{\odot}$, but the polynomial fits of the MLR
  are gentle enough to allow some extrapolation, see, e.g., Figure~2
  on \cite{hemc93}.}. A few objects in our list are not on the main
sequence, for them, of course, the adopted MLR relationship (and
therefore, the implied dynamical parallax) is not valid, the most
striking case being HIP 109908 which is further discussed in
Section~\ref{sec:comments} (see also Figure~\ref{fig:hrdiag}). The
quoted uncertainty values for the dynamical parallax come exclusively
from the range of solutions of our MCMC simulations, and not from
uncertainties on either the photometry or the width of the MLR
relationship.

\floattable
\begin{deluxetable}{cccccccc}
\tablecaption{Trigonometric and dynamical parallaxes (visual binaries). \label{tab:para}}
\tablecolumns{8}
\tablewidth{0pt}
\tablehead{
\colhead{HIP} & \colhead{Sp. Type} & \colhead{Trig. Parallax} & \colhead{Dyn. parallax} & Mass$^{\mbox{\tiny{dyn}}}_{\mbox{\tiny{P}}}$ & Mass$^{\mbox{\tiny{dyn}}}_{\mbox{\tiny{S}}}$ & Mass$^{\mbox{\tiny{dyn}}}_{\mbox{\tiny{T}}}$ & Mass$_{\mbox{\tiny{T}}}$\tablenotemark{a}\\
& & (mas) & (mas) & ($\it{M}_{\odot}$) & ($\it{M}_{\odot}$) & ($\it{M}_{\odot}$) & ($\it{M}_{\odot}$)
}
\startdata
79337 & F0IV   &  5.13\tablenotemark{b} & 5.02& 2.35 & 2.19 & 4.53 & 4.24\\
$~$   & $~$    & $\pm 0.70$ & ${5.13}_{-0.07}^{+0.05}$    & ${2.32}_{-0.01}^{+0.01}$ & ${2.16}_{-0.01}^{+0.01}$ & ${4.49}_{-0.02}^{+0.03}$ & ${4.48}_{-0.15}^{+0.10}$    \\
85679 & F0V    & 5.06 & 5.72& 1.69 & 1.37 & 3.05 & 4.41\\
$~$   & $~$    & $\pm 0.97$ & ${5.74}_{-0.10}^{+0.11}$    & ${1.68}_{-0.01}^{+0.01}$ & ${1.36}_{-0.01}^{+0.01}$ & ${3.05}_{-0.02}^{+0.02}$ & ${4.45}_{-0.20}^{+0.22}$    \\
85740 & A5     & 2.24 & 5.00& 1.72 & 1.72 & 3.44 & 38   \\
$~$   & $~$    & $\pm 1.32$ & ${4.91}_{-0.38}^{+0.56}$    & ${1.73}_{-0.08}^{+0.06}$ & ${1.73}_{-0.08}^{+0.06}$ & ${3.46}_{-0.16}^{+0.12}$ & ${37}_{-7}^{+12}$ \\
87567 & B3/5III\tablenotemark{c} & 3.68 & 3.87& 2.86 & 2.82 & 5.68 & 6.61\\
$~$   & $~$    & $\pm 0.54$ & ${3.89}_{-0.04}^{+0.05}$    & ${2.85}_{-0.02}^{+0.02}$ & ${2.81}_{-0.02}^{+0.02}$ & ${5.67}_{-0.03}^{+0.03}$ & ${6.67}_{-0.19}^{+0.20}$    \\
1566$-$1708$-$1\tablenotemark{d} & G0     & $-$  & 11.49   & 1.18 & 0.85 & 2.03 & $-$ \\
$~$   & $~$    & $-$  & ${12.46}_{-0.53}^{+0.74}$ & ${1.14}_{-0.02}^{+0.02}$ & ${0.83}_{-0.02}^{+0.01}$ & ${1.97}_{-0.04}^{+0.03}$ & $-$ \\
89076 & G3V    & 9.88 & 5.51& 1.32 & 1.32 & 2.64 & 0.46\\
$~$   & $~$    & $\pm 1.43$ & ${5.90}_{-0.36}^{+0.64}$    & ${1.28}_{-0.05}^{+0.03}$ & ${1.28}_{-0.05}^{+0.03}$ & ${2.57}_{-0.10}^{+0.06}$ & ${0.55}_{-0.08}^{+0.17}$    \\
89766 & K3+Vk\tablenotemark{e}   & 31.35      & 24   & 0.87 & 0.80 & 1.66 & 0.7\\
$~$   & $~$    & $\pm 1.25$ & ${35}_{-5}^{+10}$ & ${0.77}_{-0.06}^{+0.04}$ & ${0.71}_{-0.05}^{+0.03}$ & ${1.47}_{-0.11}^{+0.07}$ & ${2.1}_{-0.7}^{+2.1}$    \\
91159 & G2V\tablenotemark{f}     & 29.63\tablenotemark{g}      & 32.8   & 1.12 & 1.09 & 2.21 & 2.99\\
$~$   & $~$    & $\pm 0.83$ & ${30.5}_{-3.1}^{+3.8}$ & ${1.15}_{-0.05}^{+0.05}$ & ${1.12}_{-0.05}^{+0.05}$ & ${2.27}_{-0.10}^{+0.09}$ & ${2.47}_{-0.60}^{+0.89}$    \\
92726 & G5V    & 12.99      & 11.2   & 1.14 & 0.96 & 2.10 & 1.36\\
$~$   & $~$    & $\pm 1.65$ & ${12.2}_{-1.0}^{+1.2}$ & ${1.10}_{-0.04}^{+0.04}$ & ${0.94}_{-0.03}^{+0.03}$ & ${2.04}_{-0.07}^{+0.06}$ & ${1.67}_{-0.33}^{+0.49}$    \\
92909 & A3IV\tablenotemark{h}    & 6.99 & 6.70& 1.84 & 1.74 & 3.58 & 3.15\\
$~$   & $~$    & $\pm 0.83$ & ${6.69}_{-0.09}^{+0.10}$    & ${1.84}_{-0.01}^{+0.01}$ & ${1.74}_{-0.01}^{+0.01}$ & ${3.58}_{-0.02}^{+0.02}$ & ${3.14}_{-0.11}^{+0.12}$    \\
93519 & G3/5V  & 9.48\tablenotemark{i} & 13.86   & 1.19 & 1.09 & 2.28 & 7.14\\
$~$   & $~$    & $\pm 0.25$ & ${14.11}_{-0.31}^{+0.36}$ & ${1.18}_{-0.01}^{+0.01}$ & ${1.09}_{-0.01}^{+0.01}$ & ${2.27}_{-0.02}^{+0.02}$ & ${7.47}_{-0.43}^{+0.51}$    \\
96317 & A0     & 6.42 & 4.77& 1.74 & 1.73 & 3.47 & 1.43\\
$~$   & $~$    & $\pm 1.33$ & ${4.51}_{-0.14}^{+0.16}$    & ${1.79}_{-0.03}^{+0.02}$ & ${1.77}_{-0.03}^{+0.03}$ & ${3.56}_{-0.05}^{+0.05}$ & ${1.23}_{-0.09}^{+0.11}$    \\
99114 & F2IV   & 3.90\tablenotemark{j} & 3.77& 1.87 & 1.82 & 3.69 & 3.34\\
$~$   & $~$    & $\pm 0.65$ & ${3.90}_{-0.14}^{+0.25}$    & ${1.84}_{-0.05}^{+0.03}$ & ${1.79}_{-0.05}^{+0.03}$ & ${3.64}_{-0.10}^{+0.06}$ & ${3.63}_{-0.33}^{+0.63}$    \\
102945      & F6V\tablenotemark{k}     & 16.47      & 16.59   & 1.62 & 1.33 & 2.95 & 3.02\\
$~$   & $~$    & $\pm 0.59$ & ${16.59}_{-0.09}^{+0.09}$ & ${1.618}_{-0.004}^{+0.004}$ & ${1.331}_{-0.003}^{+0.003}$ & ${2.95}_{-0.01}^{+0.01}$ & ${3.02}_{-0.04}^{+0.04}$    \\
103620      & K0Vq   & 23.56\tablenotemark{l}      & 24.34   & 0.88 & 0.82 & 1.70 & 1.88\\
$~$   & $~$    & $\pm 0.31$ & ${24.39}_{-0.13}^{+0.22}$ & ${0.883}_{-0.003}^{+0.002}$ & ${0.818}_{-0.002}^{+0.001}$ & ${1.701}_{-0.005}^{+0.003}$ & ${1.89}_{-0.03}^{+0.04}$    \\
107806      & G6V    & 24.09      & 17.18   & 1.08 & 0.99 & 2.06 & 0.75\\
$~$   & $~$    & $\pm 1.03$ & ${17.35}_{-0.22}^{+0.25}$ & ${1.07}_{-0.01}^{+0.01}$ & ${0.98}_{-0.01}^{+0.01}$ & ${2.06}_{-0.01}^{+0.01}$ & ${0.77}_{-0.03}^{+0.03}$    \\
109908      & G8III+G& 11.87      & 15.22   & 2.14 & 1.59 & 3.73 & 7.88\\
$~$   & $~$    & $\pm 0.43$ & ${15.20}_{-0.26}^{+0.22}$ & ${2.15}_{-0.01}^{+0.02}$ & ${1.59}_{-0.01}^{+0.01}$ & ${3.74}_{-0.02}^{+0.03}$ & ${7.84}_{-0.34}^{+0.30}$    \\
114962      & F(8)w\tablenotemark{m}   & 15.08      & 13.80   & 1.13 & 1.07 & 2.19 & 1.68\\
$~$   & $~$    & $\pm 1.80$  & ${13.80}_{-0.14}^{+0.15}$ & ${1.126}_{-0.005}^{+0.004}$ & ${1.068}_{-0.004}^{+0.004}$ & ${2.19}_{-0.01}^{+0.01}$ & ${1.68}_{-0.04}^{+0.05}$   \\
\enddata
\tablenotetext{a}{ Using the solution from Table~\ref{tab:orbel1}, and the published trigonometric parallax
  indicated on the third column of this table}
\tablenotetext{b}{ This is the revised parallax from Gaia DR1. The Hipparcos value was $6.16 \pm 0.57$~[mas]}
\tablenotetext{c}{ B8V according to WDS}
\tablenotetext{d}{ Tycho number}
\tablenotetext{e}{ K3V according to WDS}
\tablenotetext{f}{ F8V according to WDS}
\tablenotetext{g}{ This is the revised parallax from Gaia DR1. The Hipparcos value was $30.41 \pm 0.90$~[mas]}
\tablenotetext{h}{ A5V according to WDS}
\tablenotetext{i}{ This is the revised parallax from Gaia DR1. The Hipparcos value was $14.95 \pm 3.80$~[mas] - note the large difference! See also Section~\ref{sec:comments} for further comments on this object}
\tablenotetext{j}{ This is the revised parallax from Gaia DR1. The Hipparcos value was $3.64 \pm 1.02$~[mas]}
\tablenotetext{k}{ F5IV-V according to WDS}
\tablenotetext{l}{ This is the revised parallax from Gaia DR1. The Hipparcos value was $24.59 \pm 1.14$~[mas]}
\tablenotetext{m}{ F8IV according to WDS}
\end{deluxetable}

\begin{figure}
\plotone{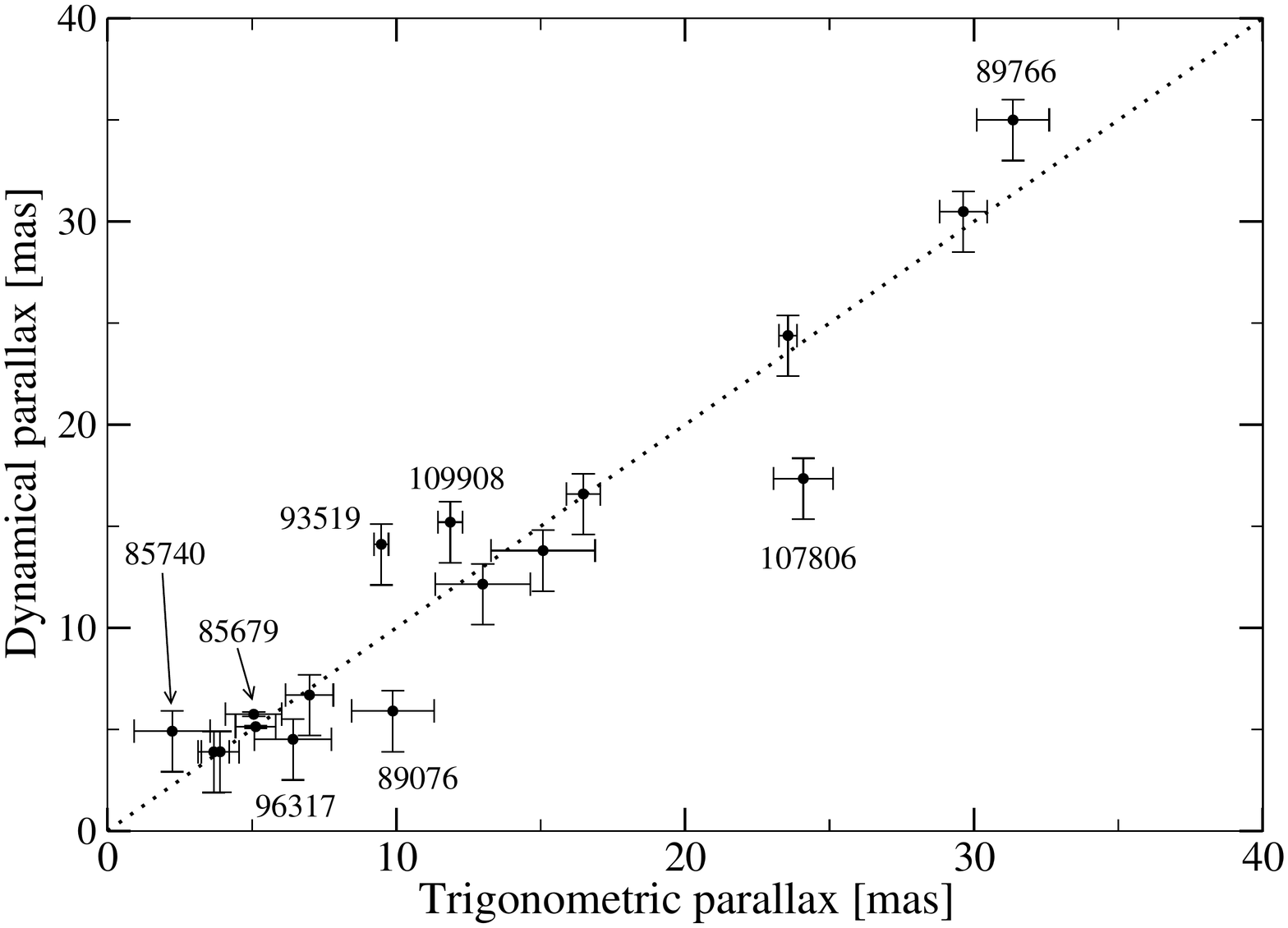}
\caption{Comparison of the dynamical {\it vs.} trigonometric
  parallaxes from the values presented in Table~\ref{tab:para}. We
  indicate the objects (Hipparcos number) that exhibit a very large
  difference between the dynamical and astrometric values, or those
  that appear as discrepant in Figure~\ref{fig:mass}. The dotted line
  is not a fit, it only shows the expected one-to-one relationship in
  both panels. In the ordinate, the quantity shown is the $2^{nd}$
  quartile from Table~\ref{tab:para}.\label{fig:para}}
\end{figure}

\begin{figure}
\plotone{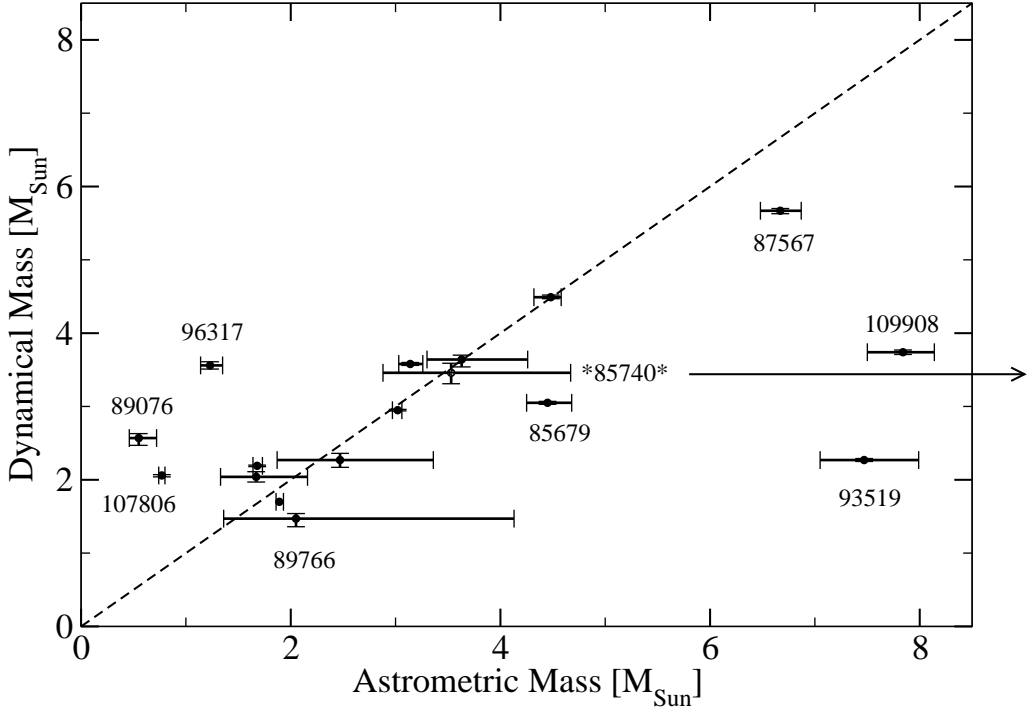}
\caption{Comparison of the dynamical {\it vs.} astrometric mass sums
  from the values presented in Table~\ref{tab:para}. We indicate the
  objects (Hipparcos number) that exhibit a very large difference
  between the dynamical and astrometric values. The dotted line is not
  a fit, it only shows the expected one-to-one relationship in both
  panels. In both axes the quantity shown is the $2^{nd}$ quartile
  from Table~\ref{tab:para}.  HIP 85740 falls out of scale
  to the right in the abscissa in this plot, and it is indicated as an
  arrow at the level of its dynamical mass (this object is further
  discussed in Section~\ref{sec:comments}). If we assume instead the
  trigonometric parallax value indicated in Section~\ref{sec:comments}
  for HIP 85740, it falls in this plot as indicated by its HIP number
  between asterisks.\label{fig:mass}}
\end{figure}

In Figures~\ref{fig:para} and \ref{fig:mass} we show the values of
Table~\ref{tab:para} in graphical form. Generally speaking, there is
good agreement between the dynamical and astrometric parallaxes and
masses, with some notable exceptions that can be attributed to either
a poor orbit determination, a large parallax uncertainty, poor
photometry, or a combination of these. This is discussed on an
object-by-object basis in more detail in
Section~\ref{sec:comments}.

We note that, given the relatively small distances for all our
targets, our dynamical parallaxes have been calculated assuming no
interstellar absorption. Using the reddening model by \citet{meva98},
we can demonstrate that this is indeed a reasonable assumption: If we
take the last point plotted in Figure~\ref{fig:para} (HIP 89766, at a
distance of about 32~pc, see Table~\ref{tab:para}), the model predicts
an extinction in the V-band of 0.027~mag at that Galactic
location. With that extinction, the ML and median dynamical parallaxes
do not change with respect to the values given in
Table~\ref{tab:para}.  For the smallest parallax object in our whole
sample, HIP 85740, with a distance of ~446~pc, the \citet{meva98}
model predicts 0.181~mag of extinction, and the corresponding ML and
median dynamical parallaxes change from 5.00, 4.91~mas (no extinction)
to 4.93, 4.84~mas (extincted) respectively, i.e., completely within
the computed inter-quartile range reported in
Table~\ref{tab:para}. Finally, according to the \citet{meva98}
reddening model, our most-extincted target is the third more distant
of our list, namely, HIP 87567, at a distance of only 272~pc, but
located towards the Galactic center ($(l,b)=(355\degr.9, -4\degr.4)$),
with $A_V = 0.4$~mag. In this case, the corresponding ML and median
dynamical parallaxes change from 3.87, 3.89~mas (no extinction) to
3.73, 3.75~mas (extincted) respectively, a difference of only 0.14~mas
(almost four times smaller than the quoted uncertainty for the
trigonometric parallax of this target), and within $2 \sigma$ of the
inter-quartile range. For all other targets in our list, interstellar
absorption effects can be safely ignored in the calculation of the
dynamical parallaxes.

\newpage
\subsection{Spectroscopic binaries} \label{sec:specbin}

As mentioned in Section~\ref{sec:sample}, two of our speckle targets
turned out to be SB2, HIP 89000, and HIP 111170. A combined solution
for the astrometric data, plus radial velocity data\footnote{To be
  precise, for HIP 89000 we used the radial velocity data published by
  \cite{gr99}, while for HIP 111170 that of \cite{po00}.} was
performed using our MCMC code.

The spectroscopic binaries were analyzed by means of the more
traditional Gibbs sampler introduced by \cite{gege84}, instead of the
DE-MC approach explained in Section~\ref{subsectionMCMC}. A number of
practical reasons support this decision: first, as the dimensionality
of the feature space is larger than that of the visual binary problem
-- ten dimensions, or seven dimensions with the dimensionality
reduction (see Appendix~\ref{sec:dimen}) --, a larger number of chains
must be run within the DE-MC algorithm (at least two times the size of
the feature space, and preferably more), increasing the computational
costs too much; secondly, unlike the visual binaries in our sample, a
tighter exploration range for the period can be proposed from a
simple visual inspection of the observations. Finally, although a raw
Metropolis-Hastings MCMC would be a simpler approach, the dimension of
the problem is large enough, and the location of the solutions
concentrated enough (see Figure~\ref{fig:posterior}), making it highly
probable to fall in zones of low likelihood after a multidimensional
random jump is applied on a sample, possibly reaching pathologically
low values of acceptance probability.

The Gibbs sampler relies on sequentially sampling each component of
the feature space according to the conditional distributions. On the
long run, this scheme is equivalent to draw samples from the joint
posterior distribution. Although the pseudo-code (Appendix~C,
Figure~\ref{alg_MH_Gibbs}) shows each component being sampled
individually, components can also be grouped in blocks if it has some
advantage (for example, if a sub-set of the parameter vector has a
known and easy-to-sample distribution). As the conditional posteriors
do not have a standard form in this problem, we used a
Metropolis-within-Gibbs approach, that is, generating a new sample
according to a proposal distribution (modifying one component or block
of components of the parameter vector at once), and rejecting or
accepting it according to the Metropolis-Hastings ratio. The Gibbs
sampler has been used in the past in the study of exoplanet orbits,
see e.g. \cite{fo05}.

Under the assumption that individual errors of both astrometric and
radial velocity sources are Gaussian, in this case the likelihood
function has the following form (compare to
Equation~(\ref{likelihood_function_astro})):

\begin{eqnarray} \label{likelihood_function_comb}
f(\vartheta_i) \propto \displaystyle \exp \biggl(-\frac{1}{2}& \Bigl(\sum_{k=1}^{N_{x}} \frac{1}{\sigma_x^2(k)} [X(k) - X^{model}(k,i)]^2 + \sum_{k=1}^{N_y
} \frac{1}{\sigma_y^2(k)} [Y(k) - Y^{model}(k,i)]^2 + \nonumber \\
&  \sum_{k=1}^{N_{VP}} \frac{1}{\sigma_p^2(k)} [V_P(k) - V_P^{model}(k,i)]^2 + \sum_{k=1}^{N_{VS}} \frac{1}{\sigma_S^2(k)} [V_S(k) - V_S^{mode
l}(k,i)]^2\Bigr)\biggr), \nonumber
\end{eqnarray}
where $(V_P(k),V_S(k))$ are the primary (with $N_{VP}$ measurements)
and secondary (with $N_{VS}$ measurements) Heliocentric radial
velocity observations with uncertainties $(\sigma_P(k), \sigma_S(k))$
respectively, $(V_P^{model}(k,i), V_S^{model}(k,i))$ are the model
radial velocities, and the remaining parameters have been defined
earlier. This $f_i$ is used to calculate the ratios within the
Metropolis-Hastings steps. We choose $N_{steps} = 2\times10^6$, with
burn-in periods of $2\times10^5$ for both HIP 89000 and HIP 111170. On
each Metropolis-Hastings step, an additive Gaussian ``noise'' was used
to propose new samples. Parameters of the proposal distributions, as
well as the boundaries of the initial uniform distributions used for
these two objects are shown in Table \ref{spec_alg_params}.

\floattable
\begin{deluxetable}{ccccccccc}
\tablecaption{Algorithm-related parameters for our two SB2 and
  astrometric binaries HIP 89000 and HIP
  111170.\label{spec_alg_params}} \tablecolumns{9}
\tablewidth{0pt}
\tablehead{ \colhead{HIP} & \colhead{Alg-related} &
  \colhead{$P\tablenotemark{a}$} & \colhead{$T^{\prime}$} & \colhead{$e$} &
  \colhead{$\Omega$} & \colhead{$i$} & \colhead{$\varpi$} &
  \colhead{$q$}\\
& parameters & $(yr)$ & $~$ & $~$ &
  $(^{\circ})$ & $(^{\circ})$ & $(mas)$ & $~$ }
\startdata
89000 & $\sigma\tablenotemark{b}$ &$0.01$ & $0.01$ & $0.01$ &
$1$ & $1$ & $1$ & $0.01$\\
$~$ & Range & $(0.1, 1.0)$ & $(0, 1)$ & $(0, 0.99)$ & $(0, 360)$ &
$(0, 180)$ & $-$\tablenotemark{c} & $(0, 1)$\\
111170 & $\sigma\tablenotemark{b}$& $0.05$ & $0.01$ & $0.01$ & $1$ &
$1$ & $1$ & $0.01$\\
$~$ & Range & $(0.2, 3.0)$ & $(0, 1)$ & $(0, 0.99)$ & $(0, 360)$ &
$(0, 180)$ & $(32.35, 46.35)$ & $(0, 1)$\\
\enddata
\tablenotetext{a}{ As in the case of visual binaries, the search was done in $\log P$ space.} 
\tablenotetext{b}{ In MCMC it is not essential that this ´evolution´
  noise be smaller than the final uncertainty of the parameter to be
  estimated, there is a rather
  wide range for $\sigma$ in which the algorithm works well and is stable in the
  solution.}
\tablenotetext{c}{ Gaussian prior with mean and standard deviation
  values indicated in the third column of Table~\ref{tab:masspec}
  (trigonometric parallax).}
\end{deluxetable}

In the case of HIP 89000, a Gaussian prior for the parallax was
included in the fitness function $f$, since infeasible values of
$\varpi$ were explored if the parallax was set free (this simply means
that the available data is not yet informative enough to give an
estimate for this parameter). For HIP 111170, instead, $\varpi$ is
uniformly sampled in the wide range indicated in
Table~\ref{spec_alg_params}, and it converges to a value close to the
dynamical parallax and not far from the trigonometric published
parallax.

\begin{figure}[!hbt]  
      \begin{minipage}[!h]{0.5\linewidth}
        \centering
        \includegraphics[height=.28\textheight]{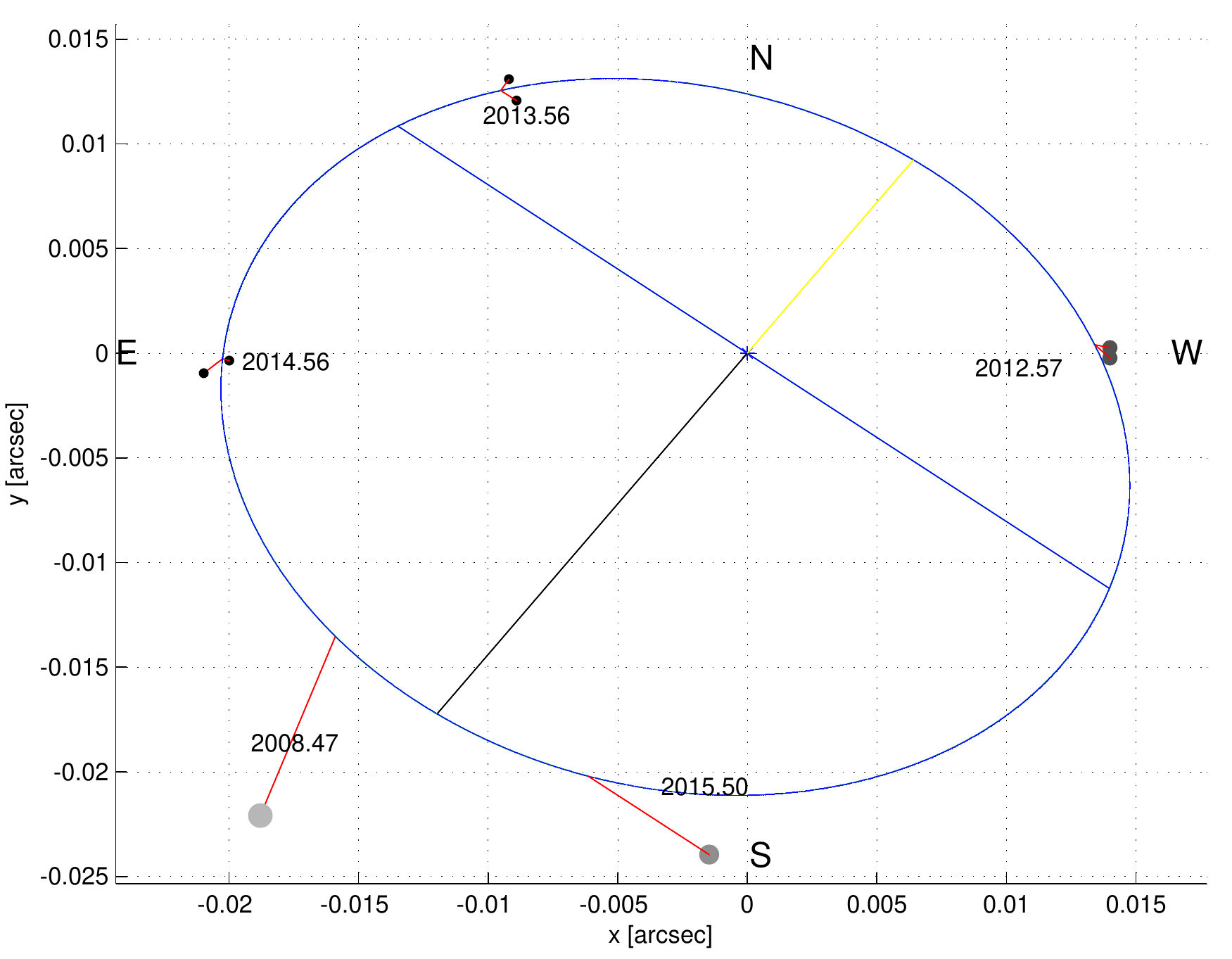}\\
        \includegraphics[height=.28\textheight]{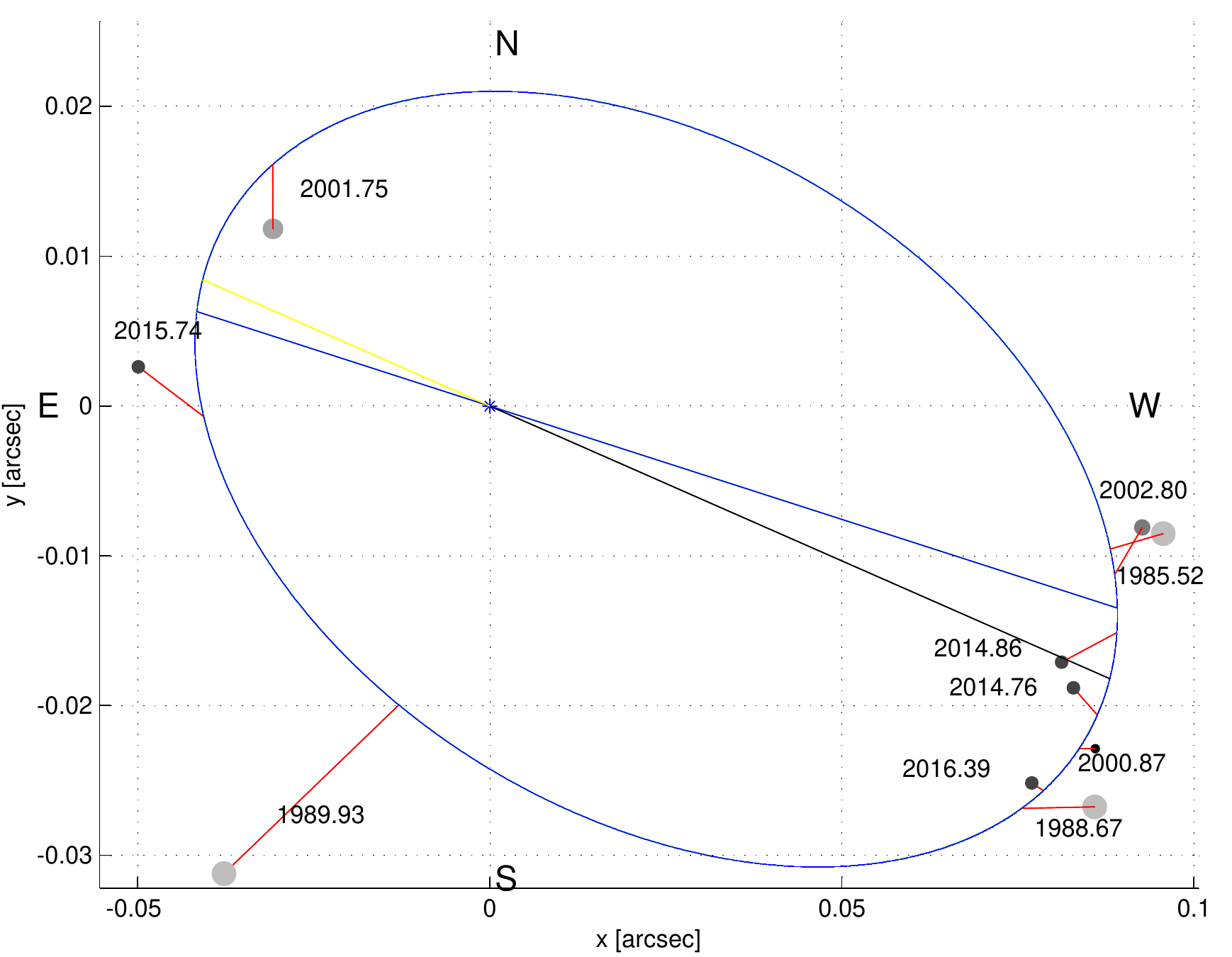}
      \end{minipage}
      \begin{minipage}[!h]{0.5\linewidth}
        \centering
        \includegraphics[height=.28\textheight]{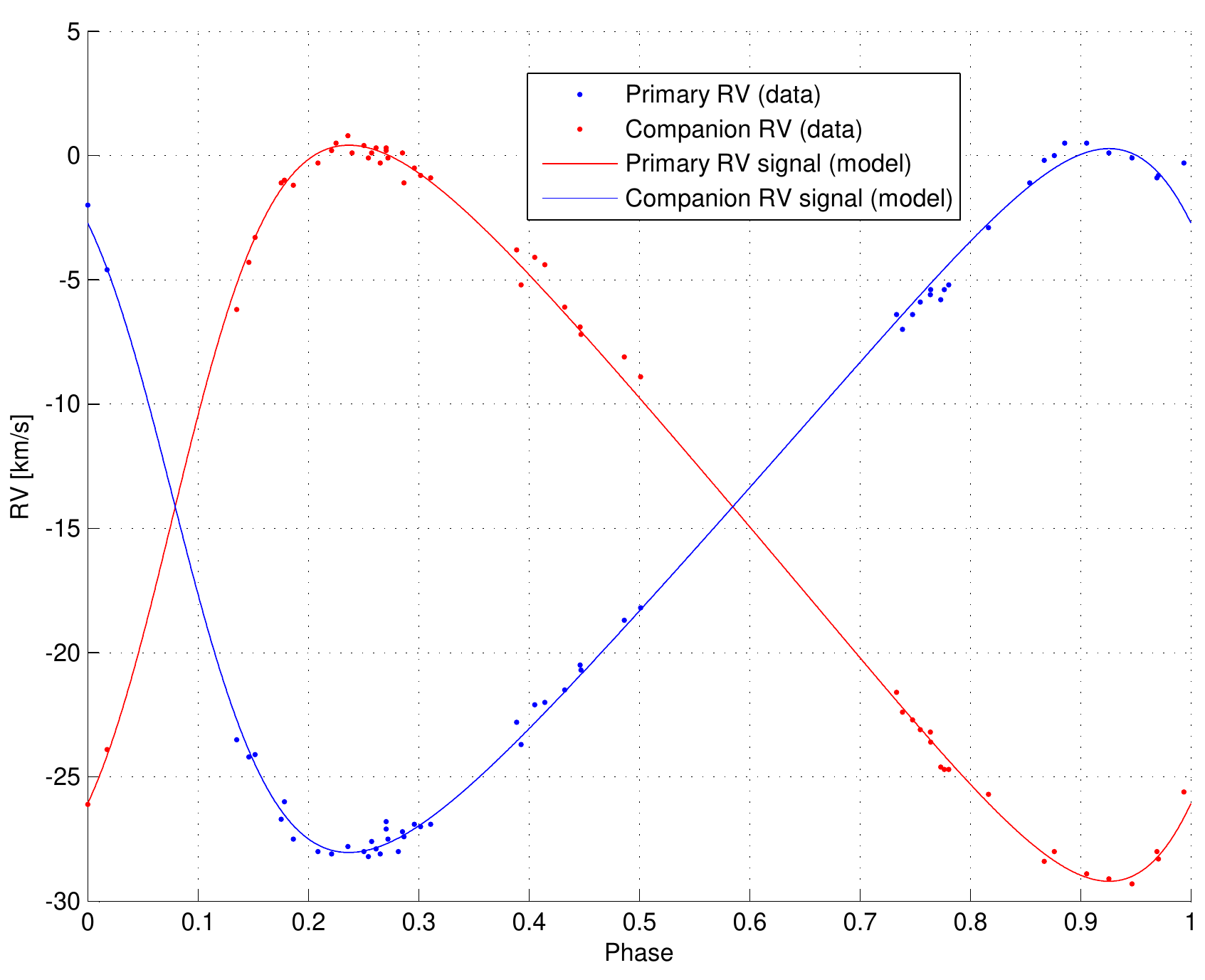}\\
        \includegraphics[height=.28\textheight]{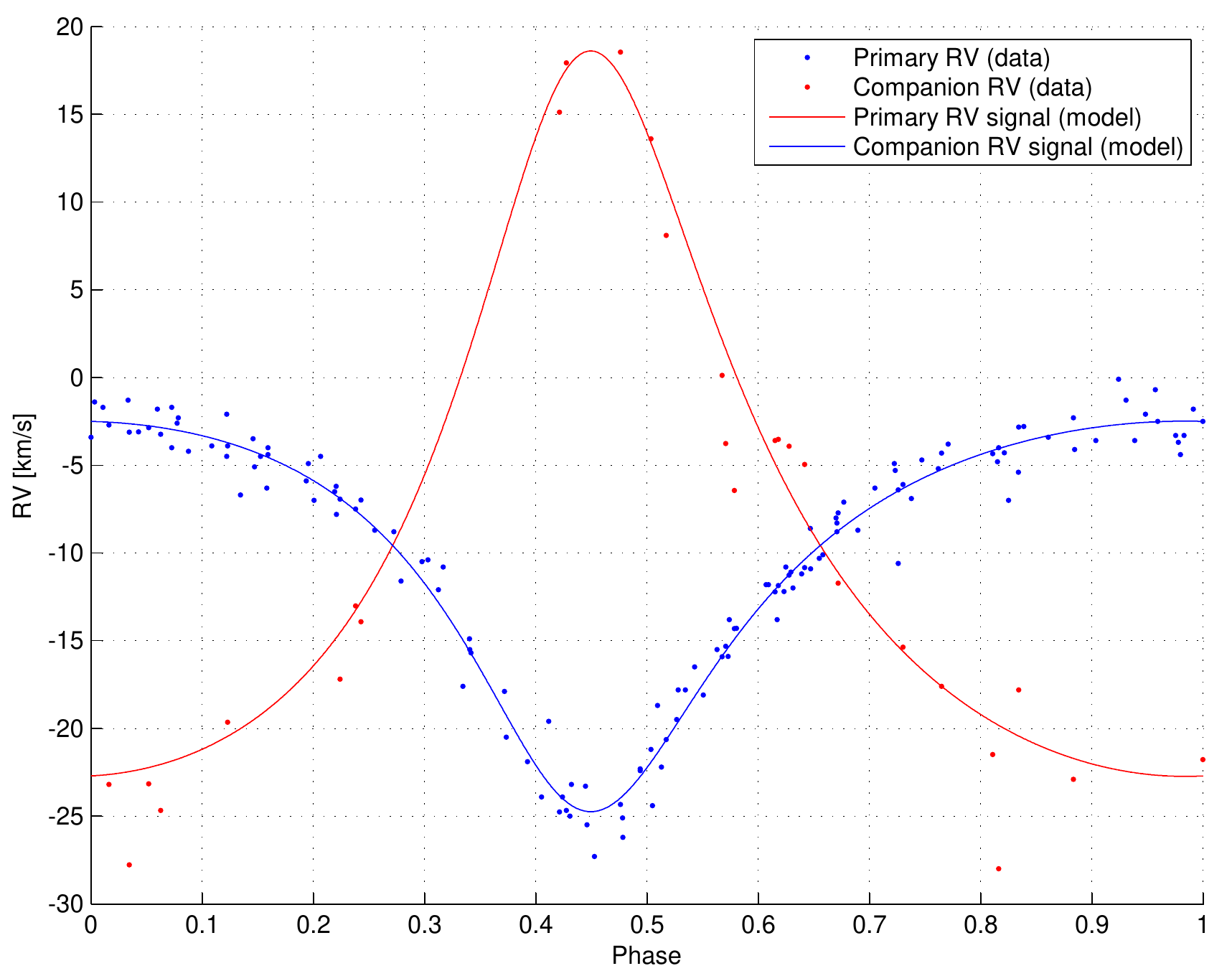}
      \end{minipage} 
      \caption{MCMC fits to HIP 89000 (upper panel) and HIP 111170
        (lower panel). The left panel shows the astrometric data and
        its fit, the right panel shows the fit to the radial velocity
        for both components.\label{fig:orbvrad}}
    \end{figure}

\begin{figure}[!hbt] 
\centering
\includegraphics[height=.48\textheight]{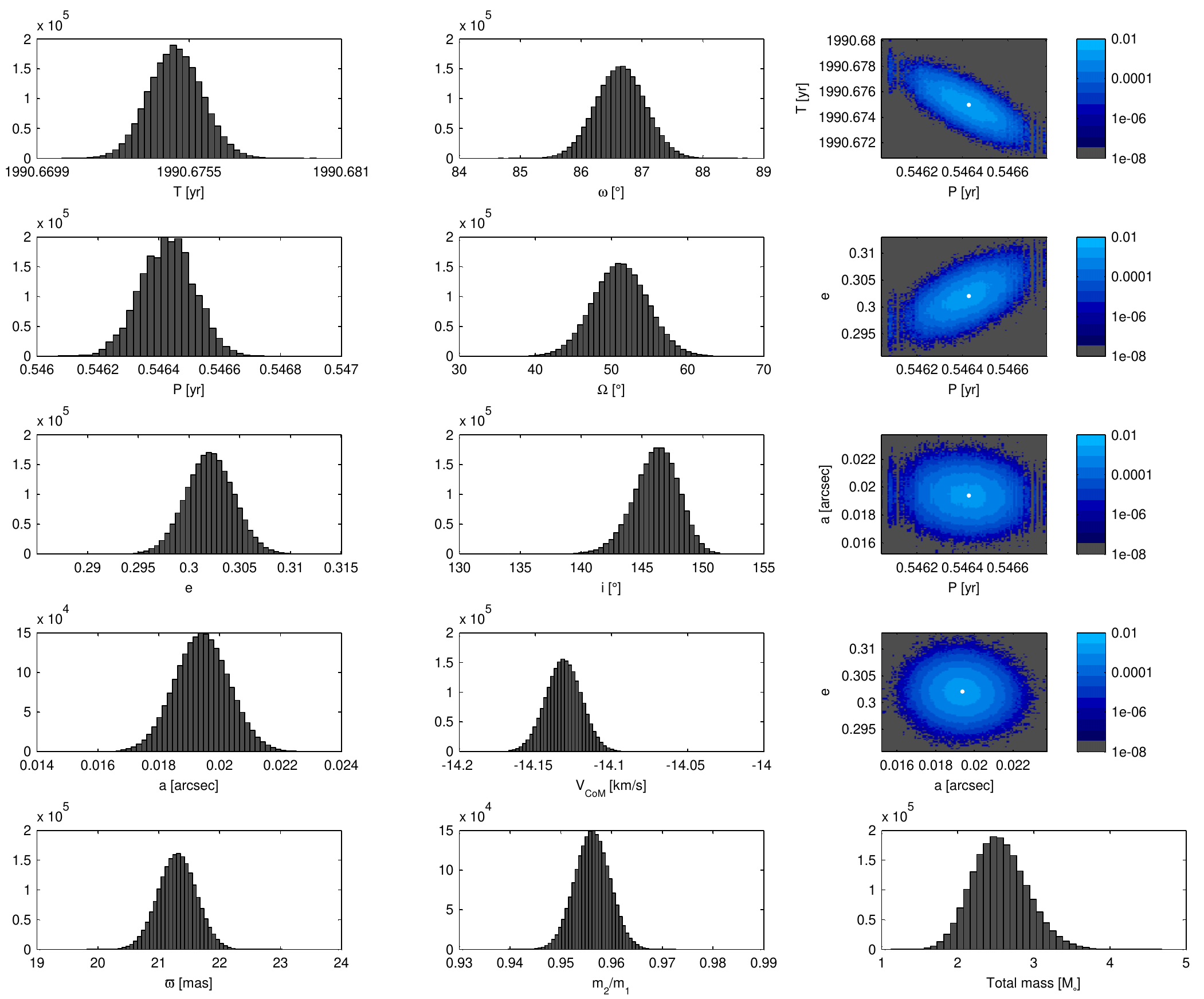}\\
$~$\\
\includegraphics[height=.48\textheight]{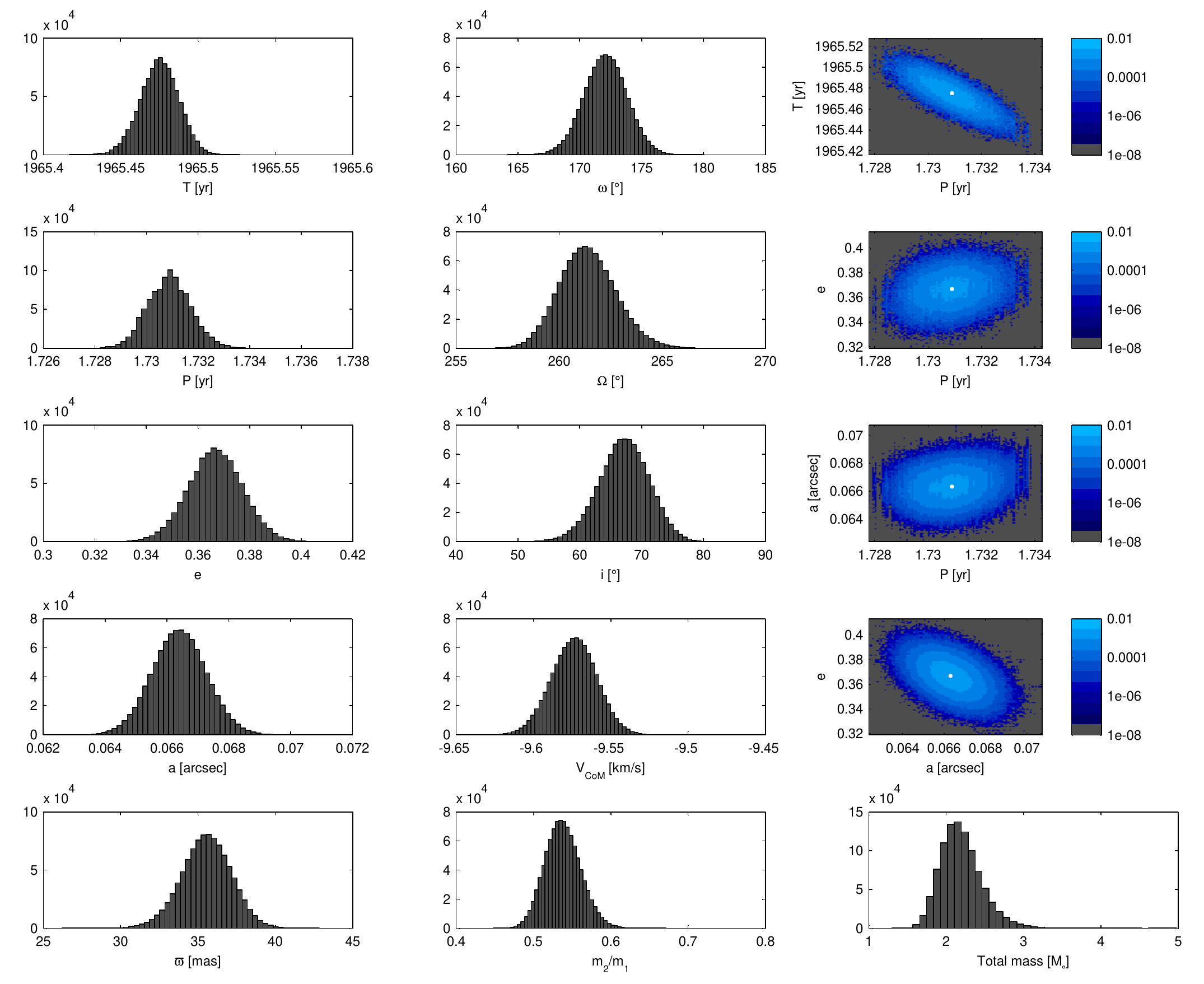}
\caption{Posterior distributions for spectroscopic binaries: HIP 89000
  (top five rows), HIP 111170 (bottom five
  rows).\label{fig:posterior}}
\end{figure}

The resultant orbital elements, as well as the mass ratio and mass sum
(with their derived uncertainties) are shown in
Table~\ref{tab:orbel2}. Since the posterior PDFs obtained here are
tighter and more Gaussian-like than those obtained for our visual
binaries, the expected value offers a good estimate of the target
parameter vector, and is the estimator of choice in this section. In
Figure~\ref{fig:orbvrad} we show the joint fit to the orbit and the
radial velocity curves. As it can bee seen from the table and figure,
even in the case of a rather poor coverage of the astrometric orbit as
is the case HIP 89000, the combined solutions produce very precise
orbital parameters. This point is also highlighted in
Figure~\ref{fig:posterior}, where we present the posterior PDFs, which
exhibit tight and well-constrained distributions. In particular,
judging from the quartile ranges, we can see that for HIP 89000 the
mass ratio is determined with a 0.3\% uncertainty, while the
uncertainty on the mass sum is 8\%. For HIP 111170 these value are 3\%
and 7\% respectively.

\floattable
\rotate
\begin{deluxetable}{cccccccccccc}
\tablecaption{Orbital elements for our two SB2 and astrometric
  binaries HIP 89000 and HIP 111170.\label{tab:orbel2}}
\tablecolumns{12}
\tablewidth{0pt}
\tablehead{
\colhead{HIP} & \colhead{P} & \colhead{T$_0$} &
\colhead{e} & \colhead{a} & \colhead{$\omega$} & \colhead{$\Omega$} &
\colhead{i} & \colhead{$V_{CoM}$} & \colhead{$m_S/m_P$} & \colhead{$\varpi$} & \colhead{$m_P+m_S$} \\
 & \colhead{(yr)} & \colhead{(yr)} &
 & \colhead{($\arcsec$)} & \colhead{($^{\circ}$)} & \colhead{($^{\circ}$)} & \colhead{($^{\circ}$)} & \colhead{($km~s^{-1}$)} & & \colhead{(mas)} & \colhead{($\it{M}_{\odot}$)}
}
\startdata
89000 & $0.54643$ & $1990.675$ & $0.302$ &
$0.0190$ & $86.65$ & $51.2$ & $146.2$ & $-14.131$ & $0.956$ &
$21.31$ & $2.52$\\
& ${0.54643}_{-0.00006}^{+0.00006}$ &
${1990.675}_{-0.001}^{+0.001}$ & ${0.302}_{-0.002}^{+0.002}$ &
${0.0190}_{-0.0004}^{+0.0006}$ & ${86.65}_{-0.27}^{+0.27}$ &
${51.2}_{-2.4}^{+2.4}$ & ${146.3}_{-1.3}^{+1.2}$ &
$-14.131_{-0.008}^{+0.007}$ & ${0.956}_{-0.002}^{+0.003}$ &
${21.31}_{-0.21}^{+0.21}$ & ${2.54}_{-0.23}^{+0.25}$\\
111170& $1.7309$ & $1965.48$ & $0.367$ & $0.0663$ &
$172.1$ & $261.39$ & $67.1$ & $-9.573$ & $0.538$ & $35.5$ &
$2.17$ \\
& ${1.7309}_{-0.0006}^{+0.0006}$ & ${1965.47}_{-0.01}^{+0.01}$ &
${0.367}_{-0.007}^{+0.007}$ & ${0.0664}_{-0.0006}^{+0.0006}$ &
${172.1}_{-1.2}^{+1.2}$ & ${261.34}_{-0.90}^{+0.94}$ &
${67.2}_{-2.7}^{+2.7}$ & $-9.574_{-0.010}^{+0.010}$ &
${0.537}_{-0.015}^{+0.016}$ & ${35.6}_{-1.1}^{+1.1}$ &
${2.17}_{-0.16}^{+0.18}$\\
\enddata
\end{deluxetable}

In Table~\ref{tab:masspec} we present a comparison of the masses for
the system as well as the individual component masses obtained from
the joint fit of the orbit to the astrometric and radial velocity data
shown in Table~\ref{tab:orbel2}. The format of the table is similar to
that of Table~\ref{tab:para} in that it includes, for comparison
purposes, dynamical parallaxes (fourth column) and individual and
total dynamical masses (fifth to seventh columns) calculated in the
same fashion as described in Section~\ref{sec:massum} for the visual
binaries. The eighth column gives the total mass using the orbital
elements given in Table~\ref{tab:orbel2}, but adopting the published
trigonometric parallax given in the third column of the table, whereas
the ninth to eleventh columns gives the individual masses and the
total mass, letting the parallax of the system be a free parameter of
the MCMC code (i.e., adopting instead the parallax given in the
twelfth column of Table~\ref{tab:orbel2}\footnote{As explained
  earlier, since the astrometric coverage for HIP 89000 is rather
  poor, the parallax was not calculated independently of the published
  value, rather, the published parallax was used as prior, albeit the
  reported value in Table~\ref{tab:orbel2} is the outcome of the MCMC
  calculation.}). We note that the quartiles on
Mass$_{\mbox{\tiny{T}}}$ in Table~\ref{tab:masspec} {\it do not}
include any contribution from trigonometric parallax errors (just as
in Table~\ref{tab:para}), while
Mass$^{\mbox{\tiny{comb}}}_{\mbox{\tiny{P}}}$,
Mass$^{\mbox{\tiny{comb}}}_{\mbox{\tiny{S}}}$, and
Mass$^{\mbox{\tiny{comb}}}_{\mbox{\tiny{T}}}$, being derived from MCMC
simulations that have the parallax as a free parameter, {\it do}
include the extra variance from this parameter.

In the case of HIP 89000 the agreement between all estimates of the
mass is excellent as well as between the dynamical and published
parallax. A preliminary astrometric orbit was published by
\cite{hoet15}, and our combined orbital parameters agree quite well
with theirs. We note that our SOAR data on 2008.47 and 2015.50 (see
Figure~\ref{fig:orbvrad}) are uncomfortably discrepant, probably due
to the small separation (below our diffraction limit), and which might
require further observations on an 8m class telescope. For HIP 111170
the dynamical and orbital parallaxes ($\sim 36$~mas) agree quite well
with each other, but are smaller by about 5~mas with respect to the
published ($39.35 \pm 0.70$~mas) trigonometric parallax (or more than
$5\sigma$ of the published parallax uncertainty), thus leading to a
larger total mass than that obtained by adopting the published
parallax directly, as it can be readily seen in
Table~\ref{tab:masspec}. This discrepancy could be due either to the
rather poor coverage of the astrometric orbit (in comparison with the
radial velocity curves), or to a biased Hipparcos parallax due to the
orbital motion of the system, as shown by \cite{so99} (see, in
particular his Section~3.1, and Table~2). We also note that the
published photometry on SIMBAD for this object does not agree very
well with that in the Hipparcos catalogue (see
Figure~\ref{fig:photom}), but there is very good agreement between
$V_{Hip}$ and $V_{Sys}$ (see Table~\ref{tab:photom}).  We note that
\cite{pola99} performed a detailed comparison of Hipparcos
trigonometric parallaxes with orbital parallaxes from the SB2s
available at that time and found, in general, good correspondence
between them with a few discrepant cases but at a less than $3 \sigma$
level. The more precise Gaia parallaxes will probably shed some light
into this issue.

As it can be seen from Table~\ref{tab:masspec}, the mass of the
individual components for both binaries are determined with a formal
uncertainty of $\sim 0.1 M_\odot$, but this could possibly be improved
by further speckle observations on an 8~m class telescope by providing
a better-constrained astrometric orbit.

We finally note the good agreement for the triad ($V_{CoM}$, $K_P$,
$K_S$) reported by the 9th Catalogue of Spectroscopic Binary Orbits
and our calculation, namely: ($-14.12 \pm 0.04$, $14.20 \pm 0.07$,
$14.80 \pm 0.07$) {\it vs.}  ($-14.13 \pm 0.08$, $14.158 \pm 0.026$,
$14.806 \pm 0.015$)~km~s$^{-1}$ for HIP 89000 and ($-9.716 \pm 0.097$,
$11.44 \pm 0.16$, $20.96 \pm 0.61$) {\it vs.}  ($-9.57 \pm 0.01$,
$11.114 \pm 0.080$, $20.68 \pm 0.55$)~km~s$^{-1}$ for HIP 111170
respectively. This is particularly interesting, since it validates the
mathematical formalism developed in Appendix~\ref{sec:dimen}, in
particular in what matters to our extension of the proposal by
\cite{wrho09} to the case of binary stars (see
Equation~(\ref{conditionsBinStars}) and the paragraph that follows
it).

\floattable
\begin{deluxetable}{ccccccccccc}
\tablecaption{Trigonometric and dynamic parallaxes (spectroscopic binaries) \label{tab:masspec}}
\tablecolumns{6}
\tablewidth{0pt}
\tablehead{
\colhead{HIP} & \colhead{Sp. Type} & \colhead{Trig. Parallax} & \colhead{Dyn. parallax} & Mass$^{\mbox{\tiny{
dyn}}}_{\mbox{\tiny{P}}}$ & Mass$^{\mbox{\tiny{dyn}}}_{\mbox{\tiny{S}}}$ & Mass$^{\mbox{\tiny{dyn}}}_{\mbox{\tiny{T}}}$ & Mass$_{\mbox{\tiny{T}}}$\tablenotemark{a} &
Mass$^{\mbox{\tiny{comb}}}_{\mbox{\tiny{P}}}$ &
Mass$^{\mbox{\tiny{comb}}}_{\mbox{\tiny{S}}}$ &
Mass$^{\mbox{\tiny{comb}}}_{\mbox{\tiny{T}}}$\\
& & (mas) & (mas) & ($\it{M}_{\odot}$) & ($\it{M}_{\odot}$) & ($\it{M}_{\odot}$) & ($\it{M}_{\odot}$) & ($\it{M}_{\odot}$)
 & ($\it{M}_{\odot}$) & ($\it{M}_{\odot}$)
}
\startdata
89000 & F6V\tablenotemark{b} & $21.31$ & $20.40$ &
$1.57$ & $1.30$ & 2.88 & $2.52$ & $1.29$ & $1.23$ & $2.52$\\
$~$ & $~$ & $\pm 0.31$ & ${20.45}_{-0.72}^{+0.73}$ &
${1.57}_{-0.02}^{+0.02}$ & ${1.30}_{-0.02}^{+0.02}$ & ${2.87}_{-0.04}^{+0.04}$ &
${2.54}_{-0.23}^{+0.24}$ & ${1.30}_{-0.12}^{+0.13}$ & ${1.24}_{-0.11}^{+0.12}$ &
${2.54}_{-0.23}^{+0.25}$ \\
111170 & F8V\tablenotemark{c} & $39.35$ & $36.37$ &
$1.20$ & $0.82$ & $2.02$ & $1.60$ & $1.41$ & $0.76$ & $2.17$\\
$~$ & $~$ & $\pm 0.70$ & ${36.42}_{-0.37}^{+0.37}$ &
${1.20}_{-0.01}^{+0.01}$ & ${0.822}_{-0.003}^{+0.003}$ & ${2.02}_{-0.01}^{+0.01}$ &
${1.60}_{-0.04}^{+0.04}$ & ${1.41}_{-0.11}^{+0.13}$ & ${0.76}_{-0.05}^{+0.06}$ &
${2.17}_{-0.16}^{+0.18}$\\
\enddata
\tablenotetext{a}{ Using the solution from Table~\ref{tab:orbel2}, and the published trigonometric parallax on the third column of this table}
\tablenotetext{b}{ F7V+F7.5V according to WDS}
\tablenotetext{c}{ F7V according to WDS}                                                                                        

\end{deluxetable}

\newpage

\section{H-R diagram and comments on individual objects} \label{sec:comments}

In this section we provide comments regarding individual objects and
their orbital fits, and we put them on an H-R diagram for an overall
discussion.

\begin{figure}
\plotone{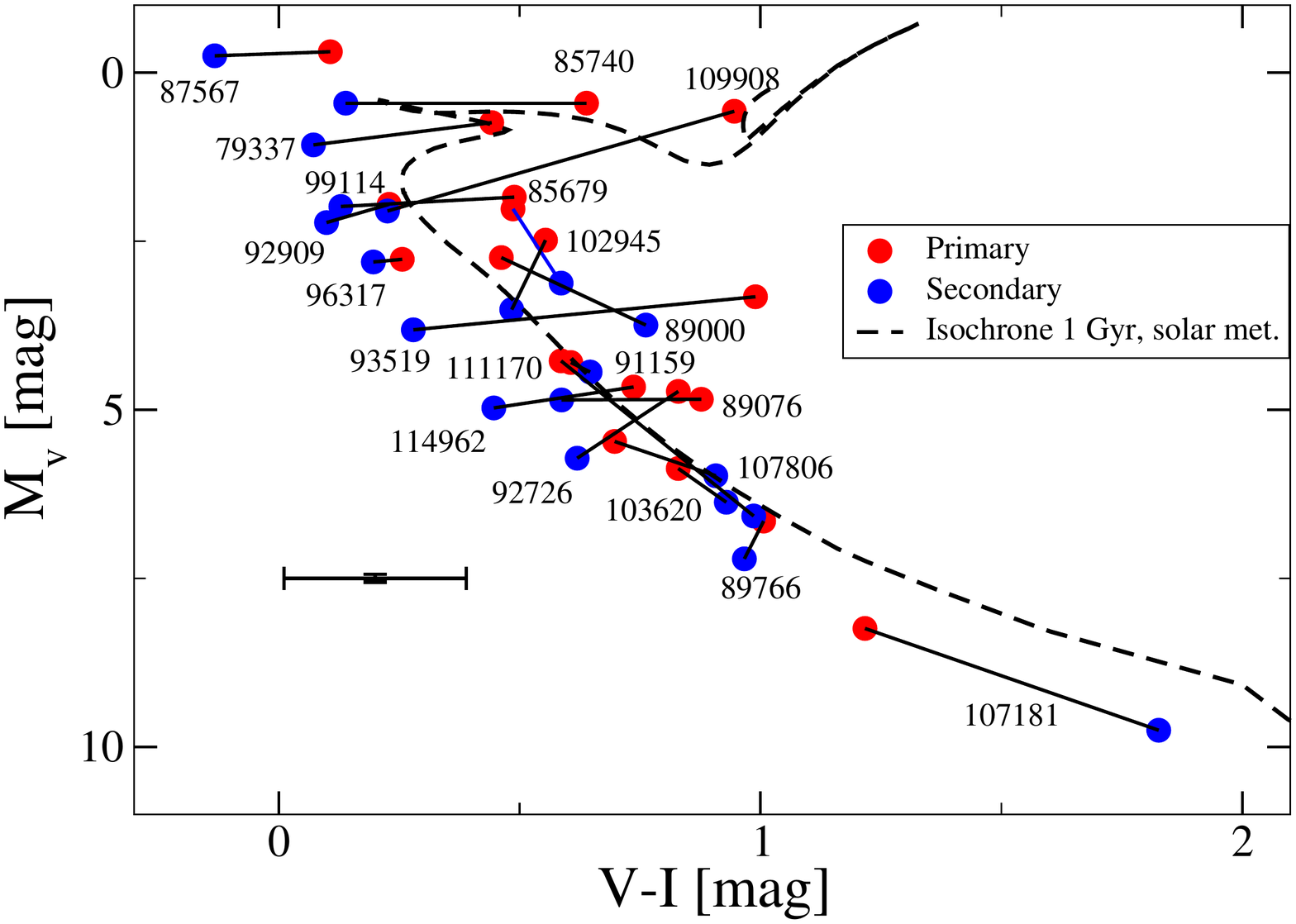}
\caption{H-R diagram for our visual and spectroscopic
  binaries. Binaries have been joined by a line, and their Hipparcos
  number noted. The point at (0.25,7.5) represents the estimated error
  on our photometry as discussed in Section~\ref{sec:sample}. We have
  also superimposed a 1~Gyr isochrone of solar metallicity from
  \cite{maet17}, which are available for download from
  http://stev.oapd.inaf.it/cgi-bin/cmd. \label{fig:hrdiag}}
\end{figure}

In Figure~\ref{fig:hrdiag} we present an observational H-R diagram for
all the objects in our sample, including the two spectroscopic
binaries described in Section~\ref{sec:specbin}. To derive individual
colors for each component, we used the individual magnitudes for the
primary and secondary in the $V$-band from Table~\ref{tab:photom}, and
the $\Delta I$ for the system from our own measurements indicated in
the same table. The (combined) magnitude for the system was computed
from $I_{Sys} = V_{Sys} - (V-I)_{Hip}$. With these values, we computed
the individual magnitudes as (primary) $I_P = I_{Sys} + 2.5 \times
\log \left( 1.0 + 10^{-0.4 \cdot \Delta I} \right) $ and (secondary)
$I_S = I_P + \Delta I$. Note that to compute $I_{Sys}$ we used
$V_{Sys}$ rather than $V_{Hip}$ so that the derived pairs
$(V,I)_{P,S}$ are self-consistent (albeit, in general, as noted in
Section~\ref{sec:sample}, there is good agreement between $V_{Sys}$
and $V_{Hip}$). Regarding distances, we adopted the published
trigonometric parallaxes shown in Tables~\ref{tab:para} and
\ref{tab:masspec}.

\begin{figure}[!hbt] 
\begin{minipage}[!h]{0.5\linewidth}
\centering
\includegraphics[height=.25\textheight]{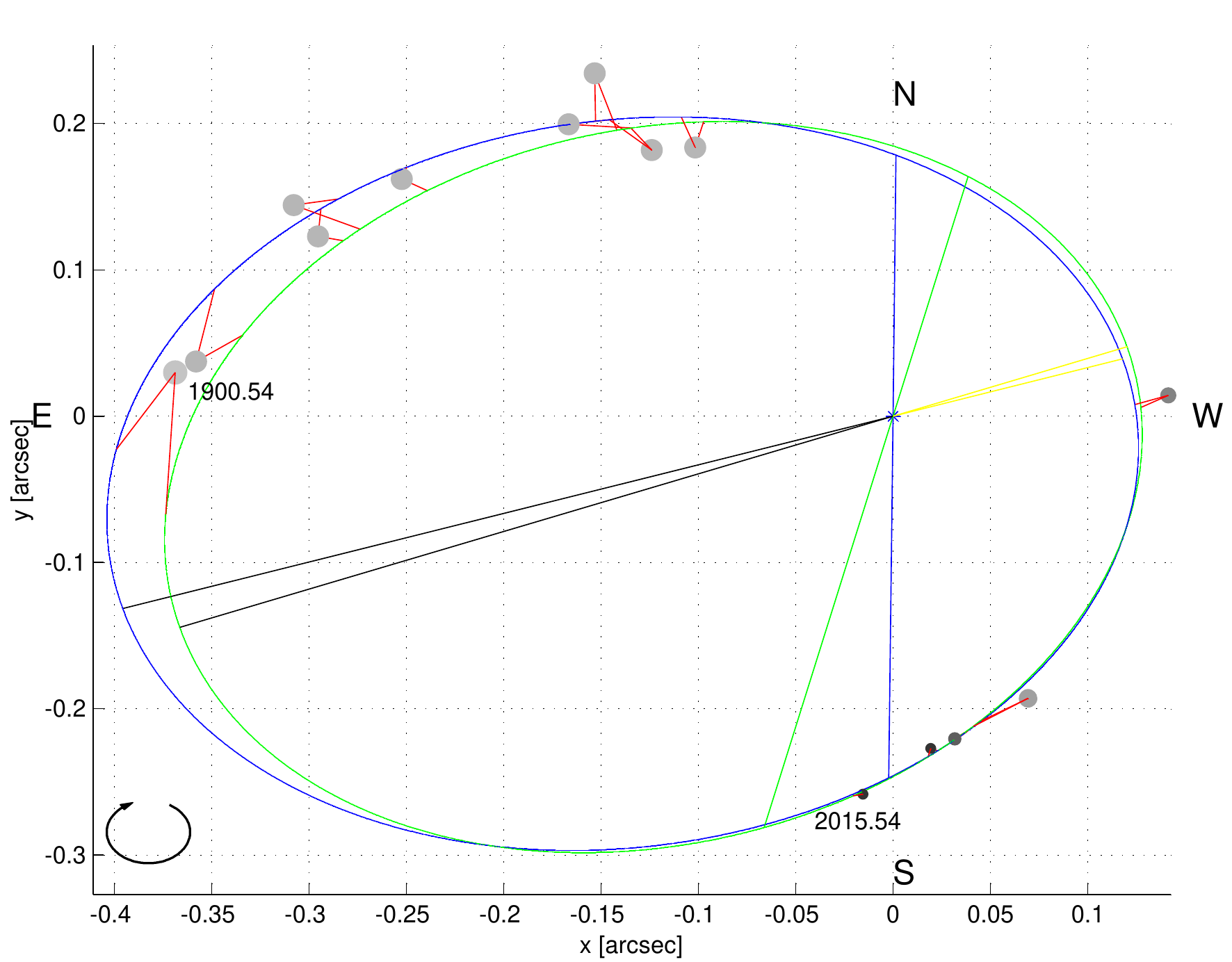}
\end{minipage}
\begin{minipage}[!h]{0.5\linewidth}
\centering
\includegraphics[height=.25\textheight]{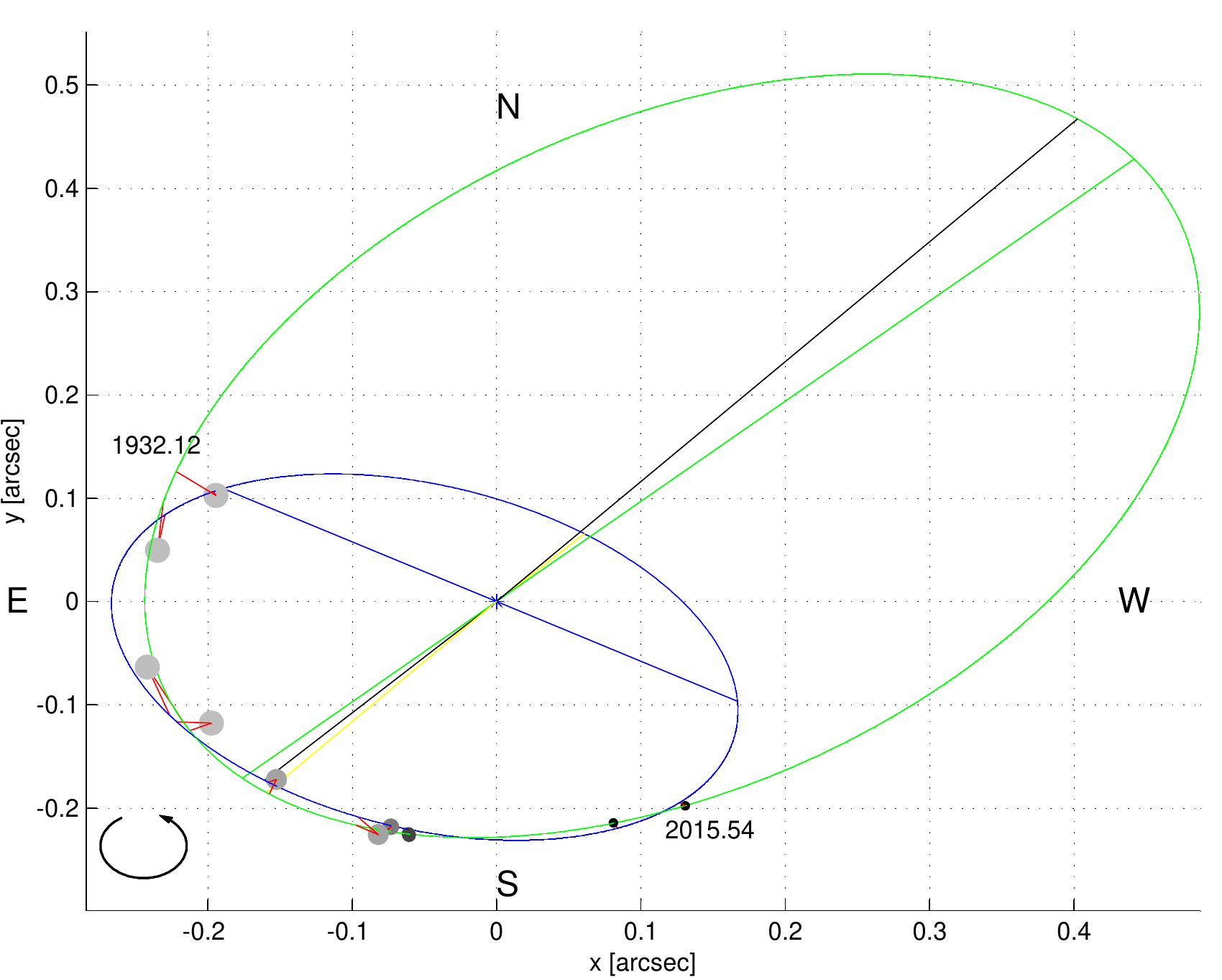}
\end{minipage} 
\caption{Orbits for two of our objects, compared to recently published
  results. Left panel is for HIP 85679, right panel is for HIP
  89076. Blue orbit is from our MCMC solution, green orbit is from IAU
  Double Stars circular \# 191. The adoption of different weights to
  each datum might be responsible for the differences in the
  fits.\label{fig:compa}}
\end{figure}

\begin{trivlist}
\item {\bf HIP 79337:} It seems to be a
  nearly circular orbit, with a degeneracy between the parameters
  $T_0$ and $\omega$. Quadrant flips were required in earlier
  data. Still, we think this is a substantial revision, and
  improvement, over the latest orbits for this object, published by
  \citet{doan13}. The inclination will be better defined by
  observations when it closes down again, in a decade.

\item {\bf HIP 85679:} The orbit coverage and overall quality of the
  current fit and data seems to merit to promote it from Grade 5 to
  Grade 3 in the USNO orbit catalogue. This object was recently
  reported by Miles \& Mason in the IAU Double Stars circular \#
  191\footnote{Available at
    http://www.usno.navy.mil/USNO/astrometry/optical-IR-prod/wds/dsl}. The
  listed orbital parameters are $P=220.41$, $T_0=1987.21$, $e=0.541$,
  $a=0.293$, $\omega=252.6$, $\Omega= 179.5$, $i= 156.3$ (no
  uncertainties are quoted). These values are in good agreement with
  our own parameters in Table~\ref{tab:orbel1}, {\it despite} the fact
  that there is a rather large discrepancy between the dynamical and
  astrometric mass (see Figure~\ref{fig:mass}), which could indicate a
  dubious orbital solution. The (dimensionless), mean square error
  value for our solution\footnote{Computed as $\frac{1}{N}
    \sum_{k=1}^{N} \Bigl( \frac{1}{\sigma_x^2(k)} [X(k) -
      X^{bestfit}(k)]^2 + \frac{1}{\sigma_y^2(k)} [Y(k) -
      Y^{bestfit}(k)]^2\Bigr)$, see also
    Equation~(\ref{likelihood_function_astro}).} is 2.5, while the
  mean square error for their solution is 8.3 (see the left panel of
  Figure~\ref{fig:compa}, where we compare both orbits), and their
  mass sum leads to 4.0$M_{\odot}$ which is slightly smaller than ours
  at 4.4$M_{\odot}$, and in the right direction (albeit still too
  large) to agree with an F0V primary. We also note the good agreement
  between the dynamical and trigonometric parallax for this object on
  Figure~\ref{fig:para} (within $1\sigma$ of the parallax error), so
  its large astrometric mass sum does not seem to be a consequence of
  an erroneous parallax. However, the astrometric mass depends very
  strongly on the assumed parallax, even small changes on the parallax
  have a big impact on the derived mass, e.g., if we adopt the
  dynamical parallax instead of the published parallax, the
  astrometric mass becomes 3.0$M_{\odot}$. Finally, we note the the
  large difference between $V_{Hip}$ (and $V_{Sim}$) {\it vs.}
  $V_{Sys}$, see Table~\ref{tab:photom} and Figure~\ref{fig:photom}.

\item {\bf HIP 85740:} Long and
  undetermined period, very small orbit coverage. Mass sum is too
  large. However, trigonometric parallax is small and has large error,
  so mass sum could be reduced importantly by considering a parallax
  larger by 2$\sigma$ of its error in Table~\ref{tab:para}. Indeed,
  increasing the parallax by 2$\sigma$ leads a value of 4.88~mas in
  consistency with the dynamical parallax, and with a very reasonable
  astrometric mass of 3.5$M_{\odot}$. Consistency with the dynamical
  parallax should however be viewed in this case with caution due to
  the discrepancy noted in Section~\ref{sec:sample} between $V_{Sys}$
  and $V_{Hip}$, $V_{Sim}$ (see also Figure~\ref{fig:photom}). Also, WDS
  reports an equal magnitude system ($V=8.7$), whereas our own Speckle
  measurements indicate a $\Delta I=0.5$ (see Table~\ref{tab:photom}),
  which casts some doubts on the the reported values by WDS, and about
  the true location of this object on the H-R diagram (see
  Figure~\ref{fig:hrdiag}). Overall, this is a tentative orbit which
  could be improved by new observations in a couple of decades.

\item {\bf HIP 87567:} Less than half the
  orbit is covered, so the period is rather uncertain, but the current
  orbit seems reasonable as well as the derived mass sum.

\item {\bf TYC 1566-1708-1:} Triple system for which we would need the
  inner orbit to further improve on the solution. Given its northern
  declination it is a challenging target for SOAR. Poor orbit coverage
  (less than half an orbit), leading to a large uncertainty in $P$. No
  trigonometric parallax available for this target, the eventual
  addition of a Gaia parallax will be of significant help to study
  this system further.

\item {\bf HIP 89076:} This object was recently reported by Miles \&
  Mason in the IAU Double Stars circular \# 191. The listed orbital
  parameters are $P=123.84$, $T_0=2039.29$, $e=0.450$, $a=0.257$,
  $\omega=81.9$, $\Omega= 240.0$, $i= 51.4$ (no uncertainties are
  quoted). These values are {\it not} in agreement with our orbital
  parameters in Table~\ref{tab:orbel1}. In Figure~\ref{fig:compa}
  (right panel) we compare the two solutions, where we can clearly see
  that both orbits are reasonable fits to the data points, and that
  only future observation will allow us to determine a more firm sets
  of parameters. We also note that the formal mean square error of
  both solutions are quite different: Ours has 0.098, while theirs
  (using our weights) has a mean square error of 0.81. Most likely the
  difference between these two solutions is due to the incomplete
  orbit coverage and/or due to a choice of different weights per
  observation, specially on the older data. Their mass sum leads to
  1.15$M_{\odot}$, which seems reasonable for a G3V primary. On the
  other hand, our astrometric mass sum is too small for the spectral
  type (0.46$M_{\odot}$, see Table~\ref{tab:para}), but the large
  discrepancy between the trigonometric and dynamical parallaxes (see
  Figure~\ref{fig:para}), added to the rather large parallax
  uncertainty (of 1.43~mas), implies that we could, e.g., accommodate
  with our solution, a much larger mass sum (up to 3$M_{\odot}$), for
  a parallax exactly 3$\sigma$ below the published value, note also
  that this parallax would be consistent with the computed dynamical
  parallax. But, even with a parallax smaller than the published value
  by $1.3\sigma$ the mass sum quartiles for our orbit increase to
  $(0.85,1.01,1.32)M_{\odot}$. It is interesting to note that we had
  to apply several quadrant flips to the earlier data, and it is
  reassuring to see that these are the same flips adopted
  independently by Miles \& Mason, judging from the fit of their orbit
  to (our) data points.

\item {\bf HIP 89766:} First orbit. Unlike
  well constrained orbits that have localized solutions, this object
  exhibits entangled posterior distributions (e.g., in $e$ {\it vs.}
  $T$). Our solution should be considered a surrogate orbit, but
  otherwise quite uncertain (see also the large quartile mass range in
  Table~\ref{tab:para}).

\item {\bf HIP 91159):} Partial coverage
  of the orbit, rather uncertain period. Due to large period, and
  despite newer observations, should probably remain in Grade 4, as in
  the current WDS catalogue. Its orbit is shown in
  Figure~\ref{fig:examples}.

\item {\bf HIP 92726:} Small coverage of the
  orbit, long period, but relatively small range in mass quartiles
  from our MCMC simulation, and the agreement between the ML and
  quartile solutions warrants promotion to orbit of Grade 4.

\item {\bf HIP 92909:} Orbit seems well defined.

\item {\bf HIP 93519:} The mass sum is too large for its spectral type
  using the Gaia DR1 parallax ($9.48 \pm 0.25$~mas), as can be seen
  from Figure~\ref{fig:mass}. Interestingly, the Hipparcos parallax
  ($14.95 \pm 3.80$~mas) is much closer to the the dynamical parallax
  (see Figure~\ref{fig:para}). For the Hipparcos parallax the mass sum
  would be 1.9~$M_\odot$. However, note that $V_{Sys}$ is suspicious
  (see Figure~\ref{fig:photom}) and that the system´s color is
  uncertain (see Table~\ref{tab:photom}), which renders doubts about
  the dynamical parallax too, and about its true location in the H-R
  diagram. Its orbit is shown in Figure~\ref{fig:examples}.

\item {\bf HIP 96317:} Poor orbital coverage,
  long and rather indeterminate period. The highly deviant (speckle)
  point at 2006.5723 from \cite{hama09}, acquired with the Mount
  Wilson 2.5 m Hooker telescope, can not be explained (even with
  quadrant flips). Indeed, several quadrant flips were required (note
  the small $\Delta m$), but those were relatively easy to identify by
  looking at the $PA$ in reverse chronological order starting from the
  more recent data. For a trigonometric parallax smaller than
  $2\sigma$ the quoted uncertainty, the dynamical and astrometric mass
  sums would however agree at 3.5$M_\odot$.

\item {\bf HIP 99114:} First orbit. Based
  on our solution, in particular, the bounded quartile range for the
  mass sum and the agreement between dynamical and trigonometric
  parallaxes in Table~\ref{tab:para}, it is likely that $3 \times \log
  a - 2 \times \log P$ should not be grossly erroneous, and so it
  qualifies as a Grade 4 orbit in the WDS grading system.

\item {\bf HIP 102945:} Well defined
  orbit. Primary seems to have evolved off MS. Its orbit is shown in
  Figure~\ref{fig:examples}.

\item {\bf HIP 103620:} First orbit, highly
  inclined, but well defined. Its orbit is shown in
  Figure~\ref{fig:examples}.

\item {\bf HIP 107806:} First orbit. Small
  magnitude difference of the pair ($\Delta V = 0.51$, $\Delta I =
  0.2$, with several plausible quadrant flips. Quadrant
  ($PA=122.0$~deg) firmly determined from lucky imaging in last data
  point at 2015.4971 helps resolve earlier ambiguities. Large (more
  than $6\sigma$) discrepancy between the trigonometric and dynamical
  parallax. A parallax smaller by $3\sigma$ gives an astrometric mass
  sum of $1.1M_\odot$.

\item {\bf HIP 109908:} The primary seems
  to have evolved off the MS, as suggested in
  Figure~\ref{fig:hrdiag}. The spectral type for the primary is indeed
  listed as G8III in SIMBAD (hence the dynamical mass - which assumes
  class V - would be erroneous, see Figure~\ref{fig:para}), while the
  computed color and absolute magnitude for the secondary imply a
  spectral type of about A6. Still the astrometric mass seem too large
  despite of an orbit that appears relatively well determined.

\item {\bf HIP 114962:} Residuals show a
  hint of a possible sub-system, but current data does not warrant a
  solution for that. This is the only object for which we have a
  published $I_{Sim}=7.37$. The primary and secondary $I$-band
  magnitudes computed in the way described at the beginning of this
  section, leads to $I_P=8.03$ and $I_{S}=8.63$, or an equivalent
  $I_{Sys}=7.54$, which compares well with the literature value
  indicated above, considering our estimated uncertainty of~0.18~mag
  for $\Delta I$, as explained in Section~\ref{sec:sample} (and it
  even suggests that perhaps our $I$-band (and colors) errors are
  somewhat overestimated).

\end{trivlist}

We finally note that tables with the adopted weights for each data
entry (see Section~\ref{sec:sample}), the adopted quadrant flips (see
Section~\ref{sec:orbits}), and the $O-C$ values for all our solutions
can be requested from the principal author.

\newpage

\section{Conclusions} \label{sec:conc}

In 2014 we started a systematic campaign with the speckle camera HRCAM
on the 4m SOAR telescope at CTIO to observe Hipparcos binaries and
spectroscopic binaries from the Geneva-Copenhagen spectroscopic survey
in the Southern sky with the purpose of computing their orbits, and
determining their masses. This work will complement and significantly
extend the WIYN Northern sky speckle program lead by Horch and
collaborators, allowing us to compile an all-sky, volume-limited
speckle survey of these two primary samples, allowing us to
investigate effects such as metallicity and age on the MLR.

In this paper we have presented orbital elements and mass sums for
eighteen visual binary stars of spectral types B to K (five of which
are new orbits) with periods ranging from 20 to more than 500~yr, and
individual component masses with a formal uncertainty of $\sim 0.1
M_\odot$ for two double-line spectroscopic binaries with no previous
orbits using combined astrometric plus radial velocity data. Using
published optical photometry and trigonometric parallaxes, plus our
own measurements, we put these objects on an H-R diagram, and briefly
discuss their evolutionary status. Cases where one (or both)
components have evolved off of the main sequence are particularly
interesting, since for them an age determination is also possible (a
hint of this can be seen for HIP~109908, see
Figure~\ref{fig:hrdiag}). However, to do this properly, it is critical
to have not only reliable parallaxes (which Gaia will provide), but
also good multi-color photometry for the individual components - which
is challenging, specially for the tighter systems.

To compute the orbital elements we have developed a MCMC algorithm
that produces maximum likelihood estimates as well as posterior PDFs
of the parameters given the measurements that allow us to evaluate the
uncertainty of our derived orbital elements in a robust way. In the
case of the spectroscopic binaries, and inspired by the work of
\cite{wrho09} in the context of exoplanets (where the primary is
considered basically at rest), we present a mathematical formalism in
which we generalize their approach to the case of binary systems
(where both components have a sizeable motion) to achieve a
significant dimensionality reduction from seven to three dimensions in
the case of visual binaries, and from ten to seven dimensions
(including orbital parallax) in the case of spectroscopic binaries
with astrometric data. Our self-consistent solution for orbital
parallax will be particularly useful when comparing to Gaia's high
precision trigonometric parallaxes. Furthermore, this dimensionality
reduction implies that we only need to explore a reduced subset of the
parameter space, thus reducing significantly the computational
cost. The remaining parameters are determined by a simple
least-squares linear fit to the data and the significant
parameters. Although in this case we have chosen to use an MCMC
approach for parameter exploration, our formalism for dimensionality
reduction is completely general, and can be used with other parameter
exploration-based methods.

In a future paper we will apply the MCMC approach outlined here to
interesting cases where partial data (see e.g., \cite{clet16}),
non-resolutions, or other sources of information (e.g., spectral type)
are available, information which can be easily incorporated as
constraining priors into our Bayesian code, and which become crucial
specially in cases of objects with very limited observational coverage
which prevents us from estimating an orbit yet with adequate
precision, but for which one might desire to have more reliable
tentative ephemerides for observational planning.

\newpage

\section{Acknowledgments}

We acknowledge Dr. Andrei Tokovinin from CTIO and Dr. Elliot Horch
from Southern Connecticut State University for all their support
throughout this entire research, including the stages of telescope time
application, data acquisition, calibration and analysis, as well as
their suggestions for improvement to the original manuscript. We also
acknowledge Dr. Jose Angel Docobo (Universidad de Santiago de
Compostela, Spain) and Venu Kalari (FONDECYT/CONICYT Postdoctoral
Fellow Universidad de Chile) for their reading and suggestions to the
original manuscript, and the referee Dr. Dimitri Pourbaix (Institute
of Astronomy and Astrophysics, Universit\'e Libre de Bruxelles,
Brussels) for many suggestions and corrections that have significantly
improved the readability of the paper. This research has made use of
the Washington Double Star Catalog maintained at the U.S. Naval
Observatory and of the SIMBAD database, operated at CDS, Strasbourg,
France. This work has made use of data from the European Space Agency
(ESA) mission {\it Gaia} (\url{https://www.cosmos.esa.int/gaia}),
processed by the {\it Gaia} Data Processing and Analysis Consortium
(DPAC,
\url{https://www.cosmos.esa.int/web/gaia/dpac/consortium}). Funding
for the DPAC has been provided by national institutions, in particular
the institutions participating in the {\it Gaia} Multilateral
Agreement. Based on Chilean telescope time under programs CN2014B-27,
CN2015B-6, and CN2016A-4.

RAM acknowledges support from the Chilean Centro de Excelencia en
Astrof\ia sica y Tecnolog\ia as Afines (CATA) BASAL PFB/06, from the
Project IC120009 Millennium Institute of Astrophysics (MAS) of the
Iniciativa Cient\ia fica Milenio del Ministerio de Econom\ia a,
Fomento y Turismo de Chile, and from CONICYT/FONDECYT Grant
Nr. 1151213. RMC has been supported by a MSc scholarship from CONICYT,
Chile (CONICYT-PCHA/Magister Nacional/2016-22162232). MEO acknowledges
support from CONICYT/FONDECYT Grant Nr. 1170044. JSF acknowledges
support from CONICYT/FONDECYT Grant Nr. 1170854. MEO and JSF also
acknowledge support from the Advanced Center for Electrical and
Electronic Engineering, Basal Project FB0008, and from CONICYT PIA
ACT1405.

\newpage

\vspace{5mm}
\facilities{CTIO:SOAR 4.0 m}

\software{IRAF,IDL,Matlab,XGrace}

\newpage

\appendix

\section{Keplerian model equations} \label{kepler_appendix}

Calculating the position $(\rho, \theta)$ of the relative orbit (or
the equivalent Cartesian coordinates) at a certain instant of time
$\tau$ involves a sequence of steps, as described as follows:

\begin{itemize}
\item Solving Kepler's Equation\footnote{We used a Newton-Raphson
  routine to solve this equation numerically.} in order to obtain the
  eccentric anomaly $E$.
\begin{equation}
\label{eqKepler1}
2 \pi (\tau - T)/P = E - e \sin E.
\end{equation}
\item Computing the auxiliary values $x$, $y$, referred to as
  \emph{normalized coordinates} hereinafter.
\begin{eqnarray}
\label{normcoordinates}
x(E) = \cos E - e,\\
y(E) = \sqrt{1-e^2} \sin E. \nonumber
\end{eqnarray}
\item Determining the Thiele-Innes constants.
\begin{eqnarray}
A =& ~a (\cos\omega~\cos\Omega - \sin \omega~\sin\Omega~\cos i), \label{TI_constants}\\
B =& ~a (\cos\omega~\sin\Omega + \sin \omega~\cos\Omega~\cos i), \nonumber\\
F =& ~a (-\sin\omega~\cos\Omega - \cos \omega~\sin\Omega~\cos i), \nonumber\\
G =& ~a (-\sin\omega~\sin\Omega + \cos \omega~\cos\Omega~\cos i). \nonumber
\end{eqnarray}
\item Calculating the position in the apparent orbit as:
\begin{eqnarray}
\label{proj_TI}
X = B x + G y\\
Y = A x + F y. \nonumber
\end{eqnarray}
\end{itemize}

In the case of spectroscopic binaries, one aims to adjust the
Keplerian model to radial velocity data as well. This is accomplished
by a somewhat different sequence of steps:

\begin{itemize}
\item Use the $E$ value (Equation \ref{eqKepler1}) to calculate the true anomaly $\nu$ at certain epoch of observation $\tau$:
\begin{equation}
\label{eqForNu}
\tan \frac{\nu}{2} = \sqrt{\frac{1+e}{1-e}} \tan \frac{E}{2}.
\end{equation} 
\item Calculate the model's radial velocity through the following equations:
\begin{eqnarray}
V_{primary} = V_{CoM} + \frac{2 \pi a_P \sin i}{P\sqrt{1 -
    e^2}}\left[\cos(\omega + \nu) + e \cos \omega \right] = V_{CoM} +
K_P \left[\cos(\omega + \nu) + e \cos \omega
  \right], \label{eqRadVelx1}\\
V_{secondary} = V_{CoM} - \frac{2 \pi a_S \sin i}{P\sqrt{1 -
    e^2}}\left[\cos(\omega + \nu) + e \cos \omega \right] = V_{CoM} -
K_S \left[\cos(\omega + \nu) + e \cos \omega
  \right], \label{eqRadVelx2}
\end{eqnarray}
where $a_P$ is calculated as $a^{\prime\prime}/\varpi \cdot q/(1+q)$
and $a_S = a^{\prime\prime}/\varpi \cdot 1/(1+q)$, where $q\le 1$ is
the mass ratio $m_S/m_P$.
\end{itemize}

\subsection{On the dimensionality of $\vec \vartheta$} \label{sec:dimen}

Since the set of objects studied in this work makes up a relatively
long list, it seems reasonable to devote some effort to reduce the
computational costs involved in the analysis. In exploration-based
methods such as the MCMC technique, the computer time required to
obtain \emph{good} results (in terms of convergence, precision and
accuracy of the estimates) grows as the dimension of the feature space
increases. For that reason, and at the expense of not exploring the
whole seven-dimensional feature space of orbital parameters
(ten-dimensional space in the case of spectroscopic binaries), we
propose a dimensionality reduction based on the separation of the
parameter vector into two lower dimension vectors: one containing
components whose least-squares solution cannot be determined
analytically ($\vec \vartheta_{1}$); and the other containing the
components whose linear dependency\footnote{With respect to quantities
  determined by $\vec \vartheta_{1}$.} makes it possible to calculate
their least-squares solution with simple matrix algebra ($\vec
\vartheta_{2}$).

In the case of binaries with astrometric measurements only, one
exploits the linear dependency of the well-known Thiele-Innes
constants ($A$, $B$, $G$, $F$) with respect to the \emph{normalized
  coordinates} $x$, $y$ (which in turn depend on $P$, $T$, $e$ and the
collection of epochs of observation, $\{\tau_i\}_{i = 1, \dots,
  N}$). The procedure to obtain the least-squares solution of
Thiele-Innes is detailed in Appendix \ref{Appendix1}. Thus, instead of
exploring the whole 7D space, the search is focused on $\vec
\vartheta_{1} = [P, T, e]$, with $\vec \vartheta_{2} = [A, B, F, G]$
determined individually from each combination of the free parameters
in $\vartheta_{1}$. The Campbell elements $a$, $\omega$, $\Omega$, $i$
can be recovered by using equations \ref{biuni} (the detailed
procedure is shown in Appendix \ref{Appendix2}). We followed the
convention of choosing solutions with $\Omega \in (0^{\circ},
180^{\circ})$ in absence of information about the real orientation of
the orbit.

\begin{eqnarray}
tan(\omega + \Omega) = \frac{B-F}{A+G}, \label{biuni}\\
tan(\omega - \Omega) = \frac{-B-F}{A-G},\nonumber\\
a^2(1+\cos^2i) = A^2 + B^2 + F^2 + G^2, \nonumber\\
a^2 \cos^2i = AG - BF \nonumber.
\end{eqnarray}

Some definitions must be introduced before describing the approach
adopted by us for binaries with spectroscopic data. In addition to the
four parameters A, B, F, G, the Thiele-Innes representation uses
parameters $C$ to $H$ to compute the coordinates in the Z-axis (along
the line-of-sight\footnote{With some algebra, it can be verified that
  $\dot Z$ leads to Equations.~\ref{eqRadVelx1} or \ref{eqRadVelx2}.}):

\begin{equation}
\label{Z_component}
    Z= C x + H y.
\end{equation}

These quantities are defined as follows:

\begin{eqnarray}
    \label{ThieleInnesZ}
    &C = a ~\sin \omega ~\sin i, \\
    &H = a ~\cos \omega ~\sin i. \nonumber
\end{eqnarray}

In \cite{wrho09}, the authors take advantage of this representation to
propose an efficient method to fit multi-Keplerian models to purely
spectroscopic, purely astrometric, and combined data sets. The core of
their approach is the reformulation of equations \ref{eqRadVelx1} and
\ref{eqRadVelx2} in a manner such that $V_P$ and $V_S$ are linear in
the parameters, allowing for analytic calculation of least-square
solutions. Making use of some trigonometric identities, the radial
velocity equation can be expressed as:

\begin{equation}
\label{equationWright}
    V(\tau) = h \cos \nu(\tau) + c \sin \nu(\tau) + \gamma,
\end{equation}
where $h= K \cos \omega$, $c = - K \sin \omega$, $\gamma = V_{CoM} + K
\cdot e \cdot \cos \omega$. Thus, $h = H/\varpi \cdot 2\pi /
(P\sqrt{1-e^2})$, $c = -C/\varpi \cdot 2\pi / (P\sqrt{1-e^2})$.

Since that paper was targeted at exoplanet research, each body
involved is modeled with an independent Keplerian orbit, omitting the
influence that each component of the system exerts on the
other. However, that influence is not negligible when analyzing
objects with masses of similar order of magnitude, and therefore that
approach is not directly applicable to binary stars. Concretely, when
analyzing binary stars, the conditions shown below must be met, making
the orbital parameters of the primary and those of the secondary
interdependent.

\begin{eqnarray} \label{conditionsBinStars}
a~=& a_P + a_S,\\
\frac{a_P}{a_S} =& \displaystyle \frac{m_S}{m_P} = q, \nonumber
\end{eqnarray}

The equalities above impose constraints on the parameters being
estimated: if we reformulate equations \ref{eqRadVelx1},
\ref{eqRadVelx2} according to the parameterization presented in
Equation \ref{equationWright}, then $h_P = H_P/\varpi \cdot 2\pi /
(P\sqrt{1-e^2})$, $c_P = -C_P/\varpi \cdot 2\pi / (P\sqrt{1-e^2})$,
being $H_P = a_P \cos \omega \sin i = \frac{q}{1+q} H$, $C_P = a_P
\sin \omega \sin i = \frac{q}{1+q} C$ (analogous equations for $V_S$:
specifically, $H_S = \frac{1}{1+q} H$, $C_S = \frac{1}{1+q} C$). The
strict relations that $(H_P, C_P)$ and $(H_S, C_S)$ must comply
(namely, $H = H_P + H_S$, $C = C_P + C_S$, $H_P/H_S = C_P/C_S = q$) do
not stem naturally when calculating these quantities as free
parameters. Therefore, those conditions must be enforced as an
additional mathematical restriction in the model.

Although one could address the interdependence problem raised in the
previous paragraph using Lagrange multipliers, there is no guarantee
that the resulting set of non-linear equations will be analytically
tractable (it may even have no unique solution). However, one can
manipulate the formulae in a manner such that both orbital and radial
velocity values of the Keplerian model are expressed as a linear
combination of parameters, and meet the restrictions mentioned in the
paragraph above at the same time:

\begin{itemize}
\item In an approach similar to that used in \cite{wrho09}, the first
  step is to use a combination of H and C -- which are simpler
  expressions -- to reconstruct parameters A, B, F and G (this
  requires the aim of trigonometric functions of $\Omega$ and $i$):
    \begin{equation}
    \includegraphics[width=.5\textwidth]{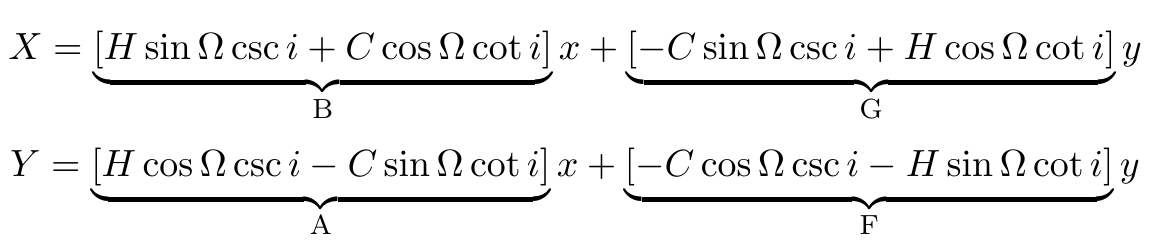}
    \end{equation}

\item Grouping the terms multiplying $C$ and $H$ yields:

    \begin{equation}
    \includegraphics[width=.5\textwidth]{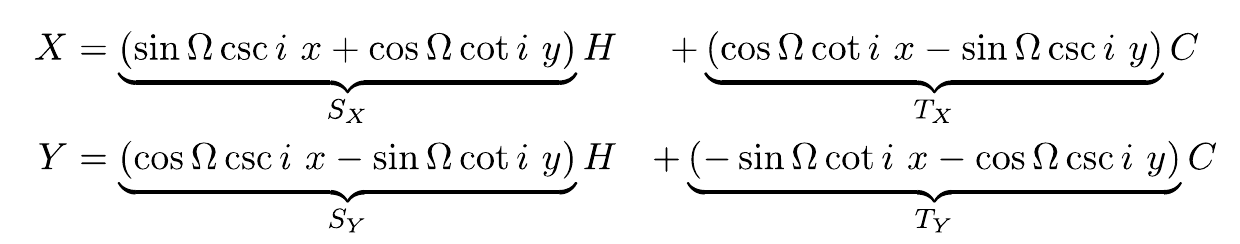}
    \end{equation}

Thus, the coordinates X, Y can be written as a linear combination of
terms $S_X$, $T_X$, $S_Y$, $T_Y$ (which can be easily computed from
$x$, $y$, $\Omega$ and $i$), being $H$ and $C$ their accompanying
constants.

\item Finally, by using $\lambda_P = \frac{q}{1+q}\cdot\frac{2
  \pi}{\varpi P \sqrt{1 - e^2}}$, $\lambda_S =
  \frac{1}{1+q}\cdot\frac{2 \pi}{\varpi P \sqrt{1 - e^2}}$ to
  transform $H$, $C$ into $h_P$, $h_S$, $c_P$, $c_S$, one can express
  both the astrometric coordinates (Equation \ref{proj_TI}) and radial
  velocity values (equations \ref{eqRadVelx1}, \ref{eqRadVelx2}) in
  terms of a vector of parameters $\vec \vartheta_{2} = [H, C,
    V_{CoM}]$:

\begin{equation}
\vec \vartheta_{2} \cdot \mathbf{F} = [\vec{X}^{model}, \vec{Y}^{model},
  \vec{V_P}^{model}, \vec{V_S}^{model}],
\end{equation}

where $\mathbf{F}$ is:

    \begin{equation}
    \includegraphics{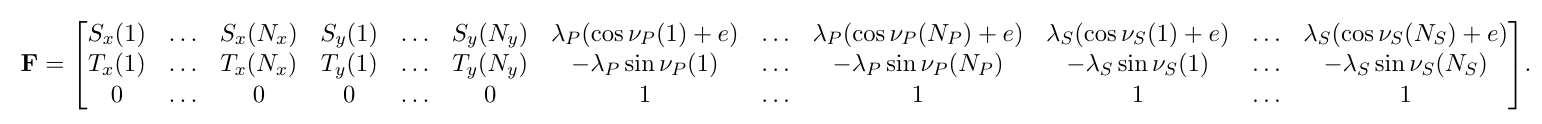}
    \end{equation}

This allows for the calculation of the least-squares solution for
$\vec \vartheta_{2}$ as (see, e.g., \cite{kay93}):

    \begin{equation}
    \label{eqBeta}
    \vec \vartheta_{2} = \vec{x} ~ \mathbf{W} \mathbf{F}^T
    (\mathbf{F}\mathbf{W}\mathbf{F}^T)^{-1},
    \end{equation}

where $\vec{x} = [\vec{X}^{obs}, \vec{Y}^{obs}, \vec{V_P}^{obs},
  \vec{V_S}^{obs}]$ is the data vector and $\mathbf{W}$ is a diagonal
matrix with the weight of each observation. From the resulting
$\hat{H}$ and $\hat{C}$ values (the values with a hat represent a
particular estimate of that quantity, based on the current data), the
parameters $a$ and $\omega$ can be recovered as follows:

    \begin{eqnarray}
    \label{recovering_a_and_w}
    \hat{a} &= \displaystyle\sqrt{\frac{\hat{C}^2 + \hat{H}^2}{\sin^2 i}},\\
    \hat{\omega} &= \displaystyle\tan^{-1} \left(\frac{\hat{C}}{\hat{H}}\right),
    \end{eqnarray}

The third component of $\vec \vartheta_{2}$ ($V_{CoM}$) has direct
physical meaning and does not need to be transformed. Under this
scheme, only seven parameters ($P$, $T$, $e$, $\Omega$, $i$, $q$,
$\varpi$) must be explored and estimated, whereas $a$, $\omega$ and
$V_{CoM}$ are calculated analytically. Although in this work we use
the MCMC technique, the representation developed here -- and the
dimensionality reduction that it involves -- can be applied to other
sorts of methods as well, even if they are not strictly
exploration-based, such as the Levenberg-Marquardt algorithm.
\end{itemize}

\section{On Thiele-Innes and Campbell elements}

\subsection{Least-squares estimate} \label{Appendix1}

The starting point is the sum of individual errors:

\begin{equation} \label{MSEAppendix}
 \sum_{k=1}^{N_{x}} \frac{1}{\sigma_x^2(k)} [X(k) - X^{model}(k)]^2 +
 \sum_{k=1}^{N_y} \frac{1}{\sigma_y^2(k)} [Y(k) - Y^{model}(k)]^2
\end{equation}

Equation \ref{proj_TI} enables us to replace $X_{model}$, $Y_{model}$
with their analytic expression for any epoch (indexed by $k$):

\begin{eqnarray}
X_{obs}(k) - X_{model}(k) &= X_{obs}(k) - [B\cdot x(k) + G \cdot
  y(k)] \label{eqDifferenceObsComp}\\
Y_{obs}(k) - Y_{model}(k) &= Y_{obs}(k) - [A\cdot x(k) + F \cdot
  y(k)]. \nonumber
\end{eqnarray}

Given the linear dependency of $X_{model}$, $Y_{model}$ with respect
to the normalized coordinates $x$, $y$, it is possible to calculate a
least-squares estimate for the unknown variables $B$, $G$, $A$, $F$ in
a non-iterative way. Moreover, the first term of Equation
\ref{MSEAppendix} depends only on the pair $(B, G)$, whereas the
second term depends on the pair $(A, F)$. Therefore, the estimate for
$(B, G)$ is obtained by minimizing the first term and the estimate for
$(A, F)$ by minimizing the second one, independently. The problem is
thus reduced to a pair of uncoupled linear equations. By calculating
the derivatives of the expression of the error with respect to each of
the Thiele-Innes constants and making the results equal to zero, one
can obtain the following formulae (for the sake of briefness, a set of
auxiliary terms is introduced first):

\begin{eqnarray}\nonumber
\alpha = \sum_i~w_i~x(i)^2 ~~~~~~& \beta = \sum_i~w_i~y(i)^2 & \gamma
= \sum_i~w_i~x(i)~y(i)\nonumber\\
~ & r_{11} = \sum_i~w_i~X_{obs}(i)~x(i) & r_{12} =
\sum_i~w_i~X_{obs}(i)~y(i)\\ \label{auxConstantsLS}\nonumber
~ & r_{21} = \sum_i~w_i~Y_{obs}(i)~x(i) & r_{22} =
\sum_i~w_i~Y_{obs}(i)~y(i) \nonumber
\end{eqnarray}

Then, the least-squares estimate for the Thiele-Innes set of
parameters is calculated as follows:

\begin{eqnarray}
\hat{B} = \frac{\beta \cdot r_{11} - \gamma \cdot r_{12}
}{\Delta},~~~~~~& \hat{G} = \frac{\alpha \cdot r_{12} - \gamma \cdot
  r_{11} }{\Delta},\\
\hat{A} = \frac{\beta \cdot r_{21} - \gamma \cdot r_{22}
}{\Delta},~~~~~~& \hat{F} = \frac{\alpha \cdot r_{22} - \gamma \cdot
  r_{21} }{\Delta}, \nonumber \nonumber
\end{eqnarray}

where $\Delta = \alpha \cdot \beta - \gamma^2$.

\subsection{Conversion from Thiele-Innes to Campbell constants} \label{Appendix2}

Once the estimates ($\hat{B}$, $\hat{G}$, $\hat{A}$, $\hat{F}$) for
the Thiele-Innes constants are obtained, it is necessary to recover
the equivalent representation in terms of the Campbell elements ($a,
\omega, \Omega, i$). For $\omega$ and $\Omega$, one must solve the
following set of equations:

\begin{eqnarray}
  \omega + \Omega &= \arctan\left( \frac{B-F}{A+G}\right),\\
  \omega - \Omega &= \arctan\left( \frac{-B-F}{A-G}\right), \nonumber
\end{eqnarray}
choosing the solution that satisfies that $\sin (\omega + \Omega)$ has
the same sign as $B - F$ and that $\sin (\omega - \Omega)$ has the
same sign as $-B-F$. If that procedure outputs a value of $\Omega$
that does not satisfy the convention that $\Omega \in (0, \pi)$, it
must be corrected in the following way: if $\Omega < 0$, values of
$\omega$ and $\Omega$ are modified as $\omega = \pi + \omega$, $\Omega
= \pi + \Omega$; whereas if $\Omega > \pi$, values of $\omega$ and
$\Omega$ are modified as $\omega = \omega - \pi$, $\Omega = \Omega -
\pi$.

For semi-major axis $a$ and inclination $i$, the following auxiliary
variables must be calculated first:

\begin{eqnarray}
    k &= \frac{A^2 + B^2 + F^2 + G^2}{2},\nonumber\\
    m &= A \cdot G - B \cdot F, \\
    j &= \sqrt{k^2-m^2}. \nonumber
\end{eqnarray}

Then, $a$ and $i$ are determined with the following formulae:

\begin{eqnarray}
    a &= \sqrt{j+k}\\
    i &= \arccos \left(\frac{m}{a^2}\right) \nonumber
\end{eqnarray}

\section{Algorithms for parameter estimation} \label{appendix_DEMC}

\begin{figure}[!hbt] 
    \centering
    \includegraphics[height=.4\textheight]{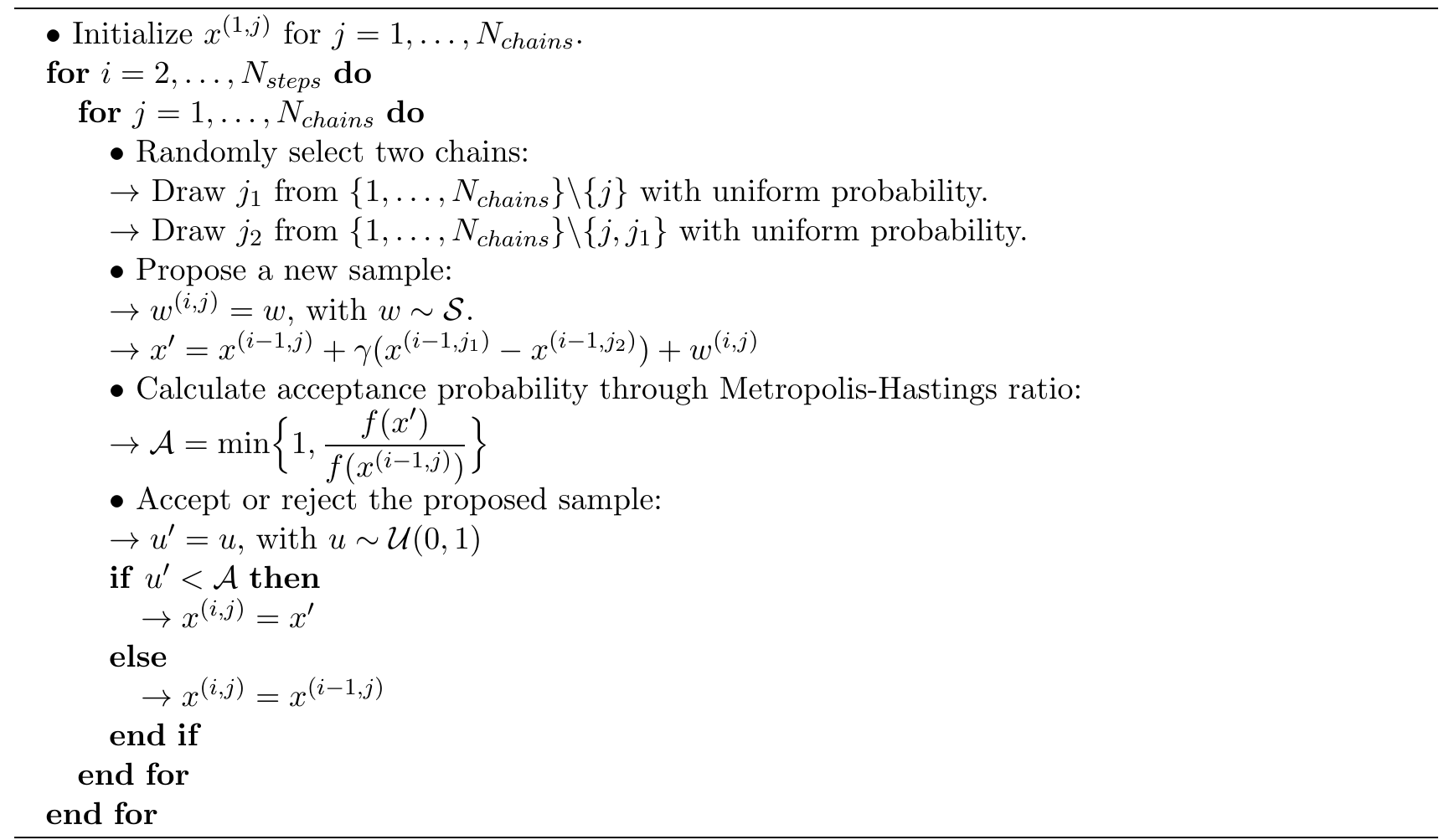}
\caption{Differential Evolution Markov Chain: This algorithm relies on
  running several chains in parallel, and letting them learn from each
  other. To do so, on each iteration $i$ and for each chain $j$, two
  different chains $j_1 \neq j$ and $j_2 \neq j$ are chosen at random,
  and the difference between their current states, $x^{(i-1,j_1)} -
  x^{(i-1,j_2)}$, is used to propose new samples $x^{\prime}$, which
  are accepted (or rejected) according to the Metropolis-Hastings
  criterion. Since we perform a dimensionality reduction in the
  parameter vector of visual binaries, $x \equiv \vec
  \vartheta_1 = [P, T^{\prime}, e]$ in our implementation. The
  pseudo-code in this figure details the procedure. \label{alg_DEMC}}
\end{figure}

Figure \ref{alg_DEMC} outlines DE-MC procedure used for visual
binaries, whereas the pseudo-code in Figure \ref{alg_MH_Gibbs}
describes the Gibbs sampler used for spectroscopic binaries.

If the parameter vector is $x = [x_1, \dots, x_d]$, then Gibbs
sampler operates as follows:

\begin{figure}[!hbt] 
\centering
\includegraphics[height=.3\textheight]{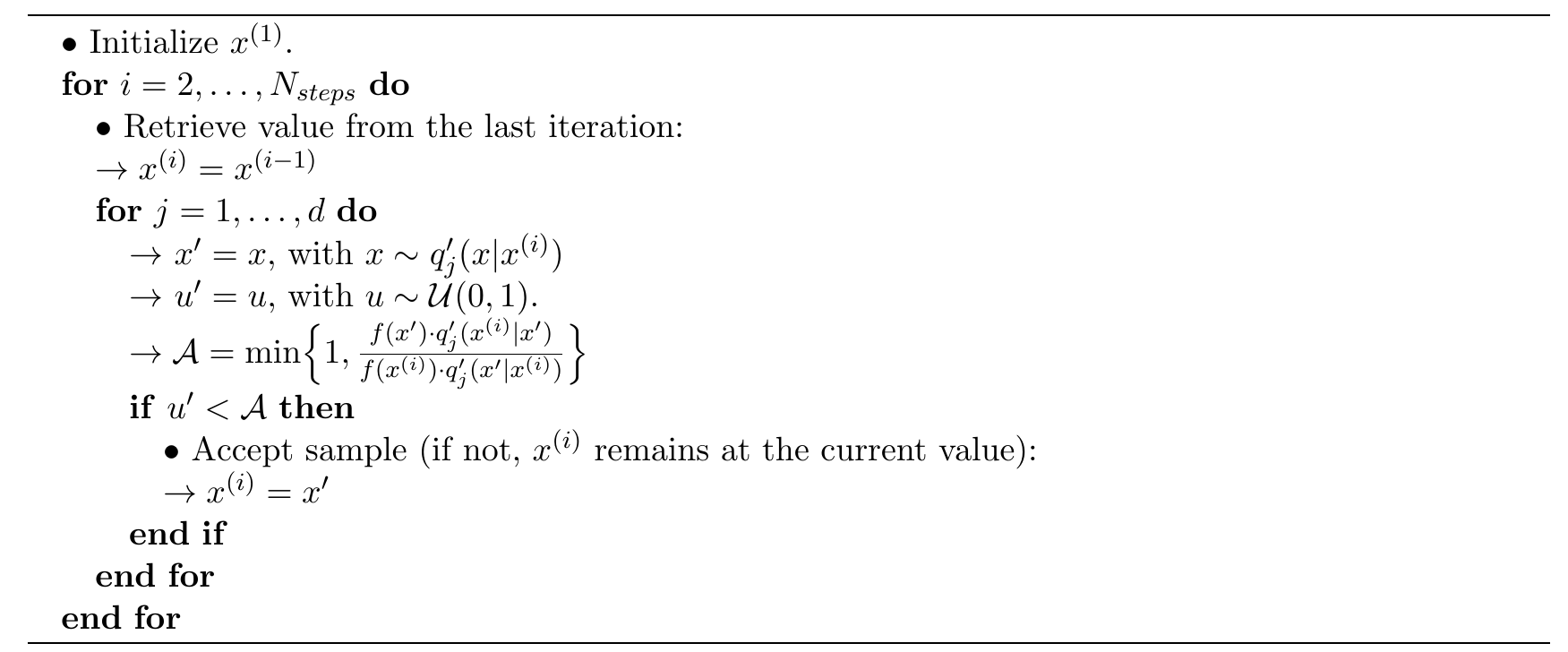}
\caption{To sample the posterior distribution of spectroscopic
  binaries, we have used a Metropolis-Hastings-within-Gibbs
  approach. A proposal distribution $q_j^{\prime}$ is defined: let
  $x^{(i)}_{-j}$ be $[x_1^{(i)}, \dots, x_{j-1}^{(i)}, x_{j+1}^{(i)},
    \dots, x_d^{(i)}]$ (all components other than $j$), then
  $q_j^{\prime}(x|x^{(i)})$ is a distribution that induces leaps only
  on the $j$-th component (i.e., $x_{-j}$ remains equal to
  $x_{-j}^{(i)}$, while $x_{j}$ is the result of a random variation on
  $x_{j}^{(i)}$). Thus, the Metropolis-Hastings-within-Gibbs algorithm
  operates as shown in this pseudo-code, where $x \equiv \vec
  \vartheta_1 = [P, T^{\prime}, e, \omega, i, q, \varpi]$ as our
  vector of interest. \label{alg_MH_Gibbs}}
\end{figure}

\allauthors

\end{document}